\documentclass{ws-rv961x669}

\usepackage[square]{ws-rv-van}

\usepackage{url}

\usepackage{hyperref}

\def\cropnote{\relax}

\usepackage{bm}
\usepackage{slashed}
\usepackage{subfigure}

\newcommand{\Tdu}[5]
{{#1}_{#2 \phantom{#3} #4 \phantom{#5}}^{\phantom{#2} #3 \phantom{#4}  #5}}

\begin{document}


\chapter*{Flavor effects in leptogenesis}
\author[]{P.~S.~B.~Dev${}^{\ast}$, P.~Di Bari${}^{\dagger}$, B.~Garbrecht${}^{\ddagger}$, S.~Lavignac${}^{\S}$,\\ P.~Millington${}^{\P}$\footnote{Corresponding Author.}, D.~Teresi${}^{\parallel}$}
\address{${}^{\ast}$ Department of Physics and McDonnell Center for the Space Sciences,\\ Washington University, St.~Louis, MO 63130, USA\\[3pt]
${}^{\dagger}$ Physics and Astronomy, University of Southampton,\\ Southampton, SO17 1BJ, UK\\[3pt]
${}^{\ddagger}$ Physik Department T70, Technische Universit\"{a}t M\"{u}nchen,\\
James-Franck-Stra\ss e, 85748 Garching, Germany\\[3pt]
${}^{\S}$ Institut de Physique Th\'eorique,
Universit\'e Paris Saclay,\\ CNRS, CEA,
F-91191 Gif-sur-Yvette, France\\[3pt]
${}^{\P}$ School of Physics and Astronomy, University of Nottingham,\\ Nottingham NG7 2RD, UK\\[3pt]
${}^{\parallel}$ Service de Physique Th\'{e}orique, Universit\'{e} Libre de Bruxelles,\\[-2pt] Boulevard du Triomphe, CP225, 1050 Brussels, Belgium
\\[3pt]
${}^{1}$ p.millington@nottingham.ac.uk}


\begin{abstract}
\textbf{Abstract}: Flavor effects can have a significant impact on the final estimate of the lepton (and therefore baryon) asymmetry in scenarios of leptogenesis. It is therefore necessary to account fully  for this flavor dynamics in the relevant transport equations that describe the production (and washout) of the asymmetry. Doing so can both open up and restrict viable regions of parameter space relative to the predictions of more approximate calculations. In this review, we identify the regimes in which flavor effects can be relevant and illustrate their impact in a number of phenomenological models. These include type I and type II seesaw embeddings, and low-scale resonant scenarios. In addition, we provide an overview of the semi-classical and field-theoretic methods that have been developed to capture flavor effects in a consistent way.
\end{abstract}


\newpage

\body

\tableofcontents


\section{Introduction}

The realization of the importance of flavor effects~\cite{Abada:2006fw, Nardi:2006fx, Abada:2006ea, Blanchet:2006be, Pascoli:2006ie, DeSimone:2006nrs} represents one of the most significant developments in leptogenesis since its original proposal~\cite{Fukugita:1986hr} as a viable mechanism for generating the observed baryon asymmetry of the Universe. The flavor effects to which we refer can be associated with either of the following:
\begin{itemize}

\item [(i)] Non-vanishing off-diagonal elements in the charged-lepton Yukawa couplings and their couplings to the mediator of the relevant $L$-violating Weinberg operator.

\item [(ii)] Non-vanishing coherences in the off-diagonal elements of the particle number densities of species carrying flavor quantum numbers.

\end{itemize}
The former are a property of the renormalized Lagrangian of the model and arise from misalignment of the flavor and mass eigenbases; the latter are a property of the primordial plasma and arise from the quantum statistical mechanics of a system with particle mixing. Throughout this review, we will refer to flavor effects arising from the contribution of additional  heavy, right-handed (RH) neutrino species as {\em heavy-neutrino flavor effects} and to those related to charged-lepton flavors as  {\em charged-lepton flavor effects}, and we will see that a general description must  take both into account.  For earlier reviews that discuss the issue of flavor effects in leptogenesis, see, e.g., Refs.~\cite{Pilaftsis:1998pd, Davidson:2008bu, Blanchet:2012bk, Fong:2013wr}.

Coherences amongst the charged-lepton flavors play an important role in the dynamics of the washout of the asymmetry, and this is of particular importance for high-scale scenarios such as thermal leptogenesis. On the other hand, coherences amongst the heavy-neutrino flavors have an important effect on the source of CP asymmetry due to oscillations. Whilst oscillations are suppressed for hierarchical heavy-neutrino mass spectra, they become important when the heavy-neutrino masses become quasi-degenerate, and this has significant implications for scenarios of resonant leptogenesis, discussed further in Chapter~\cite{leptogenesis:A03} of this review. Successful leptogenesis can, in fact, be driven entirely by oscillations through the ARS mechanism~\cite{Akhmedov:1998qx}, and these scenarios are discussed in detail in Chapter~\cite{leptogenesis:A02} of this review. In certain regimes, accounting systematically for all relevant flavor effects can both enhance and suppress the final asymmetry compared to treatments in which they are only partially captured. Moreover, aside from their impacts upon the generated asymmetry, flavor effects can be key to realising scenarios of leptogenesis that are directly testable at current and near-future experiments both at the energy and intensity frontiers.

There have been significant efforts in the literature to develop theoretical frameworks and calculational techniques that allow flavor effects to be captured in a systematic way. These efforts span both first-principles field-theoretic and more phenomenologically-inspired semi-classical approaches. The former are based on the Kadanoff-Baym formalism~\cite{Baym:1961zz, KadanoffBaym}, itself embedded within the Schwinger-Keldysh~\cite{Schwinger:1960qe, Keldysh:1964ud} closed-time-path approach of non-equilibrium field theory. The latter --- often referred to as the density matrix formalism~\cite{Dolgov:1980cq, Stodolsky:1986dx, Raffelt:1992uj, Sigl:1992fn} --- can be derived at the operator level by means of the Liouville-von Neumann and Heisenberg equations. A more comprehensive overview of recent developments in field-theoretic approaches is provided in the companion Chapter~\cite{leptogenesis:A03}.

The outline of this review is as follows. In~\sref{sec:methods}, we discuss the regimes in which flavor effects are relevant. We then provide a brief overview of calculational methods that can account for these effects in the relevant transport equations that describe the production of the asymmetry. Having summarized the necessary theoretical tools, we proceed to illustrate the importance of flavor effects in the context of a number of phenomenological models. In~\sref{sec:typeI}, we consider thermal leptogenesis in the type I seesaw scenario; in~\sref{sec:RL}, we move on to low-scale scenarios of resonant leptogenesis; and finally, in~\sref{sec:typeII}, we discuss type II seesaw models. We briefly outline the relevance of flavor effects in other models in~\sref{sec:other}, and our conclusions are presented in~\sref{sec:conclusions}.


\section{Flavor effects and calculational methods}
\label{sec:methods}

In this section, and before proceeding to discuss the role of flavor effects in particular scenarios of leptogenesis, we first review the regimes in which flavor effects are important. We will also briefly outline the frameworks that allow these flavor effects to be captured fully in the Boltzmann-like equations that describe the generation of the asymmetry.  We will discuss two in particular: semi-classical methods based on the so-called density matrix formalism~\cite{Sigl:1992fn} and field-theoretic approaches based on the Kadanoff-Baym formalism~\cite{Baym:1961zz, KadanoffBaym}.


\subsection{Flavored regimes}
\label{sec_regimes}

The Lagrangian
\begin{align}
\label{eq:1_lagrangian_general}
  \mathcal{L} \ &=\ \mathcal{L}_{{\rm SM},h_\beta=0} \: +\:i\overline{N_{\!Rk}}\slashed{\partial}N_{Rk}\nonumber\\&\qquad - \: \left( h_\beta \,\overline{\ell}_{\beta}\,\phi\, e_{R\beta} \: 
    + \: \lambda_{\alpha k}\,\overline{\ell}_{\alpha}\,\phi^cN_{Rk}\:
    + \:\frac{1}{2}\,\overline{N_{\!Rk}^{c}}M_kN_{Rk}\: 
    + \:\text{h.c.}\right)
\end{align}
selects the mass eigenstates of the charged leptons as a preferred basis. However, in order to understand flavor effects in leptogenesis and how they can be neglected at very high temperatures, we would like to use the freedom of basis transformations among the lepton doublets $\ell$. Therefore, we promote the Standard Model (SM) Yukawa couplings to a matrix, viz.~$h_\beta \,\overline{\ell}_{\beta}\,\phi\, e_{R\beta} \to h_{\alpha\beta} \,\overline{\ell}_{\alpha}\,\phi\, e_{R\beta}$, where $h_{\alpha\beta}$ is diagonal in the flavor basis. Whilst we use the same symbol for the flavor-covariant matrix and the vector in the fixed flavor basis, it will be clear from the context to which object is referred. In addition, we have explicitly identified the chirality of the right-handed singlets $N_{Rk}$ in order to distinguish them from the physical Majorana fields $N=N^c$, discussed later (see \sref{sec:typeI}).

\begin{figure}[t!]
\centering
\includegraphics[width=1.\textwidth]{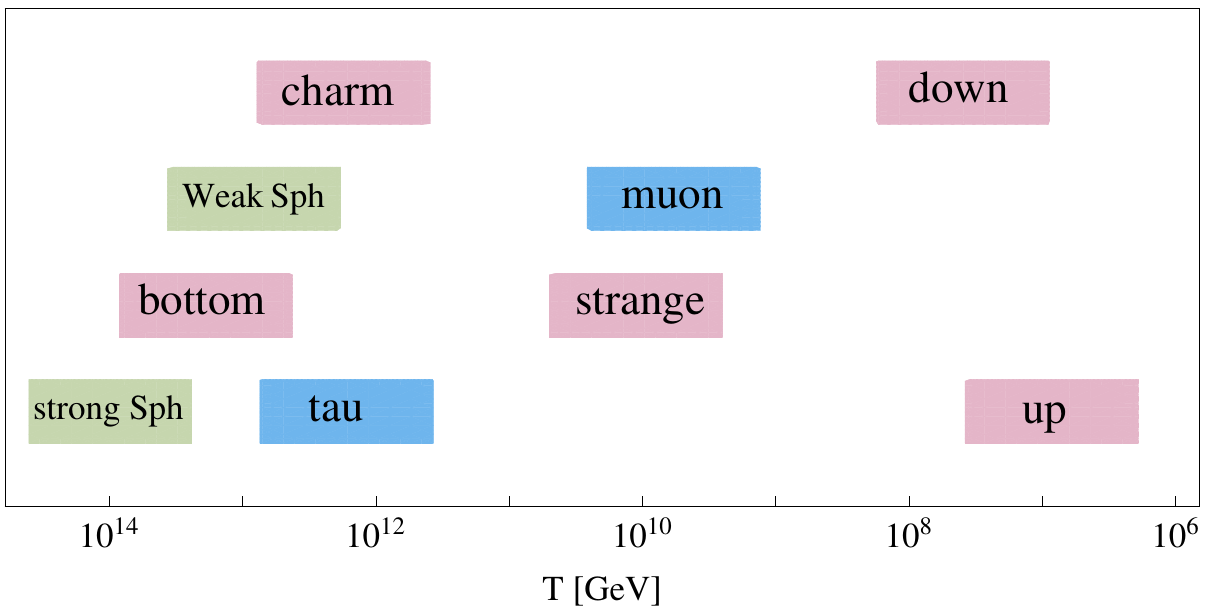}
\caption{Ranges of equilibration temperature for various SM processes, i.e.~for the strong and weak sphalerons (green), as well as quark (red) and lepton (blue) Yukawa interactions. The bands range from $T_X$ to $20\,T_X$, with $T_X$ denoting the equilibration temperature, at which the particular rate coincides with the Hubble rate. Figure taken from Ref.~\cite{Garbrecht:2014kda}.\label{fig:regions}}
\end{figure}

Flavor-sensitive rates in the early Universe should scale as $|h_{\alpha\alpha}|^2 T$, where $T$ is the temperature. These are suppressed by a phase-space factor also involving gauge couplings~\cite{Garbrecht:2013urw} because the leading processes at high temperature are two-by-two scatterings involving gauge-boson radiation, cf.~\eref{Gamma:fl} and~\eref{gammafl}. These rates are to be compared with the Hubble rate $H$, which scales as $H\sim T^2/M_{\rm Pl}$, where $M_{\rm Pl}$ is the Planck mass. Doing so implies that  flavor-sensitive processes are out of equilibrium above and in equilibrium below a certain temperature. The equilibration temperatures for various SM processes, relevant for flavor and spectator effects in leptogenesis, as well as in other cosmological scenarios, are shown in~\fref{fig:regions}. It should be noted, however, that the ranges are only indicative because loopholes can easily be found. For example, and as discussed in~\sref{sec3}, a scenario with largely hierarchical RH-neutrino Yukawa couplings can be constructed where the partial decoherence of correlations involving the $\tau$ flavor is important even when leptogenesis occurs at a low temperature due to comparably light RH neutrinos and a resonantly-enhanced CP asymmetry.

However, barring extra symmetries or tuning in the type I seesaw-model, the standard picture of flavored regimes is as follows: Suppose first that leptogenesis occurs at temperatures below $10^9\,{\rm GeV}$ from the decay of the lightest RH neutrino $N_1$ (see~\sref{subsec:N1dominated} for more details). In general, the decay creates a coherent superposition of all three lepton-doublet flavors $e$, $\mu$ and $\tau$. These superpositions can be described by off-diagonal elements that appear either in a description based on two-point correlation functions in the Schwinger-Keldysh formalism or within a matrix of number densities based on an operator formalism. Nevertheless, the flavor-sensitive rates will lead to a rapid decay of these off-diagonal correlations such that they can be ignored. It is therefore most suitable to simply remain in the mass eigenbasis where the Yukawa couplings of the charged leptons are diagonal.

Next, consider the opposite regime, where leptogenesis occurs at temperatures above $10^{14}\,{\rm GeV}$. If we remain in the mass eigenbasis, we can no longer ignore the flavor correlations, which amounts to a calculational inconvenience. The latter can, however, be removed by a flavor transformation of the doublet leptons, such that $N_1$ only couples to one of the doublet leptons in the new basis:
\begin{align}
  \left(
  \begin{array}{c}
  u_{\perp 1}\\
  u_{\perp 2}\\
  u_\parallel
  \end{array}
  \right)
  \left(
  \begin{array}{ccc}
  \lambda_{e 1} & \lambda_{e 2} &\lambda_{e 3}\\
  \lambda_{\mu 1} & \lambda_{\mu 2} &\lambda_{\mu 3}\\
  \lambda_{\tau 1} & \lambda_{\tau 2} &\lambda_{\tau 3}\\
  \end{array}
  \right)\
  = \
  \left(
  \begin{array}{ccc}
  0 & \times &\times\\
  0 & \times &\times\\
  \times & \times &\times\\
  \end{array}
  \right)
  \;,
\end{align}
where $\times$ denotes a non-vanishing entry,
\begin{align}
  u_\parallel\ =\ \frac{\left(
  \begin{array}{ccc}\lambda_{e 1}, & \lambda_{\mu 1}, &  \lambda_{\tau 1}
  \end{array}
  \right)}{\sqrt{\sum|\lambda_{\alpha 1}|^2}}
\end{align}
and $u_{\perp 1,2}$ are unit vectors perpendicular to $u_\parallel$, as well as to one another. In this description, we only need to consider the flavor aligned with $u_\parallel$ and can ignore the $\perp$ flavors altogether because no asymmetry is generated within these in the first place.

Finally, consider the narrow regime between $\tau$ and $\mu$ equilibration (around $10^{11}\,{\rm GeV}$), where we suitably transform
\begin{align}
  \left(
  \begin{array}{c}
  u_{\perp}\\
  u_{\parallel}\\
  (\begin{array}{ccc}0&0&1\end{array})
  \end{array}
  \right)
  \left(
  \begin{array}{ccc}
  \lambda_{e 1} & \lambda_{e 2} &\lambda_{e 3}\\
  \lambda_{\mu 1} & \lambda_{\mu 2} &\lambda_{\mu 3}\\
  \lambda_{\tau 1} & \lambda_{\tau 2} &\lambda_{\tau 3}
  \end{array}
  \right)\
  =\
  \left(
  \begin{array}{ccc}
  0 & \times &\times\\
  \times & \times &\times\\
  \lambda_{\tau 1} & \lambda_{\tau 2} &\lambda_{\tau 3}\\
  \end{array}
  \right)
  \;,
\end{align}
in which
\begin{align}
  u_{\parallel}\ =\ \frac{(
  \begin{array}{ccc}
  \lambda_{e 1}, & \lambda_{\mu 1}, & 0
  \end{array})}
  {\sqrt{|\lambda_{e 1}|^2+|\lambda_{\mu 1}|^2}}\;,\quad
  u_{\perp}\ =\ \frac{(
  \begin{array}{ccc}
  \lambda_{\mu 1}, & -\,\lambda_{e 1}, & 0
  \end{array})}
  {\sqrt{|\lambda_{e 1}|^2+|\lambda_{\mu 1}|^2}}\;.
\end{align}
In this setup, asymmetries are produced within the $\tau$ flavor and the flavor aligned with $u_\parallel$, and there are no correlations amongst these because any such correlations are destroyed by interactions mediated by $h_\tau$. No asymmetries are generated in the flavor aligned with $u_\perp$, which can therefore be ignored.

This leaves open the questions of how to deal with intermediate regimes and whether the above procedures can be obtained as limiting cases of a more general approach that allows to treat flavor effects throughout the entire temperature range. This will be addressed in~\sref{subsec:methods}.


\subsection{Calculational methods}
\label{subsec:methods}

In order to calculate the final lepton asymmetry, we need to describe the evolution of integrated particle number densities, $n\equiv n(t)$, in the expanding Universe~\cite{Kolb:1979qa,Luty:1992un}. This evolution is described semi-classically by coupled systems of Boltzmann equations, which take the general form
\begin{equation}
  \label{3_generalBoltzmann}
  \dot{n}_{A}\:+\:3 H n_{A}\ =\ \mathcal{C}_{A}[\{f\}]\;,
\end{equation}
where $\dot{}$ indicates a derivative with respect to cosmic time $t$ and $H$ is the Hubble rate. The subscript $A$ is a multi-index, which labels all species and their quantum numbers, i.e.~flavor, spin/helicity, isospin and so on. For our present discussions, the most important of these will be flavor. The terms on the left-hand side of~\eref{3_generalBoltzmann} are the so-called \emph{drift terms}, which include the effect of the cosmological expansion, and the \smash{$\mathcal{C}_A[\{f\}]$} on the right-hand side of~\eref{3_generalBoltzmann} are the \emph{collision terms}. The latter depend, in general, on the phase-space distribution functions $f_A$, which we define below. The remainder of this section will be concerned with the derivation of these collision terms in the flavored regime, where we must carefully treat the quantum-mechanical effects of particle mixing. Further discussion of the treatment of these effects in the context of resonant leptogenesis can be found in Chapter~\cite{leptogenesis:A03} of this review.

The first step in obtaining the requisite systems of Boltzmann-like equations is to determine what it is that we aim to count. These are the distribution functions $f_A\equiv f_A(\mathbf{p},\mathbf{X},t)$: the densities of particles in phase space. Throughout what follows, we assume spatial homogeneity, such that the distribution functions depend only on time $t$ and three-momentum $\mathbf{p}$. Given a single scalar degree of freedom, the distribution function is straightforwardly related to the number operator, itself built out of the canonical creation and annihilation operators $\hat{a}^{\dag}(\mathbf{p})$ and $\hat{a}(\mathbf{p})$. Working in the interaction picture, we have
\begin{equation}
  f(t,\mathbf{p})\ \equiv\ \big<\hat{n}(\mathbf{p})\big>_t\ \equiv\ \frac{1}{V}\mathrm{tr}\,\hat{\rho}(t)\hat{a}^{\dag}(\mathbf{p})\hat{a}(\mathbf{p})\;,
\end{equation}
where $\hat{\rho}(t)$ is the density operator ($\mathrm{tr}\,\rho(t)=1$) and $V=(2\pi)^3\delta^{(3)}(\mathbf{0})$ is the three-volume of the system. In the presence of multiple flavors, we might be tempted to add to these distribution functions a flavor index, $i$ say, such that
\begin{equation}
  f_i(t,\mathbf{p})\ \equiv\ \big<\hat{n}_i(\mathbf{p})\big>_t\ \equiv\ \frac{1}{V}\mathrm{tr}\,\hat{\rho}(t)\hat{a}_i^{\dag}(\mathbf{p})\hat{a}_i(\mathbf{p})\;.
\end{equation}
However, in the presence of particle mixing, such an extension is incomplete, and we must introduce matrices of distribution functions that count both the diagonal densities of individual flavors but also the coherences between those different flavors:
\begin{equation}
  f_{ij}(t,\mathbf{p})\ \equiv\ \big<\hat{n}_{ij}(\mathbf{p})\big>_t\ \equiv\ \frac{1}{V}\mathrm{tr}\,\hat{\rho}(t)\hat{a}_j^{\dag}(\mathbf{p})\hat{a}_i(\mathbf{p})\;.
\end{equation}
More generally, it may be necessary to count other individual quantum numbers, for example, helicity, as well as the corresponding coherences. The integrated number densities are of the form
\begin{equation}
  \label{eq:numbdenmatrix}
  n_{Xij}(t)\ =\ \sum_{q}\int\!\frac{\mathrm{d}^3\mathbf{p}}{(2\pi)^3}\;f_{Xij,q}(\mathbf{p},t)\;,
\end{equation}
where $X$ labels the particle species and the sum over $q$ includes all additional quantum numbers that we do not wish to track explicitly.

It is clear now what the relevant multi-indices $A$ and $B$ are in~\eref{3_generalBoltzmann}; they run over the particle species of interest and their corresponding flavor structure. Hence, the coupled Boltzmann equations for the fermionic species are
\begin{subequations}
\begin{gather}
  \label{3_generalBoltzmann2}
  \dot{n}_{Nij}\:+\:3 H n_{Nij}\ =\ \mathcal{C}_{ij}[\{f,\bar{f}\}]\;,\\
  \dot{n}_{\ell\alpha\beta}\:+\:3 H n_{\ell\alpha\beta}\ =\ \mathcal{C}_{\alpha\beta}[\{f,\bar{f}\}]\;,
\end{gather}
\end{subequations}
plus the CP-conjugate expressions, describing the evolution of the conjugate densities $\bar{n}_N$ and $\bar{n}_{\ell}$. We turn our attention now to the collision terms.

We may proceed in one of two ways: semi-classically via the Liouville-von Neumann and Heisenberg equations, or field-theoretically via the so-called Kadanoff-Baym formalism. Whilst the former approach is less technically involved, the latter has the advantage that all quantum effects are, in principle, incorporated systematically without external prescription.


\subsubsection{Semi-classical approach}

The aim of semi-classical approaches is to find consistent means for supplementing systems of Boltzmann equations with ingredients that involve some level of resummation. In this way, one intends to capture the pertinent quantum effects, whilst avoiding the technicalities of first-principles field-theoretic treatments. An introduction to semi-classical approaches for the simplest scenario of thermal leptogenesis is provided in Chapter~\cite{leptogenesis:A04} of this review.

We outline here the basics of the so-called density matrix formalism~\cite{Dolgov:1980cq, Stodolsky:1986dx, Raffelt:1992uj, Sigl:1992fn}, which yields rate equations for the integrated matrices of number densities in~\eref{eq:numbdenmatrix}. The derivation that follows is based on Ref.~\cite{Dev:2014laa}, and we will work in the interaction picture. Therein, we recall that the creation and annihilation operators evolve subject to the free part of the Hamiltonian $\hat{H}^0$ via the (interaction-picture form of the) Heisenberg equation of motion and that the density operator evolves subject to the interaction part of the Hamiltonian $\hat{H}^{\rm int}$ via the Liouville-von Neumann equation.

Introducing the matrix of number operators $\hat{n}_{ij}(t,\mathbf{p})$ corresponding to \eref{eq:numbdenmatrix}, the time-derivatives of the respective densities can be written
\begin{align}
\label{eq:densmatstart}
\frac{{\rm d}\,n_{ij}(t,\mathbf{p})}{{\rm d}t}\ =\ \frac{\rm d}{{\rm d}t}\,\mathrm{tr}\,\Big\{\hat{\rho}(t)\,\hat{n}_{ij}(t,\mathbf{p})\Big\}\ &=\ \mathrm{tr}\,\bigg\{\hat{\rho}(t)\,\frac{{\rm d}\,\hat{n}_{ij}(t,\mathbf{p})}{{\rm d}t}\:+\:\frac{{\rm d}\,\hat{\rho}(t)}{{\rm d}t}\,\hat{n}_{ij}(t,\mathbf{p})\bigg\}\;.
\end{align}
By means of the Heisenberg equation of motion, the first term on the right-hande side of \eref{eq:densmatstart} can be written
\begin{equation}
\mathrm{tr}\,\bigg\{\hat{\rho}(t)\,\frac{{\rm d}\,\hat{n}_{ij}(t,\mathbf{p})}{{\rm d}t}\bigg\}\ =\ i\langle[\hat{H}^0,\hat{n}_{ij}(t,\mathbf{p})]\rangle_t\;,
\end{equation}
and it describes flavor oscillations. For the second term on the right-hand side of \eref{eq:densmatstart}, we first recast the usual form of the Liouville-von Neumann equation
\begin{equation}
\frac{{\rm d}\,\hat{\rho}(t)}{{\rm d}t}\ =\ -\,i[\hat{H}^{\rm int}(t),\hat{\rho}(t)]
\end{equation}
as a Volterra integral equation of the second kind, i.e.
\begin{equation}
\hat{\rho}(t)\ =\ \hat{\rho}(0)\:-\:i\int_0^t{\rm d}t'\;[\hat{H}^{\rm int}(t'),\hat{\rho}(t')]\;.
\end{equation}
Proceeding by successive substitution to second order in the interaction Hamiltonian and subsequently differentiating with respect to time, we obtain
\begin{equation}
\label{eq:LvNsecondorder}
\frac{{\rm d}\,\hat{\rho}(t)}{{\rm d}t}\ =\ -\,i[\hat{H}^{\rm int}(t),\hat{\rho}(0)]\:-\:\int_0^t{\rm d}t'\;[\hat{H}^{\rm int}(t),[\hat{H}^{\rm int}(t'),\hat{\rho}(t')]]\;.
\end{equation}
For the models and particle species of interest to us, the first term on the right-hand side of \eref{eq:LvNsecondorder} is zero. The second term gives rise to the leading collision terms, and, by putting everything together, we obtain the exact evolution equation
\begin{equation}
\label{eq:fulldensmateq}
\frac{{\rm d}\,n_{ij}(t,\mathbf{p})}{{\rm d}t}\ =\ i\langle[\hat{H}^0,\hat{n}_{ij}(t,\mathbf{p})]\rangle_t\:-\:\int_0^t{\rm d}t'\;\langle[\hat{H}^{\rm int}(t'),[\hat{H}^{\rm int}(t),\hat{n}_{ij}(t,\mathbf{p})]]\rangle_{t'}\;.
\end{equation}
At this point, we emphasise the presence of the non-Markovian memory integral over $\rho(t')$, which depends on the complete history of the evolution.

By assuming (i) that the time-scales for the microscopic QFT processes and statistical evolution are well separated, and (ii) that momentum correlations built up by a collision are lost before the next collision (molecular chaos), we can make a Markovian (or Wigner-Weisskopf~\cite{Weisskopf:1930au}) approximation of \eref{eq:fulldensmateq} (see, e.g., Ref.~\cite{Dev:2014laa}). Doing so, yields the Markovian master equation
\begin{equation}
\label{eq:Markovianmaster}
\frac{{\rm d}\,n_{ij}(t,\mathbf{p})}{{\rm d}t}\ =\ i\langle[\hat{H}^0,\hat{n}_{ij}(t,\mathbf{p})]\rangle_t\:-\:\frac{1}{2}\int_{-\infty}^{+\infty}{\rm d}t'\;\langle[\hat{H}^{\rm int}(t'),[\hat{H}^{\rm int}(t),\hat{n}_{ij}(t,\mathbf{p})]]\rangle_{t}\;.
\end{equation}
Notice that the Markovian approximation has led to the extension of the limits of time-integration and the change of time argument $t'\to t$ in the density operator, thereby neglecting memory effects.

Whilst it is now a matter of course to find the explicit form of the oscillation and collision terms for a given Hamiltonian, it is clear that the right-hand side of \eref{eq:Markovianmaster} is truncated at second order in the interaction Hamiltonian. Moreover, in making the Markovian approximation, we have also neglected dispersive self-energy corrections. Hence, in order to capture any relevant non-perturbative effects in the resulting rate equations, we need to supplement the finite-order calculation with resummed quantities by some effective means. This process may be motivated by considering scattering matrix elements (in the case of the collision terms) or from finite-temperature field theory calculations (in the case of the thermal-mass corrections).

However, as is the case for any effective description, it is necessary to ensure that important field-theoretic properties are preserved, e.g.~unitarity, CPT invariance, gauge invariance and so on, and significant effort has been devoted to this in the literature (see, e.g., Ref.~\cite{Pilaftsis:1998pd}). For instance, in resonant scenarios (see~\sref{sec3} and Chapter~\cite{leptogenesis:A03}), it is necessary to resum the self-energies of the heavy neutrinos in order to regulate the resonant enhancement of the CP asymmetry. In this case, we need systematic methods for dealing with the resummation of transition amplitudes involving intermediate unstable states. Moreover, these unstable states will likely be subject to particle mixing. Lastly, we must avoid the double counting of processes contributing to the statistical evolution~\cite{Kolb:1979qa}. For example, if we include decays, inverse decays and two-to-two scatterings in the collision terms, we must be careful to deal with what happens when the scattering is mediated by an on-resonance $s$-channel exchange of the unstable particle. This problem can be evaded by employing so-called Real Intermediate State (RIS) subtraction~\cite{Kolb:1979qa} (see also Chapter~\cite{leptogenesis:A04}).

Rate equations can also be derived from first principles using the field-theoretic approaches that we will describe in the next subsection. Whilst this technology supersedes density matrix formalisms, semi-classical approaches remain of significant utility, and it is worth noting that many of the results reviewed in~\sref{sec:typeI}, \sref{sec3} and~\sref{sec:typeII} have been derived by these means.


\subsubsection{Field-theoretic approach}
\label{sec:methods_fieldtheory}

The program of \emph{field-theoretic} approaches is to derive the fluid equations that are used in phenomenological studies of leptogenesis from first principles of quantum field theory. As a starting point, we may choose the Schwinger-Dyson equations on the Schwinger-Keldysh closed time path (CTP)~\cite{Schwinger:1960qe, Keldysh:1964ud}, which contain the full content of the theory. Specifically, no truncations in the interactions or the quantum statistical state need to be made in their formulation. As a particular consequence, the evolution of the system is reversible prior to further truncations. The Schwinger-Dyson equations are formulated in terms of $n$-point functions and make no reference to an operator-based formalism. In fact, within statistical quantum field theory, they are most often derived in the functional formalism for the $n$-particle irreducible effective action~\cite{Calzetta:1986cq}. Nonetheless, it is important to keep in mind that within a perturbative expansion, the tree-level two-point functions can be straightforwardly constructed in the operator formalism via the density matrix, cf.~Refs.~\cite{Lee:2004we, Millington:2012pf, Millington:2013isa}, which may be useful in order to see how semi-classical and field-theoretic methods can be related. Further discussions of this point can be found in Chapter~\cite{leptogenesis:A03} of this review.

For the problem of leptogenesis, the following controlled approximations can be applied in order to reduce the Schwinger-Dyson equations to a system of quantum Boltzmann equations suitable for phenomenological studies:
\begin{itemize}
\item
Due to the smallness of the RH-neutrino Yukawa couplings $\lambda$, a perturbative truncation of the Schwinger-Dyson equations is appropriate for leptogenesis. Even more robust is an expansion based on the two-particle-irreducible effective action that readily resums one-loop corrections to the Green's functions that otherwise exhibit unphysical divergences, as occurs, for instance, for the fully mass-degenerate limit of resonant leptogenesis, as well as for the $t$-channel contribution to the production of relativistic RH neutrinos.
\medskip
\item
Another important truncation lies within neglecting the full higher-order quantum correlations, i.e.~those present within $n$-point functions for $n>2$, as well as among different species of particles. In principle, all higher-order correlations can be reconstructed from the two-particle-irreducible two-point Green's functions, but, in practice, the \emph{full} information is lost because the backreaction of the RH neutrinos on the lepton and Higgs doublets is neglected, up to an effective description through kinetic equilibrium distributions with chemical potentials.
\end{itemize}

In addition, the two-point functions will, in general, contain correlations between particles that share the same conserved quantum numbers, i.e.~members of a flavor multiplet. This is of relevance for leptogenesis in that it can affect the RH neutrinos, as well as the charged leptons. Flavor correlations of RH neutrinos lead to a contribution to the CP-violating source for leptogenesis (see the detailed discussions in the chapters on resonant leptogenesis~\cite{leptogenesis:A03} and ARS leptogenesis~\cite{leptogenesis:A02}), while correlations among the doublet leptons are at the core of the \emph{flavor} effects and their importance for the washout of the lepton asymmetries, which are in the main focus of the present chapter. Therefore, in this section, we account for flavor-correlations in the charged leptons only.\footnote{Correlations in the RH neutrinos are then still generated through wave-function corrections at one-loop order. For RH-neutrino correlations, particular care must be taken in order to avoid over-counting issues (see~\sref{subsec:rateequations}).}

An overview of the Schwinger-Keldysh CTP formalism is given in Sec.~3 of the accompanying Chapter~\cite{leptogenesis:A03} on resonant leptogenesis. Its application to leptogenesis is discussed in detail in Refs.~\cite{Buchmuller:2000nd, DeSimone:2007gkc, Garny:2009rv, Garny:2009qn, Beneke:2010wd, Anisimov:2010aq, Anisimov:2010dk}, and this present section relies particularly on Ref.~\cite{Beneke:2010dz}. Our present starting point is the Schwinger-Dyson equation for the flavored left-handed (LH) lepton propagator:
\begin{align}
  i\slashed{\partial}_x S_{\ell \alpha\beta}^{fg}(x,y)
  \ =\ f \delta^{fg}\delta_{\alpha\beta} \delta^{(4)}(x-y) P_{\rm R}\:
  +\:\sum\limits_h\int\!{\rm d}^4 w\;{\Sigma\!\!\!/}^{fh}_{\ell \alpha\gamma}(x,w) S_{\ell \gamma\beta}^{hg}(w,y)\;.
\end{align}
The lower Greek indices are for active lepton flavor, the upper latin indices indicate the CTP branches $\pm$, and $P_{{\rm L},{\rm R}}$ are the left- and right-chiral projectors.

Switching to Wigner space, truncating at leading order in gradients and taking appropriate linear combinations, one obtains
\begin{subequations}
\begin{align}
  \label{polemass:gradexp}
  \left(\slashed{k}\:-\:\slashed{\Sigma}_\ell^{\mathcal{H}}\:\mp\:\slashed{\Sigma}_{\ell}^{\cal A}\right)
  S_\ell^{A,R}\ &=\ P_{\rm R}\;,\\
  \label{KB:gradexp}
  \frac{i}{2}\,\slashed{\partial} S_{\ell}^{<,>}
  \:+\:(\slashed{k}\:-\:\slashed{\Sigma}_{\ell}^{\mathcal{H}})S_\ell^{<,>}\:
  -\:\slashed{\Sigma}^{<,>}_{\ell} S_{\ell}^{\mathcal{H}}
  \ &=\ 
  \frac12
  \left(
  \slashed{\Sigma}^{>}_{\ell}
  S_{\ell}^<\:
  -\:
  \slashed{\Sigma}^{<}_{\ell}
  S_{\ell}^>
  \right)
  \;,
\end{align}
\end{subequations}
where the superscripts $R$ and $A$ indicate retarded and advanced boundary conditions, respectively. We have also defined the linear combinations $\slashed{\Sigma}_{\ell}^{\mathcal A}\equiv(\slashed{\Sigma}_{\ell}^{A}-\slashed{\Sigma}_{\ell}^{R})/(2i)$, $\slashed{\Sigma}_{\ell}^{\mathcal{H}}\equiv(\slashed{\Sigma}_{\ell}^{A}+\slashed{\Sigma}_{\ell}^{R})/2$ with analogous definitions for the propagators $S_\ell$. The Wigner-space two-point functions (here, the propagators $S_\ell$ and self-energies $\slashed{\Sigma}_{\ell}$) are understood to be functions of the four-momentum $k$ and the average coordinate $X= (x+y)/2$ upon which the partial derivative is acting. In order to understand the physical content of these equations, it is useful to note that $S_\ell(k,X)$ describes particle properties for $k^0>0$ and anti-particle properties for $k^0<0$. We refer to the accompanying Chapter~\cite{leptogenesis:A03}, where more aspects of the Wigner transformation and the gradient expansion are reviewed. Note that when comparing with that reference, the definitions for the various two-point functions on the closed time path made here may differ by factors of $i$ and $2$.

It is of conceptual interest and an important consistency check to understand the solutions to this system of equations. It turns out that we may represent the tree-level propagators as
\begin{subequations}
\label{Slessgreater:elldoublets}
\begin{align}
  \label{Seq:less}
  iS^<_{\ell \alpha\beta}\ &=\ -\,2 S^{\cal A}_{\ell}
  \left[
  \theta(k^0)f_{\ell \alpha\beta}(\mathbf k)\:
  -\:\theta(-k^0)({1\!\!1}_{\alpha\beta}-\bar{f}_{\ell \alpha\beta}(-\mathbf k))
  \right]\;,
  \\
  \label{Seq:greater}
  iS^>_{\ell \alpha\beta}\ &=\ -\,2 S^{\cal A}_{\ell}
  \left[
  -\,\theta(k^0)({1\!\!1}_{\alpha\beta}-f_{\ell \alpha\beta}(\mathbf k))
  \:+\:\theta(-k^0) \bar{f}_{\ell \alpha\beta}(-\mathbf k)
  \right]\;,
\end{align}
\end{subequations}
where
\begin{align}
  \label{S^A:singular}
  S_\ell^{\cal A}\ =\ 
  \pi
  P_{\rm L} \slashed{k} P_{\rm R}
  \delta \!\left(k^2\right)\;,
\end{align}
and $f_{\ell \alpha\beta}$ and $\bar{f}_{\ell \alpha\beta}$ are the elements of the matrices of distribution functions for the charged leptons (unbarred) and  anti-leptons (barred). At this point, one may wonder how finite-width effects from absorptive corrections, as well as the dispersive shifts to the various pole masses in the flavor-mixing system at finite temperature, come into the game. In principle, in order to recover these effects, one has to resum the gradients to all orders~\cite{Garbrecht:2011xw, Fidler:2011yq}. Fortunately, since the lepton doublets are weakly coupled, this only amounts to perturbatively-suppressed kinematic corrections for the individual reactions.

Assuming spatial homogeneity and taking $i$ times the Hermitian part of the Kadanoff-Baym equation,~\eref{KB:gradexp}, we find that the remaining relevant information
can be isolated in the kinetic equation
\begin{align}
\label{eq:kinetic}
  &i\partial_\eta i\gamma^0 S^{<,>}_\ell
  \:-\:\left[
  \mathbf k\cdot{\bm \gamma}\gamma^0
  \:+\:{\Sigma}^{\mathcal{H}}_\ell\gamma^0,i\gamma^0 S^{<,>}_\ell
  \right]\nonumber\\&\qquad\qquad
  -\:\left[i{\Sigma}^{<,>}_\ell\gamma^0, \gamma^0 S^{\mathcal{H}}_\ell\right]\
  =\ -\:\frac12\left(i{\cal C}_\ell\:+\:i{\cal C}_\ell^\dagger\right)\;,
\end{align}
with the collision term
\begin{align}
  \label{collision:term}
  {\cal C}_\ell \ & =\  i{\Sigma}^>_{\ell}  
  iS^<_\ell\:-\: i{\Sigma}^<_\ell iS^>_\ell \;. 
\end{align}
For brevity, we have used a fixed flavor basis where the charged leptons are mass diagonal in the electroweak symmetry-broken phase. The flavor-covariant generalization can be found in Ref.~\cite{Beneke:2010dz}. Moreover, we assume here spatial homogeneity, such that there is no dependence on $X^i$ for $i=1,2,3$. In addition, to account for the expansion of the Universe, we use a parametrization where $X^0=\eta$ is the conformal time.

It turns out that in the parametric regime relevant for leptogenesis, oscillations among the charged-lepton flavors are effectively frozen in. In order to explain this effect, we decompose the fluid equations into particle and anti-particle distributions, as well as number densities
\begin{subequations}
\label{relate:nu:S}
\begin{align}
  n_{\ell \alpha\beta}\ &=\
  \int\!\frac{{\rm d}^3 \mathbf{k}}{(2\pi)^3}\;
  f_{\ell \alpha\beta}(\mathbf k)
  \ =\ -
  \int\!\frac{{\rm d}^3 \mathbf{k}}{(2\pi)^3}
  \int_{0}^{\infty}
  \frac{{\rm d} k^0}{2\pi}\;
  {\rm tr}\left[
  i\gamma^0 S_{\ell \alpha\beta}^{<}
  \right] \;,
  \\
  \bar{n}_{\ell \alpha\beta} \ &=\
  \int\!\frac{{\rm d}^3 \mathbf{k}}{(2\pi)^3}\;\bar{f}_{\ell \alpha\beta}(\mathbf k)\
  =\ 
  \int\!\frac{{\rm d}^3 \mathbf{k}}{(2\pi)^3}
  \int_{-\infty}^{0}
  \frac{{\rm d} k^0}{2\pi}
  {\rm tr}\left[
  i\gamma^0 S_{\ell \alpha\beta}^{>}
  \right] \;.
\end{align}
\end{subequations}
Note that in view of including flavor effects, $n_\ell$ counts the charge density within one component of the ${\rm SU}(2)_{\rm L}$ doublet of SM leptons only (in contrast to, e.g., the quantity $n_L$ used in the accompanying Chapters~\cite{leptogenesis:A03} and~\cite{leptogenesis:A04}). This way, compensating factors that would appear in the equations describing the reactions with the right-handed charged leptons of the SM can be avoided.

Integrating over the four momentum of the lepton doublets brings us from a kinetic to a fluid description. Avoiding the technical details, we will simply present the resulting fluid equations:
\begin{subequations}
\label{kin:eq_nu}
\begin{align}
  \frac{\partial \delta n_{\ell \alpha\beta}}{\partial \eta}\ 
  &=\ 
  -\,i\Delta\omega^{\rm eff}_{\ell \alpha\beta} \delta n_{\ell \alpha\beta}
  \:-\: \sum\limits_{\gamma}[W_{\alpha\gamma}\delta n_{\ell \gamma\beta}\:+\:\delta n^*_{\ell \gamma\alpha}W_{\beta\gamma}^*]
  \nonumber\\&+\: S_{\alpha\beta}
  \:-\:\Gamma^{\rm bl}(\delta n_{\ell \alpha\beta}\:+\:\delta \bar{n}_{\ell \alpha\beta})
  \:-\:\Gamma_{\ell \alpha\beta}^{\rm fl}
  \;,\\
  \frac{\partial \delta \bar{n}_{\ell \alpha\beta}}{\partial \eta}
  \ &=\ 
  +\,i\Delta\omega^{\rm eff}_{\ell \alpha\beta} \delta \bar{n}_{\ell \alpha\beta}
  \:-\:\sum\limits_{\gamma}[W_{\alpha\gamma}\delta \bar{n}_{\ell \gamma\beta}
  \:+\:\delta \bar{n}^*_{\ell \gamma\alpha}W_{\beta\gamma}^*]\nonumber\\&
  -\: S_{\alpha\beta}
  \:-\:\Gamma^{\rm bl}(\delta n_{\ell \alpha\beta}\:+\:\delta \bar{n}_{\ell \alpha\beta})
  \:-\:\overline{\Gamma}_{\ell \alpha\beta}^{\rm fl}\;,
\end{align}
\end{subequations}
and discuss their physical content and relation to~\eref{eq:kinetic}. The details of the evaluation of the particular terms can be found in Ref.~\cite{Beneke:2010dz}.

First, we discuss the kinetic aspects. Notice that we have expressed this equation in terms of the deviations of the lepton and anti-lepton number densities ($\delta n_{\ell}$ and $\delta \bar{n}_{\ell}$) from their
equilibrium values. One can show that for these quantities, the commutator term involving $S^{\mathcal{H}}_\ell$ in~\eref{eq:kinetic} (which is essentially an inhomogeneous term) drops out~\cite{Garbrecht:2011xw}. The remaining commutator term involving $\Sigma_\ell^{\mathcal{H}}$ potentially gives rise to flavor oscillations due to the thermal masses of the charged leptons. Only flavor-sensitive terms are relevant here. (Specifically, there are no direct oscillation effects due to the flavor-blind gauge interactions, which give rise to a contribution to the effective mass that is proportional to the identity matrix in flavor space.) Upon momentum averaging, the oscillation effects are therefore described by
\begin{align}
  \Delta\omega_{\ell \alpha\beta}^{\rm eff}(\eta)
  \ &=\ \int\!\frac{{\rm d}^3 \mathbf{k}}{(2\pi)^3}\;
  \frac{12\, e^{|\mathbf k|/T}}{T^3(e^{|\mathbf
    k|/T}+1)^2}\,
  \left(\frac{h_\alpha h_\beta^* T^2}{16|\mathbf k|}\right)\;.
\end{align}

Next, we turn to the collisional contributions, where we can identify the washout rate
\begin{align}
  W_{\alpha\beta}\ 
  &= \ \lambda_{\alpha 1} \lambda_{\beta 1}^*
  \int\!
  \frac{{\rm d}^3 \mathbf{k}}{(2\pi)^3 2|\mathbf k|}\,
  \frac{{\rm d}^3 \mathbf{p}}{(2\pi)^3 2\sqrt{\mathbf{p}^2+(a(\eta)M_1)^2}}\,
  \frac{{\rm d}^3 \mathbf{q}}{(2\pi)^3 2|\mathbf{q}|}\;
  \notag\\
  &\times(2\pi)^4 \delta^{(4)}(p-k-q)
  k\cdot p
  \big[
  f_{N1}(\mathbf{p})+f_{\phi}(\mathbf{q})
  \big]\,
  \frac{12\,e^{|\mathbf k|/T}}{T^3(e^{|\mathbf k|/T}+1)^2}\;.
\end{align}
Here, the integration variables are understood to be conformal momenta, such that the physical momenta are, e.g.,~given by $\mathbf{k}/a(\eta)$, where $a(\eta)$ is the scale factor of the Friedmann-Lema\^{i}tre-Robertson-Walker metric. Similarly, $T$ is a conformal temperature, and the physical temperature is $T/a(\eta)$.

The CP-violating source term consists of a vertex and a wave-function contribution:
\begin{align}
  S_{\alpha\beta}\ =\ S^{({\rm v})}_{\alpha\beta}\:+\:S^{({\rm wf})}_{\alpha\beta}\;,
\end{align}
where
\begin{align}
  S^{({\rm v})}_{\alpha\beta}\ &=\ -\:i\sum\limits_{j\,\neq\,1}(\lambda_{\alpha 1}\lambda_{\gamma 1}\lambda_{\gamma j}^*\lambda_{\beta j}^*-\lambda_{\alpha j}\lambda_{\gamma j} \lambda^*_{\gamma 1} \lambda^*_{\beta 1})\notag\\
  &\times\ 
  \int\!\frac{{\rm d}^3\mathbf{k}}{(2\pi)^32|\mathbf{k}|}\,\frac{{\rm d}^3\mathbf{p}}{(2\pi)^32\sqrt{\mathbf{p}^2+M_1^2}}\,\frac{{\rm d}^3\mathbf{q}}{(2\pi)^32|\mathbf{q}|}\;(2\pi)^4\delta^{(4)}(p-k-q) \notag\\& \times\ k^\mu \frac{M_1}{16\pi M_j}K_{\mu j}(p,q)
  \big[1-f_\ell(k)+f_\phi(q)\big]\;,
  \label{S:vertex}
\end{align}
and
\begin{align}
  S^{({\rm wf})}_{\alpha\beta}\ &=\ 8\,i\sum\limits_{j\,\neq\,1}\big[\big(\lambda_{\alpha 1}\lambda_{\gamma 1}\lambda_{\gamma j}^*\lambda_{\beta j}^*-\lambda_{\alpha j}\lambda_{\gamma j} \lambda^*_{\gamma 1} \lambda^*_{\beta 1}\big)\notag\\
  &+\ \big(\lambda_{\alpha 1}\lambda^*_{\gamma 1}\lambda_{\gamma j}\lambda_{\beta j}^*-\lambda_{\alpha j}\lambda^*_{\gamma j} \lambda_{\gamma 1} \lambda^*_{\beta 1}\big)\big]\int\!\frac{{\rm d}^3\mathbf{p}}{(2\pi)^32\sqrt{\mathbf{p}^2+M_1^2}}\;\hat\Sigma_{N\mu}(p)
  \hat\Sigma_N^{\mu}(p)\;.
  \label{S:wavefunction}
\end{align}
Here, duplicate indices other than $j$ are summed over according to the Einstein convention. We have chosen to present these contributions in integral form in order to highlight the structure of the thermal cuts and the pertaining quantum statistical effects, as well as to facilitate comparison with the companion Chapters~\cite{leptogenesis:A02, leptogenesis:A03, leptogenesis:A04}. The expression for the vertex function $K_{\mu j}(p,q)$ can be found in
Chapter~\cite{leptogenesis:A04}, and
\begin{align}
  \hat\Sigma_N^\mu(p)\ =\ \frac12\int\!\frac{{\rm d}^3\mathbf{k}}{(2\pi)^3 2|\mathbf k|}
  \frac{{\rm d}^3\mathbf{q}}{(2\pi)^3 2|\mathbf q|}
  (2\pi)^4 \delta^{(4)}(p-k-q)\,p^\mu
  \big[
  1-f_\ell^{\rm eq}(\mathbf k)+f_\phi^{\rm eq}(\mathbf q)
  \big]\;,
\end{align}
which relates to the expression from Chapter~\cite{leptogenesis:A04} as $\hat\Sigma_{N\mu}(p)=L_\mu(p)/2$. We choose this different normalization in order to highlight the symmetry of the internal (cut) and external phase space
of the CTP Feynman diagrams, as well as to make connection with the discussion on ARS leptogenesis in the accompanying Chapter~\cite{leptogenesis:A02}.

It is of interest to comment on the CP-odd combinations of Yukawa couplings that appear in~\eref{S:vertex} and~\eref{S:wavefunction}. The combination in~\eref{S:vertex} and in the first term in round brackets in~\eref{S:wavefunction} arises due to lepton number violating contributions mediated by the Majorana mass $M$. In contrast, the second term in round brackets in~\eref{S:wavefunction} is lepton number conserving but lepton flavor violating, where the total lepton number conservation can be easily seen when taking the trace over the flavor indices $\alpha$ and $\beta$ of the charged leptons. Yet, lepton flavor violation in the type I seesaw model is only mediated by the RH neutrinos. Therefore, the different washout rates for the particular active lepton flavors (provided the latter are distinguishable from rates that are mediated by SM Yukawa couplings) can lead to a net lepton asymmetry even when starting only from the lepton number conserving contribution to the source. This has important consequences: Firstly, in case lepton number violation is suppressed for some reason, flavor effects can still lead to a sizable or even enhanced lepton asymmetry, as occurs for ARS leptogenesis, cf.~the accompanying Chapter~\cite{leptogenesis:A02} on this topic. Secondly, since all the active lepton flavors are summed over, the trace of the lepton number violating source is apparently independent of the weak basis transformation implied by the PMNS matrix. Therefore, unflavored leptogenesis is independent of the Dirac and Majorana phases in the PMNS matrix. In turn, once flavor effects are important, the outcome of leptogenesis depends, in general, on the PMNS phases, but we should be aware that extra ``high-energy'' phases will contribute~\cite{Nardi:2006fx, Abada:2006ea, Abada:2006fw}. For a decomposition of lepton number conserving versus lepton number violating sources in terms of effective decay asymmetries, see~\eref{veial} of the present chapter.

Finally, we turn to the last two terms in~\eref{kin:eq_nu}, which may be categorized as lepton number conserving dissipative effects. Flavor-blind contributions are mediated by gauge interactions and are described by $\Gamma^{\rm bl}\sim g^4 T$, where $g$ stands collectively for the weak and weak-hypercharge couplings. The relative signs are discussed carefully in Ref.~\cite{Beneke:2010dz}. The physical content is, however, that loss terms in, say, leptons and their flavor correlations tend to be compensated by gain terms from anti-leptons. This has an important consequence for the frustration of flavor oscillations, which we discuss below. The leading flavor-sensitive term is evaluated to be~\cite{Beneke:2010dz}
\begin{subequations}
\label{Gamma:fl}
\begin{align}
  \Gamma^{\rm fl}_{\ell \alpha\beta}\ &=\
  +\:\frac12\,
  {\rm tr}\int\limits_{0}^{\infty}\frac{{\rm d} k^0}{2\pi}\int\!\frac{{\rm d}^3\mathbf{k}}{(2\pi)^3}
  \left(
  {\cal C}_{\ell \alpha\beta}^{\rm fl}(k)
  \:+\:{\cal C}_{\ell \alpha\beta}^{{\rm fl}\dagger}(k)
  \right)
  \notag\\
  &=\ 
  \gamma^{\rm fl}
  \left(
  h_\alpha h^*_\gamma \delta n_{\ell \gamma\beta}\:+\:
  \delta n_{\ell \alpha\gamma}^{\dagger}h_\gamma h^*_\beta
  \:-\:h_\alpha \delta n_{{\rm R}\alpha} h^*_{\alpha}\delta_{\alpha\beta}
  \:-\:h_{\alpha} \delta n_{{\rm R}\alpha}^{\dagger} h^*_{\alpha} \delta_{\alpha\beta}
  \right)\;,\\
  \overline{\Gamma}^{\rm fl}_{\ell \alpha\beta}\ &=\ 
  -\:\frac12
  {\rm tr}\,\int\limits_{-\infty}^{0}\frac{{\rm d} k^0}{2\pi}\int\!\frac{{\rm d}^3\mathbf{k}}{(2\pi)^3}
  \left(
  {\cal C}_{\ell \alpha\beta}^{\rm fl}(k)
  \:+\:{\cal C}_{\ell \alpha\beta}^{{\rm fl}\dagger}(k)
  \right)
  \notag\\
  & =\
  \gamma^{\rm fl}
  \left(
  h_\alpha h^*_\gamma \delta \bar{n}_{\ell \gamma\beta}\:+\:
  \delta \bar{n}_{\ell \alpha\gamma}^{\dagger}h_\gamma h^*_\beta
  \:-\:h_\alpha \delta \bar{n}_{{\rm R}\alpha} h^*_{\alpha}\delta_{\alpha\beta}
  \:-\:h_{\alpha} \delta \bar{n}_{{\rm R}\alpha}^{\dagger} h^*_{\alpha} \delta_{\alpha\beta}
  \right)\;,
\end{align}
\end{subequations}
where no summation over $\alpha$ and $\beta$ is performed.\footnote{Here, we have taken the right-handed charged leptons to live in their flavor basis, which we can always do without the need to rotate other couplings. This is, of course, different for the doublet leptons, which have SM Yukawa coupling, as well as couplings to RH neutrinos, that cannot be simultaneously diagonalized. A flavor-covariant description of the right-handed charged leptons is presented in Ref.~\cite{Beneke:2010dz}.} This rate describes the direct damping of the off-diagonal correlations because these appear in the loss terms while the gain terms are diagonal in the flavor basis. Note that, in order to conserve baryon-minus-lepton number in the SM sector, we have to supplement our network of equations with one for the right-handed charged leptons, which can be considered as a spectator process that we omit here for brevity. The relevant fluid equations for the right-handed charged leptons are presented in Ref.~\cite{Beneke:2010dz}. The scattering processes leading to flavor decoherence are dominated by thermal effects because tree-level $1\leftrightarrow2$ reactions among massless particles mediated by the SM Yukawa couplings are kinematically suppressed. A logarithmic enhancement occurs due to $t$ channel divergences from fermion exchange that is regulated by Landau damping and Debye screening. From these considerations, one can compute the rate~\cite{Garbrecht:2013urw}
\begin{align}
\label{gammafl}
  \gamma^{\rm fl} \ & =\ 
  \gamma^{{\rm fl}(\phi)\delta\ell}\:+\:\gamma^{{\rm fl}(\ell)\delta\ell}\:+\:
  \gamma^{{\rm fl}({\rm R})\delta\ell}\:+\:\gamma^{{\rm fl}}_{{\rm vertex}}
  \notag\\
  &=\ 1.32\times 10^{-3} \times  h_t^2 T\:+\:3.72\times 10^{-3} \times G T\:+\: 8.31\times 10^{-4} \times G (\log G^{-1}) T \notag\\
	& \qquad+\:4.74 \times 10^{-3} \times g_1^2 T
  \:+\: 1.67\times 10^{-3} \times g_1^2 (\log g_1^{-2}) T
  \:+\: 1.7\times 10^{-3} G T \;,
\end{align}
where $G=\frac{1}{2} (3 g_2^2 + g_1^2) $. In the SM, one may take $\gamma^{\rm fl}=5\times 10^{-3} T$, where a mild dependence of the numerical factor on the temperature scale due to the running couplings may be neglected in view of other uncertainties. Note that this value for $\gamma^{\rm fl}$ coincides with what had been used in the literature before a detailed calculation was available~\cite{Cline:1993bd, Abada:2006fw}.

We now turn our attention to the frustration of flavor oscillations. Close to equilibrium, the term with  $\Gamma^{\rm bl}=O(g^4 T)$ imposes the constraint
\begin{align}
  \label{charge_constraint}
  \delta n_{\ell\alpha\beta}\ =\ -\:\delta \bar{n}_{\ell\alpha\beta}\;.
\end{align}
This means that gauge interactions force opposite chemical potentials, and this condition generalizes to a matrix form in the presence of flavor coherences. Now, due to the opposite sign for particles and anti-particles
in the oscillation term of the kinetic equations,~\eref{kin:eq_nu}, it turns out that a large $\Gamma^{\rm bl}$ effectively frustrates flavor oscillations. To explain this, we consider the system of equations
\begin{subequations}
\label{g_matrix_eq}
\begin{align}
  \label{g_matrix_eq1}
  \frac{{\rm d}}{{\rm d}t}\,\delta g(t)\ &=\ -\:i\Delta\omega\,\delta g(t)\:-\:
  \Gamma\big[\delta g(t)\:+\:\delta \bar{g}(t)\big]\;,
  \\
  \label{g_matrix_eq2}
  \frac{{\rm d}}{{\rm d}t}\,\delta \bar{g}(t)\ &=\ +\:i\Delta\omega\,\delta \bar{g}(t)\:-\:
  \Gamma\big[\delta \bar{g}(t)\:+\:\delta g(t)\big]\;.
\end{align}
\end{subequations}
For flavored leptogenesis, the order of magnitude of the parameters are as follows:
\begin{align}
  \Gamma\ =\ \Gamma^{\rm bl}\ \sim\ g^4 T\;,\qquad \Delta\omega\ \sim\ h_{\tau,\mu}^2 T 
  \ll \Gamma\;,
\end{align}
where we should take the $\tau$ or $\mu$ Yukawa coupling depending on which of these dominates the mass splitting of the flavors under consideration. Since $g^4 \gg h_{\tau,\mu}^2$, there are eigenmodes with short decay times
$\tau_{\rm s} = 1/(\Gamma+\sqrt{\Gamma^2-\Delta\omega^2}) \approx 1/(2 \Gamma)$ and long decay times $\tau_{\rm l} = 1/(\Gamma-\sqrt{\Gamma^2-\Delta\omega^2}) \approx 2 \Gamma / \Delta\omega^2$. The corresponding eigenvectors are
\begin{align}
  \delta g_{\rm s,l} \ =\  \delta g\:+\:\frac{-\,i
  \Delta\omega\: \pm \:\sqrt{\Gamma^2-\Delta\omega^2}}{\Gamma} \delta \bar{g}
  \ \approx\   \delta g \:\pm\: \left(1 \mp i 
  \frac{\Delta\omega}{\Gamma} \right) \delta \bar{g}\;,
\end{align}
with
\begin{align}
  \delta g_{\rm s,l}(t)\ =\ \delta g_{\rm s,l}(0) \,{\rm e}^{-t/\tau_{\rm s,l}}\;.
\end{align}
The short mode $\delta g_{\rm s} \approx \delta g +  \delta \bar{g}$ thus rapidly approaches zero due to pair annihilations, leading to an  effective constraint
\begin{align}
  \label{eq:delg}
  \delta g\  \sim\  - \left(1 - i \frac{\Delta\omega}{\Gamma}
  \right) \delta \bar{g}\;.
\end{align}
The opposite signs in front of the $\Delta \omega$ term in~\eref{g_matrix_eq} are crucial because they imply that the source of the oscillations in
\begin{equation}
  \frac{{\rm d}}{{\rm d}t}\big[\delta g(t)\:-\:\delta \bar{g}(t)\big]
  \ =\ -\,i \Delta\omega\big[\delta g(t)\:+\:\delta \bar{g}(t)\big]\;,
\end{equation}
is damped due to the flavor-blind gauge interactions.

The interplay of the flavor-blind interactions with the flavored oscillation term leads to the slow decay of the long mode. The rate for this effect is, however, much smaller than the direct damping rate from flavor-dependent scatterings,
\smash{${\Delta\omega^{\rm eff}}^2/\Gamma^{\rm bl} \sim h_{\tau,\mu}^4 g_2^{-4} T \ll \Gamma^{\rm fl}\sim g_2^2h_\tau^2 T$}, since \smash{$h_{\tau,\mu} \ll g_2^3$}. Therefore, it is a suitable approximation to neglect the oscillations and the damping due to flavor-blind interactions altogether, accounting only for the direct damping from flavor-dependent scatterings. While, for leptogenesis, we are in the parametric regime where $\Delta\omega\ll \Gamma$ and flavor oscillations are overdamped and frustrated, this is not expected to be true for general systems of flavor mixing at finite temperature, where flavor oscillations and damping due to the interplay with flavor-blind scatterings mediated by gauge interactions may be quantitatively important.

In conclusion, we have shown that the CTP framework leads to a fluid description in the form of~\eref{kin:eq_nu}, where the terms involving $\Delta \omega^{\rm eff}$ and $\Gamma^{\rm bl}$ can be neglected. In this approximation, we can then perform the obvious simplification of taking the difference between the equations for $\delta n_\ell$ and $\delta \bar{n}_\ell$ such that we obtain a single equation for $n_{\Delta \ell}=\delta n_\ell-\delta \bar{n}_\ell$. At that stage, we have obtained then fluid equations for the LH charged leptons that can be applied to the fully flavored and unflavored, as well as intermediate regimes. The flavor damping $\Gamma^{\rm fl}$ leads to the decay of off-diagonal correlations. Provided the damping is large, we obtain the commonly used fully-flavored description by simply deleting the off-diagonal components of the fluid equation.


\section[Flavor phenomenology of leptogenesis in the type I seesaw mechanism]{Flavor phenomenology of leptogenesis in the\\ type I seesaw mechanism}
\label{sec:typeI}

In this section, we discuss the importance of flavor effects in minimal scenarios of leptogenesis embedded within the type I seesaw scenario, wherein the SM Lagrangian is extended by introducing ${\cal N}_N$ RH Majorana neutrinos that are assumed to be produced thermally in the early Universe. Moreover, we highlight how leptogenesis can play an important role in testing high-energy seesaw models especially when flavor effects are taken into account. 

Assuming a hierarchical RH neutrino spectrum, if one neglects completely the flavor composition of leptons produced by the decays of heavy RH neutrinos (unflavored assumption), the dominant contribution to the final asymmetry comes from the lightest RH neutrinos ($N_1$-dominated scenario), barring a special region of parameter space where the next-to-lightest RH neutrinos' contribution dominates ($N_2$-dominated scenario). On the other hand, when charged-lepton flavor effects are taken into account, the region of  parameter space where the next-to-lightest RH neutrinos' contribution dominates  gets significantly larger ($N_2$-dominated scenario). In some cases, the heaviest of the RH neutrinos, usually $N_3$, might also give a non-negligible contribution, as long as there is not a too strong mass hierarchy suppressing their CP asymmetries.

The RH-neutrino Yukawa couplings $\lambda$ and Majorana mass term $M$ are such that, after spontaneous symmetry breaking, we can write the neutrino mass terms in a basis where both charged-lepton and Majorana mass matrices are diagonal (the flavor basis):
\begin{equation}
  -\,{\cal L}^{\nu}_{\rm m}\ =\ \overline{\nu_{L \alpha}} \, m_{D \alpha i} \, N_{R i}\:+\: 
  \frac{1}{2} \overline{N_{R i}^c} \, M_i \, N_{R i} \:+\:{\rm h.c.}\;, 
\end{equation}
where $\alpha\in\{e, \mu, \tau\}$, $i\in\{1,\dots, N\}$ and $m_D = v\, \lambda/\sqrt{2}$ is the neutrino Dirac mass matrix generated by the Higgs vev $v$. In the seesaw limit, $M \gg m_D$, the mass spectrum splits into a set of heavy (Majorana, almost RH) neutrinos $N_i = N_{R i} + N_{R i}^c+(m_D/M)(\nu_{Li}+\nu_{Li}^c)$ with masses  (almost) coinciding with the eigenvalues $M_i$ of the  Majorana mass matrix and into a set of light (Majorana, almost LH) neutrinos $\nu_{i} = \nu_{L i} + \nu_{L i}^c - (m_D/M)(N_{Ri}+N_{Ri}^c)$ with masses given by the seesaw formula
\begin{equation}
\label{seesaw}
 D_m\ =\ U_{\nu}^{\dagger} \, m_D \, M^{-1} \, m_D^{\mathsf{T}} \, U^*_{\nu}   \;,
\end{equation}
where the diagonalizing matrix $U_{\nu}$ is the leptonic mixing (PMNS) matrix and we have defined $D_m \equiv {\rm diag}(m_1, m_2, m_3)$.  

Neutrino mixing experiments measure two mass-squared differences. For the atmospheric neutrino mass scale, global analyses find~\cite{Capozzi:2017ipn} $m_{\rm atm}\equiv \sqrt{m^{\, 2}_3 - m_1^{\, 2}} = (50.5\pm 0.04)\,{\rm meV}$, and for the solar neutrino mass scale  $m_{\rm sol} = (8.6\pm 0.1)\,{\rm meV}$, defined as $m_{\rm sol} \equiv \sqrt{m^{\, 2}_2 - m_1^{\, 2}}$ for normally-ordered neutrino masses (NO) and as $m_{\rm sol} \equiv \sqrt{m^{\, 2}_3 - m_2^{\, 2}}$ for inverse-ordered neutrino masses (IO), where we are adopting the convention $m_1 \leq m_2 \leq m_3$. See the accompanying Chapter~\cite{leptogenesis:A06} for a review of the current status of the data on neutrino masses and lepton mixing.

For NO, the leptonic mixing matrix can be parametrized in the usual way in terms of three mixing angles $\theta_{12}, \theta_{23}$ and $\theta_{13}$, one CP-violating Dirac phase $\delta$, and two CP-violating Majorana phases $\alpha$ and $\beta$:
\begin{equation}
\label{eq:PMNS}
  U_{\nu}\ =\  \begin{pmatrix}
  c_{12}\,c_{13} & s_{12}\,c_{13} & s_{13}\,e^{-i\delta} \\
  -s_{12}\,c_{23}-c_{12}\,s_{23}\,s_{13}\,e^{i\delta} &
  c_{12}\,c_{23}-s_{12}\,s_{23}\,s_{13}\,e^{i\delta} & s_{23}\,c_{13} \\
  s_{12}\,s_{23}-c_{12}\,c_{23}\,s_{13}\,e^{i\delta}
  & -c_{12}\,s_{23}-s_{12}\,c_{23}\,s_{13}\,e^{i\delta}  &
  c_{23}\,c_{13}
  \end{pmatrix}
  {\rm diag}\big(1, e^{i\alpha}, e^{i\beta}
  \big)\;.
\end{equation}
In order to account for different orderings, it is convenient to relabel the neutrino masses in a way that $m_1' < m_2' < m_3'$ with
$1'=1$ , $2'= 2$ and $3' = 3$ for NO, and  $1'=3$, $2'= 1$ and $3' = 2$ for IO. In this primed basis, the leptonic mixing matrix for IO changes as
\begin{equation}
  U_{\nu}^{\rm (IO)} \ =\  
  U_{\nu}^{\rm (NO)} \, 
  \left(\begin{array}{ccc}
  0 & 1 & 0\\
  0 & 0 & 1\\
  1 & 0 & 0  
  \end{array}\right) \;.
\end{equation}
However, in order to simplify the notation, we will omit the primed indexes. Global analyses of results from the neutrino oscillation experiments find the following best fit values ($1\sigma$ errors and $3\sigma$ intervals) for the mixing angles and the leptonic Dirac phase $\delta$ in the case of NO~\cite{Capozzi:2017ipn}:
\begin{subequations}
\label{expranges}
\begin{align}
  \theta_{13} \ & = \  8.45^{\circ}\pm 0.15^{\circ} \, \;\;  [8.0^{\circ}, 9.0^{\circ}] \;  , \\
  \theta_{12} \ & = \  33^{\circ}\pm 1^{\circ} \,  \;\;  [30^{\circ}, 36^{\circ}]  \;  , \\
  \theta_{23} \ & = \  {41^{\circ}} \pm {1^{\circ}} \,  \;\;  [38^{\circ}, 51.65^{\circ}]  \;  ,  \\ 
  \delta \ & = \  {-\,0.62 \pi \pm 0.2\pi} \,  \;\;  [-\,1.24\pi, 0.17\pi] \; .
\end{align}
\end{subequations}
It is interesting that there is already an excluded interval $\delta\notin [0.17\,\pi, 0.76\pi]$ at $3\,\sigma$ and that  $\sin\delta \geq 0$ is excluded at  $2\sigma$, favouring $\sin\delta < 0$ (in Ref.~\cite{Esteban:2016qun}, a lower statistical significance is found).  A confirmation of the exclusion of $\sin \delta =0$ would imply the discovery of CP violation in neutrino oscillations, a very interesting (and favourable) result for leptogenesis; we will come back to this point. There are no experimental constraints on the Majorana phases $\alpha$ and $\beta$. 

There is no signal from neutrinoless double beta ($0\nu\beta\beta$) decay experiments, and this therefore places an upper bound on the effective $0\nu\beta\beta$ neutrino mass $m_{ee} \equiv |m_{\nu ee}|$. Currently, the most stringent reported upper bound comes from the KamLAND-Zen collaboration, finding $m_{ee} \leq (61 \mbox{--} 165)\,{\rm meV}$ at $90\%\,{\rm C.L.}$~\cite{KamLAND-Zen:2016pfg} (for other recent results, see Refs.~\cite{Agostini:2017dxu,Albert:2017owj,Aalseth:2017btx}), where the range accounts for nuclear matrix element uncertainties (see the discussion in Chapter~\cite{leptogenesis:A06}). 

Cosmological observations place an upper bound on the sum of the neutrino masses. The \emph{Planck Collaboration} obtains a robust stringent upper bound $\sum_i m_i \lesssim 170\,{\rm meV}$ at $95\% {\rm C.L.}$~\cite{Aghanim:2016yuo} that, taking into account the experimental determination of the solar and atmospheric neutrino mass scales from neutrino-oscillation experiments, translates into an upper bound on the lightest neutrino mass $m_1 \lesssim 50\,(42)\,{\rm meV}$ for NO (IO).


\subsection{Vanilla leptogenesis}
\label{sec:vanilla}

We will be particularly interested in phenomenological scenarios  where the asymmetry is produced in the so-called strong washout regime. This occurs when the RH-neutrino inverse decays are in equilibrium during a certain 
interval of temperatures $[T_{\rm in}, T_{\rm out}]$ centred approximately about $T \sim M_i$, efficiently washing out any asymmetry produced while $T \gtrsim T_{\rm out}$~\cite{Buchmuller:2004nz}. Moreover, if one assumes a hierarchical RH-neutrino spectrum or is, in any case, not in the resonant regime, and if flavor effects are neglected, one obtains an $N_1$-dominated scenario for most of the parameter space. In this case, the asymmetry can be described to a reasonable approximation by a very simple set of Boltzmann rate (i.e.~momentum-integrated) equations (see the accompanying Chapter~\cite{leptogenesis:A04} for more details):
\begin{subequations}
\begin{align} 
  \frac{{\rm d}Y_{N_1}}{{\rm d}z} \ & = \  -\:D_1\,(Y_{N_1}-Y_{N_1}^{\rm eq}) \;,
  \label{dlg1} \\ 
  \frac{{\rm d}Y_{B-L}}{{\rm d}z} \ & = \  -\:\epsilon_1\,D_1\,(Y_{N_1}-Y_{N_1}^{\rm eq})\:-\: 
  [\Delta W(z) +W_1^{\rm ID}(z)]  \, Y_{B-L}  \;   ,
  \label{dlg2}
\end{align}
\end{subequations}
written here in terms of the yields\footnote{An alternative and simplifying option to variables $Y_X$ is to normalize the abundance of any quantity $X$ to the number of RH neutrinos in ultra-relativistic equilibrium, defining $N_X \equiv n_X/n_{N}^{\rm eq}(z \ll 1)$. The two definitions are related by
\begin{equation*}
  N_X (z) \ =\  \frac{g_*}{g_{N_1}} \, \frac{8\,\pi^4}{135 \zeta(3)} \, Y_X (z)\ =\ 
  \frac{Y_X (z)}{Y_{N_1}^{\rm eq}(z=0)}\;.
\end{equation*}}
\begin{equation}
  Y_{N_1}\ \equiv\ \frac{n_{N_1}}{s}\qquad  \text{and}\qquad Y_{B-L}\ =\ \sum_{\alpha}Y_{\Delta_{\alpha}}\ \;,
\end{equation}
where
\begin{equation}
  Y_{\Delta_{\alpha}}\ \equiv\ Y_{\Delta B/3-L_{\alpha}}\ =\ \frac 1 3\, Y_{\Delta B} - Y_{\Delta \ell_{\alpha}}-Y_{\Delta e_{R\alpha}}\;,
\end{equation}
$s=2\pi^2g_*T^3/45$ is the entropy density of the $g_*$ effective degrees of freedom and we have defined $z\equiv M_1/T$.  The $N_1$ total CP asymmetry $\epsilon_1$ is defined as
\begin{equation}
  \epsilon_1\ \equiv\  \frac{\Gamma_1-\bar{\Gamma}_1}{\Gamma_1+\bar{\Gamma}_1} \;,
\end{equation}
where $\Gamma_1\equiv \sum_{\alpha} \, \Gamma_{1\alpha}$ is the $N_1$ decay rate into leptons and $\bar{\Gamma}_1\equiv\sum_{\alpha}\, \bar{\Gamma}_{1\alpha}$ is the $N_1$ decay rate into anti-leptons and we have defined $\Gamma_{1\alpha} \equiv \Gamma(N_1\to\ell_{\alpha}\phi)$ and $\bar{\Gamma}_{1\alpha} \equiv \Gamma(N_1\to\bar{\ell}_{\alpha}\bar{\phi})$. A perturbative calculation from the interference of tree-level with one-loop self-energy and vertex diagrams gives~\cite{Covi:1996wh}
\begin{equation}
\label{CPas}
  \epsilon_1 \ =\ \frac{1}{8\pi}\, \sum_{j\,\neq\, 1}\,\frac{{\rm
  Im}\big[(\lambda^{\dagger}\lambda)^2_{1j}\big]}{(\lambda^{\dagger}\lambda)_{11}}\,\xi\bigg(1,\frac{M_j^2}{M_1^2}\bigg)\; ,
\end{equation}
where
\begin{equation}
\label{xi}
  \xi(b,x)\ =\ \sqrt{x}\,
  \bigg[1+\frac{b}{1-x}-(1+x)\,\ln\left(\frac{1+x}{x}\right)\bigg] \; .
\end{equation}
The (dimensionless) decay term $D_1$ and the washout term from inverse decays $W_1^{\rm ID}$ are given respectively by
\begin{equation}
\label{DW}
  D_1(z)\ \equiv\ \frac{\Gamma_1+\bar{\Gamma}_1}{H\,z} \ =\ K_1\,z\,
  \left\langle \frac{1}{\gamma_1} \right\rangle
\end{equation}
and
\begin{equation}
  W_1^{\rm ID}(z)\  \equiv\ \frac{1}{2}\,\frac{\Gamma_1^{\rm ID}+\bar{\Gamma}_1^{\rm ID}}{H\,z} \ =\  \frac{1}{4}\,K_1 \,{\cal K}_1(z)\,z^3 \;  ,
\end{equation}
where $K_1$ is the total decay parameter  defined as
\begin{equation}
  K_1 \ \equiv\  \frac{(\Gamma_1+\bar{\Gamma}_1)_{T\,=\,0}}{H_{T\,=\,M_1}} \;  ,
\end{equation}
with $H$ being the expansion rate of the Universe. Finally, the averaged dilution factor, in terms of the modified Bessel functions of the second kind, is given by $\left\langle {1/\gamma_1} \right\rangle = {{\cal K}_{1}(z) / {\cal K}_{2}(z)}$.

The final $B-L$ asymmetry is simply given  by
\begin{equation}
Y^{\infty}_{B-L}\ =\ -\,Y_{N_1}^{\rm eq}(0)\,\epsilon_1\,\kappa^{\infty}(K_1,m_1)\;   ,
\end{equation}
where $\kappa^{\infty}(K_1, m_1)$ is the total final efficiency factor that can be calculated  in the case of an initial thermal $N_1$ abundance as 
\begin{equation}
\label{efunf}
  \kappa^{\infty}(K_1, m_1) \ \simeq\  \kappa(K_1, m_1)  \ \equiv \ 
  \kappa(K_1) \, \exp\left[-\,\frac{\omega}{z_B}\,\frac{M_1}{10^{10}\,{\rm GeV}}\,
  \frac{\sum_i m_i^2}{{\rm eV}^2} \right] \;  ,  
\end{equation}
with
\begin{equation}
  \kappa(K_1) \ \equiv\ \frac{2}{K_1 \, z_{\rm B}(x)}\left[1-{\rm exp}\left(-\,\frac{1}{2}\,K_1 \, z_{\rm B}(K_1)\right)\right]  
\end{equation}
and $\omega \simeq 0.186$. The exponential term is an effect of the $\Delta L =2$ washout term $\Delta W$. In the case of an initially vanishing $N_1$ abundance, the expression is more complicated and is the sum of a negative and a positive contribution. In any case, in the strong washout regime, realised for $K_1 \gtrsim 3$, there is no dependence on the initial $N_1$ abundance. This is because the asymmetry is generated within quite a narrow interval of temperatures centred at $T_{\rm lep}\equiv M_1/z_{B1}$, where $z_{B1}\equiv z_{B}(K_1) ={\cal O}(0.1)$, when the RH neutrinos are fully non-relativistic. All the asymmetry generated at higher temperatures, in the relativistic regime, and depending on the initial $N_2$-abundance, is efficiently washed out~\cite{Blanchet:2006ch}. This strongly reduces the theoretical uncertainties, since, in the relativistic regime, many different effects, most of which are not well under control,  have to be taken into account in the calculation of the asymmetry. 

Finally, the  baryon-to-photon number ratio can be calculated in a very simple way from the the final $B-L$ asymmetry:
\begin{equation}
  \eta_B \ =\  a_{\rm sph}\,\frac{n_{B-L}^{\infty}}{n_{\gamma}^{\rm rec}}\ \simeq\ 0.01\,\frac{Y_{B-L}^{\infty}}{Y_{N_1}^{\rm eq}(0)}\;  ,
\end{equation}
where $a_{\rm sph} \simeq 1/3$ is the fraction of $B-L$ asymmetry that goes into a baryon asymmetry when sphaleron processes~\cite{Kuzmin:1985mm} are in equilibrium (occurring approximately in the temperature range
$10^{12} \,{\rm GeV} \gtrsim T \gtrsim 100\,{\rm GeV}$). For successful leptogenesis, the result obtained for $\eta_B$ must reproduce the experimental value extracted from CMB temperature anisotropies. The \emph{Planck Collaboration} has recently found~\cite{Ade:2015xua}
\begin{equation}
  \label{etaBPlanck}
  \eta_{B}^{\rm CMB} \ =\  (6.10 \pm 0.04)\: \times\: 10^{-10}  \;   . 
\end{equation}
An interesting feature of this simple picture  is that both the RH-neutrino abundance and the washout of the asymmetry are described just by the efficiency factor. This depends only on the decay parameter $K_1$ and, quite interestingly, on the neutrino masses, which can be parametrized entirely in terms of $m_1$, when the measured values of the mass-squared differences are combined. The total decay parameter can then be re-expressed in terms of the Dirac mass matrix as 
\begin{equation}
  K_1 \ =\ \frac{(m^{\dagger}_D\,m_D)_{11}}{M_1 \, m_{\star}}\ =\ \frac{\widetilde{m}_1}{m_{\star}} \;,
\end{equation}
where $\widetilde{m}_1 \equiv (m^{\dagger}_D\,m_D)_{11}/M_1$ is the {\em effective neutrino mass} and
\begin{equation}
  m_{\star}\ \equiv \
  \frac{16\, \pi^{5/2}\,\sqrt{g_*}}{3\,\sqrt{5}}\,
  \frac{v^2}{M_{\rm Pl}}
  \simeq 1.08 \, {\rm meV}
\end{equation}
is the {\em equilibrium neutrino mass}.  For most of the seesaw parameter space and barring fine-tuned cancellations in the seesaw formula, one has $\widetilde{m}_1 \simeq m_{\rm sol}$ -- $m_{\rm atm}$ corresponding to $K_1 \sim 10$ -- $50$. For these values of $K_1$, most of the produced asymmetry is washed out, since one has $\kappa(K_1) \sim 1/K_1^{1.2} \sim 10^{-3}$ -- $10^{-2}$. However, successful leptogenesis can still be attained for $|\epsilon_1| \sim 10^{-6}$ -- $10^{-5}$. At the same time, for these large values of $K_1$, the value of $\kappa(K_1)$ is independent of the initial $N_1$ abundance. They also imply a washout of a pre-existing asymmetry $Y_{B-L}^{{\rm pre},0}$ as large as $\sim 1$, since its relic final value is given by
\begin{equation}
  Y_{B-L}^{{\rm pre},\infty} \ =\  e^{- \frac{3\pi}{8}\, K_1} \, Y_{B-L}^{{\rm pre}, 0}  \;  ,
\end{equation}
which is therefore exponentially suppressed.  This result is due to the interesting experimental finding $m_{\rm sol},m_{\rm atm} \sim 10\,m_{\star}$, a coincidence that might be regarded as a phenomenological indication of  {\em strong thermal leptogenesis}, wherein the final asymmetry is independent of the initial conditions. Notice that any asymmetry generated by the heavier RH neutrinos, and in particular by the $N_2$'s, will be exponentially washed out and can be neglected. 

Barring fine-tuned cancellations in the seesaw formula, one obtains the upper bound~\cite{Davidson:2002qv}
\begin{equation}
\label{upperbound}
  |\epsilon_1|\ \lesssim\ 10^{-6}\,\frac{M_1}{10^{10}\,{\rm GeV}}\,\frac{m_{\rm atm}}{m_1 + m_3} \;  .
\end{equation}
This upper bound on the CP asymmetry implies an upper bound on the final asymmetry, and the condition of successful leptogenesis yields a lower bound $M_1 \gtrsim 10^9\,{\rm GeV}$~\cite{Davidson:2002qv, Buchmuller:2002rq}.
A more precise value depends on the assumed initial $N_1$ abundance. In the case of strong washout, for $K_1 \gtrsim 3$, there is no such dependence, and one finds $M_1 \gtrsim 3 \times 10^9\,{\rm GeV}$. The lower bound on $M_1$ implies a lower bound on the reheat temperature of the Universe $T_{\rm reh} \gtrsim 1 \times 10^9 \,{\rm GeV}$.  Within gravity-mediated supersymmetric models, this lower bound might be incompatible with the upper bound from avoidance of gravitino over-production~\cite{Khlopov:1984pf, Ellis:1984eq, Kawasaki:2008qe}. However, the latest constraints on supersymmetric models from the LHC strongly relieve the tension, since they favor large values of the gravitino mass above a TeV, making the upper bound more relaxed, $T_{\rm reh} \lesssim 10^{10} \,{\rm GeV}$, and reconcilable with thermal leptogenesis. Allowing for very strong fine-tuning in the seesaw relation, the lower bound can be relaxed if $M_2 \neq M_3$ due to an extra term in the the total CP asymmetry that does not respect the upper bound~\eref{upperbound} and that is suppressed by a factor $(M_1/M_2)^2$~\cite{Hambye:2003rt, Blanchet:2008pw}.


\subsection[Flavor effects in the N1-dominated scenario]{Flavor effects in the $N_1$-dominated scenario}
\label{subsec:N1dominated}

The vanilla leptogenesis scenario and the rate equations in~\eref{dlg1} and~\eref{dlg2} rely on the implicit assumption that leptons produced from the decays of the RH neutrinos  do not lose their coherence in flavor space prior to inverse decays that would otherwise fully wash out the asymmetry produced by the decays. If one depicts the asymmetry produced in the decays in the flavor space of the three charged leptons, this is equivalent to saying that decays and inverse decays all occur along one definite flavor direction, and flavor effects are, therefore, absent in practice. This is the unflavored approximation.\footnote{This is sometimes called the one-flavored approximation. However, this can be misleading, especially when heavy-neutrino flavors are introduced, and we prefer to refer to it as the unflavored approximation. Also notice that in the limit of no washout, corresponding to the case when inverse decays are never in equilibrium, there is no real difference between an unflavored description and a flavored one.} 

However, this picture is highly over-simplified, and a proper account of flavor effects can strongly affect the final value of the asymmetry. Within the $N_1$-dominated scenario, the source of flavor effects is given by the interactions of the charged leptons~\cite{Barbieri:1999ma, Abada:2006fw}, described by $-\,{\cal L}^{\ell}_Y = h\,\bar{\ell}\,\phi\, e_R$. It results from the fact that the charged-lepton and neutrino Yukawa coupling matrices, respectively $h$ and $\lambda$, are, in general, not diagonal in the same basis. Therefore, charged-lepton interactions occurring between decays and inverse decays will tend to break the coherent propagation of the leptons produced in $N_1$ decays 
before their inverse decays~\cite{Blanchet:2006ch}. Charged-lepton interactions are, of course, strongly flavor-dependent, since the eigenvalues of $h$ are very hierarchical: $h_{\tau} \gg h_{\mu} \gg h_e$. This implies that tau interactions, with rate $\Gamma_{\tau} \simeq 8 \times 10^{-3}\,h^2_{\tau} \, T$ are the strongest ones and are effective when $\Gamma_{\tau} \gtrsim\Gamma^{\rm ID}$ for $M_1 \lesssim 5 \times 10^{11}\,{\rm GeV}$. On the other hand, muon interactions are effective for $\Gamma_{\mu} \simeq 10^{-3}\,h^2_{\tau} \, T  \gtrsim \Gamma^{\rm ID}$, implying $M_1 \lesssim 5 \times 10^8 \,{\rm GeV}$.  In this way, we have three important flavor regimes, determined by the mass of the lightest RH neutrino $M_1$, as follows. 


\subsubsection{Unflavored regime: $M_1 \gg 5\times 10^{11}\,{\rm GeV}$}

As discussed earlier, all charged-lepton interactions can be neglected. One then recovers the {\em unflavored regime}, where charged-lepton effects have negligible impact.


\subsubsection{Two-flavor regime: $5 \times 10^{8}\,{\rm GeV} \ll M_1 \ll 5 \times 10^{11} \,{\rm GeV}$}

Leptons of type ${\ell}_1$, produced by the $N_1$ decays, can be described in their inverse decay as an incoherent mixture of a $\tau$ component and an $e+\mu$ component, which we indicate by $\tau_1^{\bot}$. The flavor composition is then determined by the probabilities \smash{$P_{1\alpha} \equiv |\langle \ell_1 | \alpha \rangle |^2$}, with \smash{$\alpha=\tau, \tau_1^\bot$} and such that \smash{$P_{1\tau} + P_{1\tau_1^\bot} =1$}. One can do the same for the anti-leptons, introducing probabilities $\bar{P}_{1\alpha}$. At tree level, the  $\ell_1$ and $\bar{\ell}_1$ quantum states are CP-conjugates of each other.  However, when loop effects are considered, one has $P_{1\alpha} \neq \bar{P}_{1\alpha}$.

The yields for the asymmetry in the two flavors $\tau$ and $\tau_1^{\bot}$, respectively \smash{$Y_{\Delta_\tau}$} and \smash{$Y_{\Delta_{\tau_1^{\bot}}}$}, have to be tracked separately, and we enter a so-called {\em two-flavor regime}. If we indicate the tree-level probabilities with $P^0_{1\alpha}$, their inverse-decay washout term  is then reduced, compared to $W_1^{\rm ID}$, by a factor $P^0_{1\alpha} =  (\Gamma_{1\alpha}+\bar{\Gamma}_{1\alpha})/(\Gamma_1 +\bar{\Gamma}_1)$. The kinetic equation  for the total asymmetry in the unflavored regime,~\eref{dlg2}, is now replaced by two equations: one for \smash{$Y_{\Delta_\tau}$} and one for \smash{$Y_{\Delta_{\tau_1^{\bot}}}$}. The RH-neutrino kinetic equation remains unchanged, and the relevant set of Boltzmann equations is
\begin{subequations}
\label{flke}
\begin{align}
  \frac{{\rm d} Y_{N_1}}{{\rm d}z}\ & = \ -\:D_1\,(Y_{N_1}-Y_{N_1}^{\rm eq}) \; , \\
  \frac{{\rm d}Y_{\Delta_{\tau}}}{{\rm d}z} \ & = \ -\:
  \epsilon_{1\tau}\,D_1\,(Y_{N_1}-Y_{N_1}^{\rm eq}) \:-\:P_{1\tau}^{0}\,W_1\,Y_{\Delta_{\tau}} \; , \\
  \frac{{\rm d}Y_{\Delta_{\tau_1^{\bot}}}}{{\rm d}z} \ & = \ -\:
  \epsilon_{1{\tau}^{\bot}_1}\,D_1\,(Y_{N_1}-Y_{N_1}^{\rm eq})\:-\:P_{1{\tau}^{\bot}_1}^{0}\,W_1\,Y_{\Delta_{\tau_1^{\bot}}} \;   ,
\end{align}
\end{subequations}
where we have introduced the flavored CP asymmetries  ($\alpha=e,\mu,\tau$), given by~\cite{Covi:1996wh}
\begin{align}
\label{veial}
  \epsilon_{1\alpha} \ & =  \
  \frac{1}{8\pi (\lambda^{\dagger}\lambda)_{11}} \, \sum_{j\,\neq\, 1}  \bigg\{ {\rm Im}\,
  \left[ \lambda_{\alpha 1}^{*}\lambda_{\alpha j} (\lambda^{\dagger}\lambda)_{1j}\right] \xi\bigg(1,\frac{M_j^2}{M_1^2}\bigg)
  \nonumber\\
  & + \  \frac{M_1^2}{M_1^2-M_j^2} {\rm Im}\,\Big[\lambda_{\alpha 1}^*\lambda_{\alpha j} (\lambda^{\dagger}\lambda)_{j1}
  \Big] \bigg\} \; ,
\end{align}
and defined $\epsilon_{1\tau_1^{\bot}} \equiv \epsilon_{1e} + \epsilon_{1\mu}$ and $P^0_{1\tau_{1}^\bot} \equiv P^0_{1e} + P^0_{1\mu}$.\footnote{Since $\epsilon_1 = \sum_\alpha \epsilon_{1\alpha}$, one can indeed verify that the expression for $\epsilon_1$ in~\eref{CPas} is recovered after summing over $\alpha$ in~\eref{veial}.} The loop function $\xi(b,x)$ is defined in~\eref{xi} (see also Chapter~\cite{leptogenesis:A04}). If the $\ell_1$ and $\bar{\ell}_1$ quantum states were simply CP-conjugates of each other, the flavored CP asymmetries would just be given by $\epsilon_{1\alpha} = P^0_{1\alpha}\,\epsilon_1$. As mentioned above, this holds at tree level, but loop contributions\footnote{They must necessarily be considered, since the CP asymmetries are generated by the interference of tree-level and one-loop graphs. One would, of course, have  $\epsilon_1=\epsilon_{1\alpha}=0$ at tree level.} generate a mismatch~\cite{Nardi:2006fx} $\Delta P_{1\alpha} \equiv P_{1\alpha} -\bar{P}_{1\alpha}$, so that the flavored CP asymmetries get additional contributions. We then have
\begin{equation}
  \epsilon_{1\alpha}\  =\  \frac{P_{1\alpha} + \bar{P}_{1\alpha}}{2}\,\epsilon_1\:+\:\frac{\Delta P_{1\alpha}}{2}
\end{equation}
and note that $\Delta P_{1\tau} + \Delta P_{1\tau_1^\bot} =0$.

The solution for the final asymmetry is a quite trivial generalization of the result obtained in the unflavored case (see \sref{sec:vanilla}). One has
\begin{equation}
  Y_{B-L}^{\infty}\ =\ Y_{\Delta_{\tau}}^{\infty}\: +\: Y_{\Delta_{\tau_1^{\bot}}}^{\infty} \;  ,
\end{equation}
with $Y_{\Delta_{\tau}}/Y_{N_1}^{\rm eq}(0) \simeq  -\,\epsilon_{1\tau}\,\kappa(K_{1\tau})$ and $Y_{\Delta_{\tau_1^{\bot}}}/Y_{N_1}^{\rm eq}(0) \simeq  -\,\epsilon_{1\tau_1^\bot}\,\kappa(K_{1\tau})$. Barring fine-tuning in the seesaw formula, the total final asymmetry can then be written as~\cite{Blanchet:2008pw}
\begin{equation}
\label{finalas}
  Y_{B-L}^{\infty}/Y_{N_1}^{\rm eq}(0)\  \simeq \ -\,N_{\rm fl} \, \epsilon_1 \, \kappa(K_1)\: +\: \frac{\Delta P_{1 \tau}}{2} \, 
  \left[\kappa(K_{1\tau_1^\bot})-\kappa(K_{1\tau}) \right]\;  ,
\end{equation}
where $N_{\rm fl}$ is an effective number of flavors with value between 1, when there is no washout at all ($K_1 \ll 1$) and the unflavored result is recovered, and 2, the number of flavors. This expression shows that large deviations from the unflavored case can arise only  in the presence of washout, if $\Delta P_{1\alpha} \neq 0$ and $\kappa(K_{1\tau_1^\bot})-\kappa(K_{1\tau}) \neq 0$. For this reason, the lower bound on $M_1$ and on $T_{\rm reh}$ in the limit of no washout are not changed by flavor effects. It should also be said that, allowing for some fine-tuning in the seesaw formula, the flavored CP asymmetries can be enhanced by unbounded extra terms that are suppressed
by $M_1/M_2$. With some mild fine-tuning, and without a too strongly hierarchical spectrum, one can relax the lower bounds on $M_1$ and on $T_{\rm reh}$ to $\sim 10^8\,{\rm GeV}$~\cite{Blanchet:2008pw}.

The most extreme case of deviation from the unflavored case is realised when $\epsilon_1 =0$, implying conservation of total lepton number~\cite{Nardi:2006fx}. Even in this case, if the second term is large enough, one can attain successful leptogenesis~\cite{Blanchet:2006be,Pascoli:2006ie,Pascoli:2006ci,Anisimov:2007mw,Molinaro:2007uv,Molinaro:2008rg,Molinaro:2008cw}. The CP violation then stems uniquely from low-energy phases, although certain conditions on the high-energy parameters still have to be verified. Therefore, the measurement of CP-violating values of low-energy phases is not a sufficient (nor a necessary) condition for successful leptogenesis. However, the discovery of CP violation at low energies, in particular of a CP-violating value of the Dirac phase, as now supported by the data, would, of course, be a very important conceptual result, not least of all because CP violation at low energies is, in general, accompanied by CP violation at high energies.\footnote{Imposing a discrete flavor symmetry, this would not be true: one could have CP-violating values of the low-energy phases with no CP violation at high energies. However, a flavor symmetry has to be broken, and even a very small breaking would be sufficient to generate enough CP violation at high energies to produce the correct asymmetry. Implications of flavor and CP symmetries in leptogenesis are discussed in detail in Chapter~\cite{leptogenesis:A06}.}


\subsubsection{Three-flavor regime: $M_1 \ll 5 \times 10^{8} \, {\rm GeV}$}

In this case, the muon interaction rate is large enough at the asymmetry production that also the leptonic quantum states $\tau_1^\bot$ produced by the $N_1$ decays decohere before they inverse decay. One therefore has to calculate separately the electron asymmetry $Y_{\Delta_e}$ and the muon asymmetry $Y_{\Delta_\mu}$ in addition to the tau asymmetry $Y_{\Delta_\tau}$, thereby realising a {\em three-flavor regime}.

The set of kinetic equations are easily generalized and will comprise three kinetic equations: one for each flavor asymmetry $Y_{\Delta_\alpha}$. However, in this case, the asymmetries are, barring a quasi-degenerate RH-neutrino spectrum or fine-tuning in the seesaw formula, too small to have successful leptogenesis. For this reason, the two-flavor regime is, in general, more significant.\footnote{In a supersymmetric case, the transition between the two- and the three-flavor regimes occurs at $M_1 \simeq 5 \times 10^{8}\,(1 +\tan^2 \beta)\,{\rm GeV}$~\cite{Abada:2006fw}. One can then have successful leptogenesis even in the three-flavor regime.}


\subsection{Density matrix equation}

The unflavored regime and the two-(or three-)flavor regimes are asymptotic limits of a more general physical picture where, at the inverse decay, not all leptonic quantum states $| \ell_1 \rangle $ are either a coherent  superposition or an incoherent admixture but there is a coexistence of both states. In this intermediate regime, a useful statistical description  is provided by a {\em density matrix equation}~\cite{Barbieri:1999ma, Abada:2006fw, DeSimone:2006nrs, Beneke:2010dz, Blanchet:2011xq}. In this more general approach, all abundances are replaced by matrices in (charged-lepton) flavor space.  The density matrix equation is then flavor invariant upon rotations in (charged-lepton) flavor space (see the discussions in~\sref{sec:3flavourcovariance}). In the limit where one interaction dominates over all others in flavor space, the density matrix equation asymptotically reproduces the Boltzmann equations that we discussed above. In the intermediate regime, it manages to give a description of the transition between two different flavor regimes.  For example, we can consider the important transition between the unflavored and the two-flavor regimes. In this case, the only charged-lepton interactions that we can consider are the tau interactions.

When gauge interactions are taken into account, they force the matrix for the sum of leptons and anti-leptons to be given approximately by $Y^{\ell +\bar{\ell}}_{\alpha\beta} = 2 \, Y_{\ell}^{\rm eq}\,\delta_{\alpha\beta}$. This leads to
the following (closed) equation for the  $B-L$ density matrix~\cite{Barbieri:1999ma, Blanchet:2011xq}
\begin{align}
\label{fullyflavoured}
\frac{{\rm d}[Y_{B-L}]_{\alpha\beta}}{{\rm d}z}  \ =\ &-\:
\epsilon^{(1)}_{\alpha\beta}\,D_1\,(Y_{N_1}-Y_{N_1}^{\rm eq})\:-\:\frac{1}{2}\,W_1\,\left\{{\cal P}^{0(1)}, Y_{B-L}\right\}_{\alpha\beta}
\nonumber\\&-\: \frac{\Gamma_{\tau}}{H\, z} \, [\sigma_{1}]_{\alpha\beta}\,[Y_{B-L}]_{\alpha\beta} \;  ,
\end{align}
specialized in the (two) charged-lepton flavor basis \smash{$\tau-\tau_1^{\bot}$}. In this equation, \smash{$\epsilon^{(1)}_{\alpha\beta}$} is the CP asymmetry matrix for $N_1$ decays that feeds the source term, \smash{${\cal P}^{0(1)}_{\alpha\beta}$} is the tree-level flavor projector along the $\ell_1$ direction and $\sigma_1$ is the Pauli matrix. 

As expected, the two-flavor regime is recovered in the limit  $\Gamma_{\tau}/(Hz) \gg W_1$ (or, equivalently, $\Gamma_{\tau} \gg \Gamma_1^{\rm ID} + \bar{\Gamma}_1^{\rm ID}$), when all leptons ${\ell}_1$ experience a tau interaction before inverse decaying. In this limit, the third term on the right-hand side of~\eref{fullyflavoured} efficiently damps the off-diagonal terms, and one immediately recovers the kinetic equations,~\eref{flke}.

The unflavored limit is more tricky, and there is even an interesting twist.  First of all, one can neglect tau lepton interactions. This is equivalent to neglecting the term $\propto \Gamma_{\tau}$ in~\eref{fullyflavoured}. The density matrix equation in the unflavored limit then becomes
\begin{equation}
\label{dmatrixunfl}
  \frac{{\rm d}[Y_{B-L}]_{\alpha\beta}}{{\rm d}z} \  = \
  -\,\epsilon^{(1)}_{\alpha\beta}\,D_1\,(Y_{N_1}-Y_{N_1}^{\rm eq})\:-\:\frac{1}{2}\,W_1\,\left\{{\cal P}^{0(1)}, Y_{B-L}\right\}_{\alpha\beta} \;  .
\end{equation}
Taking the trace of this equation,  one immediately finds the usual equation for $Y_{B-L}$ in the unflavored regime,~\eref{dlg2}.

At the same time, after some easy steps, one can also find an equation for the difference
\begin{align}
  \frac{{\rm d}\big(Y_{\Delta_{\tau\tau}} - Y_{\Delta_{\tau_1^{\bot}\tau_1^{\bot}}}\big)}{{\rm d}z}  \ = \
  &-\:\Delta P_{1\tau}\, D_1 \, (Y_{N_1}-Y_{N_1}^{\rm eq}) \nonumber\\&-\:
  \frac{1}{2}\,W_1 \Big(Y_{\Delta_{\tau\tau}} - Y_{\Delta_{\tau_1^{\bot}\tau_1^{\bot}}}\Big) \;   ,
\end{align}
with solution 
\begin{equation}
  Y_{\Delta_{\tau\tau}} - Y_{\Delta_{\tau_1^{\bot}\tau_1^{\bot}}} \ =\  -\,Y_{N_1}^{\rm eq}(0)\,\Delta P_{1\tau} \, \kappa(K_1/2) \;  ,
\end{equation}
so that, for the leptonic asymmetries, one has
\begin{subequations}
\label{NBmLttTB2}
\begin{align}
  Y^{\infty}_{\Delta_{\tau\tau}}\ &\simeq\ P^0_{1\tau}\,Y_{B-L}^{\infty}
  \:-\: \frac{1}{2}\,Y_{N_1}^{\rm eq}(0)\,\Delta P_{1\tau} \, \kappa(K_1/2) \;  , \\ 
  Y^{\infty}_{\Delta_{\tau_1^{\bot}\tau_1^{\bot}}} \ &\simeq\  P^0_{1\tau_1^{\bot}}\,Y_{B-L}^{\infty} \:+\:\frac{1}{2}\,
  Y_{N_1}^{\rm eq}(0)\,\Delta P_{1\tau}\, \kappa(K_1/2) \; .
\end{align}
\end{subequations}
The second terms on the right-hand sides of the two expressions are the so-called {\em phantom terms}. In the $N_1$-dominated scenario, with no further dynamical stage after the $N_1$ production, they cannot leave any detectable trace since they cancel out in the final $Y_{B-L}$ and, therefore, in $\eta_B$. However, as we will discuss in the next subsection, when heavy-neutrino flavor effects are also taken into account, their exact cancellation at the production can be removed afterwards. In this case, they would give a contribution to the final expression for the baryon asymmetry.  


\subsection{Flavor coupling}

In the Boltzmann equations for the flavored regimes in~\sref{subsec:N1dominated}, the evolution of the flavored asymmetries is independent of each other. For example, in the case of the $N_1$-dominated and two fully-flavored regime, one has that the equations for $Y_{\Delta_{\tau}}$ and $Y_{\Delta_{\tau_1^{\bot}}}$ are decoupled (see~\eref{flke}). The dynamics of the two asymmetries are then independent of one another, and one can say that the two flavors are thermally uncoupled. 

There are, however, different effects (spectator processes) that are able to couple the dynamics of the two flavors~\cite{Barbieri:1999ma, Buchmuller:2001sr, Nardi:2005hs, Blanchet:2008pw}. The most important one is the {\em Higgs asymmetry}. Since the Higgs doublet carries hypercharge, the $\phi$'s  couple to leptons and the $\bar{\phi}$'s couple to anti-leptons. On the other hand, the Higgs asymmetry is clearly unflavored. 

Suppose, for example, that the asymmetry is entirely produced in the tau flavor and not in the $\tau_1^{\bot}$ flavor. The asymmetry created in the former will necessarily be accompanied by an opposite Higgs asymmetry. This, however, through inverse decays, will then necessarily induce an asymmetry also in the $\tau_1^{\bot}$ flavor, even though we have assumed that there is no source term in this flavor. Therefore, the Higgs asymmetry couples the dynamics of the two flavors, thereby realising a kind of thermal contact between them such that the asymmetry in one flavor induces  an asymmetry in the other flavor. In addition to the Higgs asymmetry, one has also to consider that sphaleron processes are able to transfer the asymmetry initially injected into lepton doublets and Higgs bosons  to all other particles, including quarks (indeed creating a baryon asymmetry). A lepton asymmetry created in a specific flavor can then induce asymmetries in the other flavors through baryon asymmetries, analogously to what we have seen for the Higgs asymmetry.  

It should be noticed how, in this case, the inverse decays, which have so far only played the role of washout processes, can actually generate an asymmetry in one flavor, although this is possible only if there is a source term injecting an asymmetry in another flavor from the start.  The Boltzmann equations in the two-flavor regime,~\eref{flke}, then get modified in the following way:
\begin{subequations}
\label{flcke}
\begin{align}
  \frac{{\rm d}Y_{N_1}}{{\rm d}z} \ & =  \ -\,D_1\,(Y_{N_1}-Y_{N_1}^{\rm eq}) \;,\\
  \frac{{\rm d}Y_{\Delta_{\tau_1^{\bot}}}}{{\rm d}z}\ & =  \
  -\:\epsilon_{1\tau_1^{\bot}}\,D_1\,(Y_{N_1}-Y_{N_1}^{\rm eq})\:-\:
  P_{1\tau_1^{\bot}}^{0}\,W_1\,\sum_{\alpha\,=\,\tau_1^\bot,\tau}\,C_{\tau_1^\bot\alpha}^{(2)}\,Y_{\Delta_\alpha} \;,\\
  \frac{{\rm d}Y_{\Delta_{\tau}}}{{\rm d}z}\ & = \
  -\:\epsilon_{1\tau}\,D_1\,(Y_{N_1}-Y_{N_1}^{\rm eq})\:-\:
  P_{1\tau}^{0}\,W_1\,\sum_{\alpha\,=\,\tau_1^\bot,\tau}\,C_{\tau\alpha}^{(2)}\,Y_{\Delta_\alpha}  \; .
\end{align}
\end{subequations}
The flavor coupling matrix $C^{(2)}$ is given by the sum of two contributions,
\begin{equation}
  C_{\alpha\beta}\ =\ C^{\ell}_{\alpha\beta}\:+\:C^{\phi}_{\alpha\beta} \;,
\end{equation}
the first one connecting the asymmetry in the lepton doublets and the second connecting the asymmetry in the Higgs bosons. It relates the asymmetries stored in the lepton doublets and Higgs bosons to the $Y_{\Delta_\alpha}$'s, and one can see how it acts in a way that the asymmetry in a flavor $\beta\neq \alpha$ influences the asymmetry $\alpha$ through the washout terms.  Imposing chemical equilibrium conditions among the different asymmetries, one finds 
\begin{equation}
  C^{\ell(2)}\ =\ \left(\begin{array}{ccc}
  417/589 & -\,120/589 \\ -\,30/589 & 390/589 \end{array}\right) \, \hspace{4mm} \mbox{\rm and}
  \hspace{4mm}
  C^{\phi(2)}\ =\ \left(\begin{array}{ccc}
  164/589 & 224/589 \\
  164/589 & 224/589
  \end{array}\right) \;,
\end{equation}
whose sum yields
\begin{equation}
  C^{(2)} \ \equiv\
  \left(\begin{array}{ccc}
  C^{(2)}_{\tau_1^\bot\tau_1^\bot} & C^{(2)}_{\tau_1^\bot\tau}  \\  C^{(2)}_{\tau\tau_1^\bot} & C^{(2)}_{\tau_1^\bot\tau_1^\bot}
  \end{array}\right) \ =\ 
  \left(\begin{array}{ccc}
  581/589 & 104/589 \\ 194/589 & 614/589 \end{array}\right) \; .
\end{equation}
In the {\em three-flavor regime}, the Boltzmann equations for each flavored asymmetry, taking into account the flavor coupling matrix, become
\begin{equation}
  \label{flkewA}
  \frac{{\rm d}Y_{\Delta_\alpha}}{{\rm d}z} \  =\ -\:
  \epsilon_{1\alpha}\,D_1\,(Y_{N_1}-Y_{N_1}^{\rm eq})
  \:-\:P_{1\alpha}^{0}\,\sum_{\beta\,=\,e,\mu,\tau}\,C^{(3)}_{\alpha\beta}\,W_1^{\rm ID}\,Y_{\Delta_\beta}\;.
\end{equation}
The flavor coupling matrices in the three-flavor regime are given by
\begin{equation}
  C^{\ell(3)}\ =\ \left(\begin{array}{ccc}
  151/179 & -\,20/179 & -\,20/179 \\ -\,25/358 & 344/537 & -\,14/537 \\ -\,25/358 & -\,14/537 & 344/537	
  \end{array}\right)
\end{equation}
and
\begin{equation}
  C^{\phi(3)}\ =\ \left(\begin{array}{ccc}
  37/179 & 52/179 & 52/179 \\
  37/179 & 52/179 & 52/179 \\
  37/179 & 52/179 & 52/179 \\
  \end{array}\right) \; ,
\end{equation}
whose sum yields
\begin{equation}
  C^{(3)} \ \equiv\ 
  \left(\begin{array}{ccc}
  C_{ee}^{(3)} & C_{e\mu}^{(3)} & C_{e\tau}^{(3)} \\
  C_{\mu e}^{(3)} & C_{\mu\mu}^{(3)} & C_{\mu \tau}^{(3)} \\
  C_{\tau e}^{(3)} & C_{\tau\mu}^{(3)} & C_{\tau\tau}^{(3)}	
  \end{array}\right) \ =\ 
  \left(\begin{array}{ccc}
  188/179 & 32/179 & 32/179 \\ 49/358 & 500/537 & 142/537 \\ 49/358 & 142/537 & 500/537	
  \end{array}\right) \;.
\end{equation}
In an $N_1$-dominated scenario, the correction to the final asymmetry from accounting for flavor coupling is at most $40\%$~\cite{JosseMichaux:2007zj}. We will see, however, that the modification introduced by flavor coupling can be much larger in an $N_2$-dominated scenario, and it can even make completely new regions of parameter space accessible. 


\subsection{Heavy-neutrino flavors}

The impact of charged-lepton flavor effects on the $N_1$-dominated scenario is quite important, but, in different cases, it  only provides a correction, as we discussed following~\eref{finalas}. For example, the lower bounds on $M_1$ and $T_{\rm reh}$ in the $N_1$-dominated scenario do not change. The reason is that they are saturated in the limit of no washout, when flavor effects are irrelevant.  However, when heavy-neutrino flavor effects are also considered, their interplay opens up many new opportunities for leptogenesis scenarios, some of which can be realised within certain categories of models embedding the type I seesaw mechanism. 

The first clear consequence of heavy-neutrino flavor effects is that  the final asymmetry receives a contribution from the decays of the different RH neutrino species. If we consider for definiteness the case of three RH neutrino species, one can simply write \smash{$\eta_B = \sum_{i=1,2,3} \, \eta_B^{(i)}$}.   The first thing to notice is that each contribution is non-vanishing only if the mass \smash{$M_i \lesssim z_B^{(i)}\, T_{\rm reh}$}, where \smash{$z_B^{(i)}$} is the particular value of \smash{$M_i/T_{\rm reh}$} about which the asymmetry is generated.  From this point of view, a straightforward condition that can be imposed for the validity of the $N_1$-dominated scenario is to have \smash{$M_2 \gtrsim T_{\rm reh}$}.  However, in general, the next-to-lightest RH-neutrino mass $M_2$ is below the reheat temperature and, in this case, the $N_2$'s are also produced in the thermal bath 
and can potentially contribute to the final asymmetry.

As we said, if charged-lepton flavor effects are neglected, the $N_2$ contribution would be exponentially suppressed by the $N_1$ washout as $\exp(-3\pi \, K_1 /8)$ and since, given the measured values of $m_{\rm sol}$ and $m_{\rm atm}$, one typically has $K_1 \gg 1$, the possibility to have an $N_2$-dominated scenario is relegated to a special region of parameters in which $K_1 \lesssim 1$~\cite{DiBari:2005st}. There is, however, an important caveat to this result. If the $N_1$ washout occurs at temperatures $T \sim M_1 \lesssim T_{\rm sph}^{\rm out}$, where $T_{\rm sph}^{\rm out}$ is the out-of-equilibrium temperature of sphaleron processes, it has no effect, since it will wash out the lepton asymmetry but not the baryon asymmetry~\cite{DiBari:2015svd}. This is a possibility to be taken into account. However, even in the case $M_1 \gtrsim T_{\rm sph}^{\rm off}$, when charged-lepton flavor effects are considered, the washout from the lightest RH neutrino does not necessarily act along the flavor where the asymmetry is produced, and some part might survive and contribute to the observed asymmetry (or even explain it). 

First, suppose that $M_1 \ll 5 \times 10^{8}\,{\rm MeV}$. In this case, the $N_1$ washout acts along the three (orthogonal) charged-lepton flavor directions. One has then to consider separately the asymmetry produced in the three
charged-lepton flavors, obtaining~\cite{Vives:2009zz}
\begin{equation}
  Y_{B-L}^{\infty}\  =\  \sum_{\alpha} Y_{\Delta_\alpha}({T \gtrsim M_1}) \, e^{-\frac{3\pi}{8}\,K_{1\alpha}}\;,
\end{equation}
where $Y_{\Delta_\alpha}({T \gtrsim M_1})$ are the flavored asymmetries produced prior to the $N_1$ washout. One can see that the exponential suppression of the three terms is given by the flavored decay parameters that can be much more easily $\lesssim 1$ than the total decay parameter $K_1 = \sum_{\alpha} K_{1\alpha}$. In this way, the asymmetry produced before the lightest RH-neutrino washout can more easily survive in a particular flavor. 

If both the production of the asymmetry and the $N_1$ washout occur in the same flavor regime and above $5 \times 10^{8}\,{\rm GeV}$, i.e.~either in the unflavored or in the two-flavor regimes, then there is another effect to be considered that reduces the effectiveness of the $N_1$ washout: the {\em projection effect}~\cite{Barbieri:1999ma, Engelhard:2006yg}. This will only act along the flavor component that is parallel either to ${\ell}_1$, 
in the unflavored regime, or to ${\ell}_{\tau^{\bot}_1}$, in the two-flavor regime. The asymmetry in the orthogonal flavor to ${\ell}_1$ or $\tau^{\bot}_1$ cannot be washed out. Both effects have then to be taken into account. 

Within a density matrix formalism, accounting for heavy-neutrino flavor effects basically corresponds to having interactions acting on additional flavor directions. The density matrix equation,~\eref{fullyflavoured}, then generalizes to~\cite{Blanchet:2011xq}
\begin{align}
  \label{denmaeqfinal}
  \frac{{\rm d}[Y_{B-L}]_{\alpha\beta}}{{\rm d}z}   \ & = \  -\:\epsilon^{(1)}_{\alpha\beta}\,D_1\,(Y_{N_1}-Y_{N_1}^{\rm eq})\:-\:\frac{1}{2}\,W_1\,\left\{{\cal P}^{(1)0}, Y_{B-L}\right\}_{\alpha\beta}\nonumber \\
  & - \  \epsilon^{(2)}_{\alpha\beta}\,D_2\,(Y_{N_2}-Y_{N_2}^{\rm eq})\:-\:\frac{1}{2}\,W_2\,\left\{{\cal P}^{(2)0}, Y_{B-L}\right\}_{\alpha\beta}\nonumber  \\
  \nonumber
  & -  \ \epsilon^{(3)}_{\alpha\beta}\,D_3\,(Y_{N_3}-Y_{N_3}^{\rm eq})\:-\:\frac{1}{2}\,W_3\,\left\{{\cal P}^{(3)0}, Y_{B-L}\right\}_{\alpha\beta}\nonumber \\
  & - \  \Gamma_{\tau} \, \left[\left(\begin{array}{ccc}
  1 & 0 & 0 \\
  0 & 0 & 0 \\
  0 & 0 & 0
  \end{array}\right),\left[\left(\begin{array}{ccc}
  1 & 0 & 0 \\
  0 & 0 & 0 \\
  0 & 0 & 0
  \end{array}\right),Y_{B-L} \right]\right]_{\alpha\beta}\nonumber  \\ 
  & - \ 
  \Gamma_{\mu}\,\left[\left(\begin{array}{ccc}
  0 & 0 & 0 \\
  0 & 1 & 0 \\
  0 & 0 & 0
  \end{array}\right),\left[\left(\begin{array}{ccc}
  0 & 0 & 0 \\
  0 & 1 & 0 \\
  0 & 0 & 0
  \end{array}\right),Y_{B-L} \right]\right]_{\alpha\beta}
   \;,
\end{align}
where we have extended the definitions of all quantities introduced for $N_1$ to the two heavier RH neutrinos $N_2$ and $N_3$. Clearly, in a general case, all terms on the right-hand side compete with each other in making lepton quantum states collapse along a particular direction in flavor space and its orthogonal one. However, assuming a hierarchical RH-neutrino spectrum, the different stages of asymmetry production and washout from each RH neutrino species occur sequentially, proceeding from the heaviest to the lightest one. 

In this case, the equation now has different possible limits described by different sets of Boltzmann equations. Each limit is realised differently, depending on how the set of values $\{M_1,M_2,M_3\}$  is arranged in the three different flavor regimes (unflavored, two-flavor and three-flavor):
\begin{itemize}

\item[(a)] There are three different cases for both $N_1$ and $N_2$ in the three-flavor regime ($M_2, M_1 \ll 5 \times 10^8 \, {\rm GeV}$).

\item[(b)] One has three more cases for only $N_1$ in the three-flavor regime ($M_1 \ll 5 \times 10^8\,{\rm GeV}$). This is the {\em $N_2$-dominated scenario} to which we will give special consideration in the next subsection.

\item[(c)] There are three cases for the lightest RH neutrino in the two-flavor regime with $5 \times 10^{11}\,{\rm GeV} \gg M_1 \gg 5 \times 10^8\,{\rm GeV}$.

\item[(d)] Finally, there is the case when all three RH neutrinos are in the unflavored regime, with $M_i \gg 5 \times 10^{11}\,{\rm GeV}$. 

\end{itemize}
%


\subsubsection[N2-dominated scenario and strong thermal leptogenesis]{$N_2$-dominated scenario and strong thermal leptogenesis}

Out of all these 10 possible mass patterns,  the three in (b) have a special interest. The asymmetry produced from $N_1$ is insufficient to reproduce the observed value and, therefore, this has to be reproduced by the next-to-lightest RH neutrinos. The two scenarios with $N_2$ in the two-flavor regime are the only ones that can realise strong thermal leptogenesis, where the final asymmetry is independent of the initial conditions. Within the unflavored assumption, as we discussed, the only condition one has to impose is simply $K_1 \gg 1$, and this is strongly supported by the neutrino mixing data, since $m_{\rm sol}, m_{\rm atm} \sim 10\,m_{\star}$. However, when flavor effects are considered, a possible large pre-existing asymmetry can now avoid more easily the washout from the RH neutrinos. An easy way to wash out a large pre-existing asymmetry in all three flavors is to have $N_1$ in the three-flavor  regime and all three $K_{1\alpha} \gg 1$~\cite{Engelhard:2006yg}. However, in this way, one cannot attain successful leptogenesis, since the lightest RH-neutrino production is insufficient and the asymmetry from the two heavier RH neutrinos is also washed out together with the pre-existing one. 

The only possibility to achieve successful strong thermal leptogenesis is within a tau $N_2$-dominated scenario~\cite{Bertuzzo:2010et}. In this case, a pre-existing tau asymmetry is washed out by $N_2$ inverse decays already in the two-flavor regime (requiring $K_{2\tau} \gg 1$), when the tau flavor is already detected. At the  end of the $N_2$-washout stage, the $N_2$ out-of-equilibrium decays produce a tau asymmetry, which is the one that must reproduce the observed asymmetry. Finally, at the $N_1$ washout, the pre-existing electron and muon asymmetries are also washed out (requiring $K_{1\mu}, K_{1e} \gg 1$), while the tau asymmetry produced by the $N_2$-decays survives (requiring $K_{1\tau} \lesssim 1$) and explains the observed baryon asymmetry. 

As we will see, this seemingly special set of conditions for successful strong thermal leptogenesis can be realised within a well-motivated class of models. Moreover, it is interesting that it implies a lower bound on the lightest neutrino mass $m_1 \gtrsim 10\,{\rm meV}$~\cite{DiBari:2014eqa}, with the precise value depending logarithmically on the initial value of the pre-existing asymmetry. 

Within the $N_2$-dominated scenario, with $5 \times 10^{11}\,{\rm GeV} \gg M_2 \gg 5 \times 10^8 \,{\rm GeV} \gg M_1$, if one neglects flavor coupling, the final asymmetry can be calculated using
\begin{equation}
  Y_{B-L}^{\infty}\ =\ \sum_{\alpha\,=\,e,\mu,\tau}Y_{\Delta_{\alpha}}^{\infty} \;  ,
\end{equation}
with
\begin{subequations}
\label{twofl} 
\begin{align}
  Y_{\Delta_e}^{\infty}\ & \simeq \ -\:Y_{N_1}^{\rm eq}(0)
  \Bigg[\frac{K_{2e}}{K_{2\tau_2^{\bot}}}\,\epsilon_{2 \tau_2^{\bot}}\kappa(K_{2 \tau_2^{\bot}})\nonumber\\&\qquad \qquad
  +\: \Bigg(\epsilon_{2e} - \frac{K_{2e}}{K_{2\tau_2^{\bot}}}\, \epsilon_{2 \tau_2^{\bot}} \Bigg)\,\kappa(K_{2 \tau_2^{\bot}}/2)\Bigg]\,
  e^{-\frac{3\pi}{8}\,K_{1 e}} \;   , \\ \nonumber
  Y_{\Delta_{\mu}}^{\infty}\ & \simeq \ -\:Y_{N_1}^{\rm eq}(0)\Bigg[\frac{K_{2\mu}}{K_{2 \tau_2^{\bot}}}\,
  \epsilon_{2 \tau_2^{\bot}}\,\kappa(K_{2 \tau_2^{\bot}})\nonumber\\&\qquad \qquad +\:
  \Bigg(\epsilon_{2\mu} - \frac{K_{2\mu}}{K_{2\tau_2^{\bot}}}\, \epsilon_{2 \tau_2^{\bot}} \Bigg)\,
  \kappa(K_{2 \tau_2^{\bot}}/2) \Bigg]
  e^{-\frac{3\pi}{8}\,K_{1 \mu}}\;  , \\
  Y_{\Delta_{\tau}}^{\infty}\ & \simeq \ -\:Y_{N_1}^{\rm eq}(0)\epsilon_{2 \tau}\,\kappa(K_{2 \tau})\,e^{-\frac{3\pi}{8}\,K_{1 \tau}} \;   .
\end{align}
\end{subequations}
This expression takes into account phantom terms but neglects flavor coupling. Including flavor coupling, two additional terms should also be taken into account, and these can become dominant in certain cases~\cite{Antusch:2010ms}. These terms contribute to an $\alpha$ flavor asymmetry despite being proportional to the $\beta \neq \alpha$ flavored CP asymmetry. Although these terms are proportional to small off-diagonal numerical coefficients in the flavor coupling matrix, they can in some models open up new regions of parameter space. Therefore, whilst flavor coupling is a correction within the $N_1$-dominated scenario, it can become crucial within the $N_2$-dominated scenario. 


\subsection{Low-energy neutrino parameters} 

Imposing successful leptogenesis is equivalent to constraining the seesaw parameter space and, very interestingly, it involves those heavy-neutrino parameters that we cannot test in low-energy neutrino experiments. If the masses $M_i$ are well above the TeV scale then they also evade all collider constraints. Therefore, leptogenesis provides a unique way to place constraints on these parameters and ideally one would like to over-constrain the seesaw parameter space by combining leptogenesis with low-energy neutrino experimental data. In this way, leptogenesis can be regarded as a very high energy ``experiment'' able to give us information on the physics at very high energies
embedding the seesaw mechanism.

This ambitious strategy encounters, however, a clear difficulty, since the number of seesaw parameters to be tested is much higher than the experimental constraints.  The seesaw parameter space contains 18 additional parameters: 3 RH-neutrino masses and 15 additional parameters in the Dirac mass matrix. A convenient way to parameterize the Dirac mass matrix in the seesaw limit  is the orthogonal parameterization~\cite{Casas:2001sr}
\begin{equation}
  m_D \ =\  U_{\nu}\,D_m^{1/2}\,\Omega\,D_M^{1/2} \;  ,
\end{equation}
following from the seesaw formula,~\eref{seesaw}. In this way, the 15 parameters in the Dirac mass matrix are re-expressed through the 9 low-energy neutrino parameters (3 light neutrino masses and 6 parameters in $U_{\nu}$), the 3 $M_i$ and 6 parameters in the orthogonal matrix $\Omega$.\footnote{The fact that on the right-hand side one has 18 parameters and on the left-hand side 15 parameters of course means that 3 parameters on the right-hand side, e.g.~the three RN-neutrino masses $M_i$, have to be regarded as independent of the 15 parameters in $m_D$.} This parametrization is model independent, meaning that it works for any model embedding the type I seesaw models and allows to take into account automatically the low-energy neutrino experimental information.

The orthogonal matrix $\Omega$ encodes information on the 3 lifetimes and the 3 total CP asymmetries of the RH neutrinos. Low-energy neutrino experiments alone cannot test the seesaw mechanism. The baryon-to-photon number ratio calculated from leptogenesis, $\eta_B^{\rm lep}$, depends on all 18 seesaw parameters, in general. Model independently, leptogenesis is then clearly insufficient to over-constrain the seesaw parameter and, in general, it does not produce testable model-independent predictions. However, a few things might help in reducing the number of independent parameters:
\begin{itemize}

\item Successful leptogenesis might be satisfied only about {\em peaks}, i.e.~only for very special regions in parameter space that can correspond to testable constraints on some  low-energy neutrino parameters.

\item Some of the parameters might cancel out in the calculation of $\eta_B^{\rm lep}$.

\item One might impose some cosmologically-motivated condition  to be respected, such as the {\em strong thermal leptogenesis} (independence of the initial conditions) or, even stronger, that one of the heavy RH neutrino 
species is the dark matter candidate.

\item We might add phenomenological constraints from particle physics, such as collider signatures, charged LFV, EDM's, etc.

\item The seesaw might be embedded within a model that implies conditions on $m_D$ and $M_i$. 

\end{itemize}
%


\subsubsection{Upper bound on neutrino masses in the unflavored regime}

In~\eref{efunf} for the efficiency factor in the unflavored regime, the exponential factor is an effect of $\Delta L =2$ washout processes. If this is combined with the upper bound in~\eref{upperbound} on the total CP asymmetry from the successful leptogenesis condition, one finds an upper bound $m_1 \lesssim 0.1 \,{\rm eV}$~\cite{Buchmuller:2002jk, Buchmuller:2004nz} in addition to the lower bound on $M_1$. Interestingly, this is now confirmed by the current cosmological upper bound placed by the \emph{Planck Collaboration}~\cite{Aghanim:2016yuo}. This upper bound is also  very interesting, since it provides an example of how, despite our starting from 18 parameters, the successful leptogenesis condition, which constrains only one combination of them, can indeed produce testable constraints. The reason is that the final asymmetry in the unflavored approximation does not depend on the 6 parameters in $U$, since this cancels out in $\epsilon_1$, or on the 6 parameters associated with the two heavier RH neutrinos. There are only 6 parameters left ($m_1, m_{\rm atm}, m_{\rm sol}, M_1, \Omega^2_{11}$), out of which two are measured, thereby leaving only 4 free parameters. The asymmetry, however, has a peak strongly suppressed by the value of $m_1$, due mainly to the exponential suppression from $\Delta L=2$ washout processes in~\eref{efunf}. The latter is the origin of the upper bound on $m_1$. 

Notice that  the upper bound is saturated at values $M_1 \sim 10^{13} \,{\rm GeV}$ and, therefore, it still holds when flavor effects are included in the unflavored regime. In the two-flavor regime, due to the fact that the flavored CP asymmetries respect a more relaxed upper bound than the total, and since the washout can be reduced, the upper bound on $m_1$ is relaxed. However, within the validity of the two-flavor regime, it is still $m_1 \lesssim {\cal O}(0.1\,{\rm eV})$. A calculation based on a density matrix formalism should merge the upper bounds calculated within the flavored regimes where Boltzmann equations hold. One expects some relaxation but not much above $0.1\,{\rm eV}$~\cite{Blanchet:2008pw}. In the $N_2$-dominated scenario, the upper bound on $m_1$ is much looser, and one can have solutions for $m_1$ as large as $1\,{\rm eV}$.


\subsection[SO(10)-inspired leptogenesis]{$SO(10)$-inspired leptogenesis}

In the unflavored case,  imposing so-called $SO(10)$-inspired conditions, which essentially corresponds to assuming that the neutrino Dirac mass matrix does not differ too much from the up-quark mass matrix, prevents successful leptogenesis, since $M_1 \ll 10^9 \,{\rm GeV}$ and, at the same time, an $N_2$ contribution is efficiently washed out. However, when flavor effects are considered, the $N_2$ asymmetry can escape the $N_1$ washout for a set of solutions that yield successful leptogenesis. Typically, the final asymmetry is in the tau flavor~\cite{DiBari:2008mp}. Interestingly, this set of solutions requires certain constraints on the low-energy neutrino parameters~\cite{DiBari:2010ux}. For example, the lightest neutrino mass cannot be below $\simeq 1\,{\rm meV}$, i.e.~one expects some deviation from the hierarchical limit, although we do not know any experimental way to test this lower bound fully at present. It should be added that $SO(10)$-inspired leptogenesis also strongly favors normally-ordered neutrino masses and that, for $m_1 \simeq m_{\rm sol} \simeq 10\,{\rm meV}$, it is allowed only for $\theta_{23}$ in the first octant. Recently, it has been noticed~\cite{DiBari:2017uka} that for the current favored values of $\delta \sim -\,\pi/2$, the effective Majorana mass $m_{ee}$ of $0\nu\beta\beta$ decay cannot be too much lower than $\sim 10\,{\rm meV}$. Scatter plots of the solutions in the plane $m_{ee}$ versus $m_1$ are shown in the left panel of Fig.~2. Yellow points indicate the dominant tau solutions (the orange points are obtained in the approximation $V_L =I$), and green points indicate some marginal muon solutions, which are now almost entirely excluded by the cosmological upper bound on $m_1$. If such values of $\delta$ are confirmed then $0\nu\beta\beta$ experiments will be able to test $SO(10)$-inspired leptogenesis fully in the coming years. 

\begin{figure}[!t]
\begin{center}
\includegraphics[width=60mm]{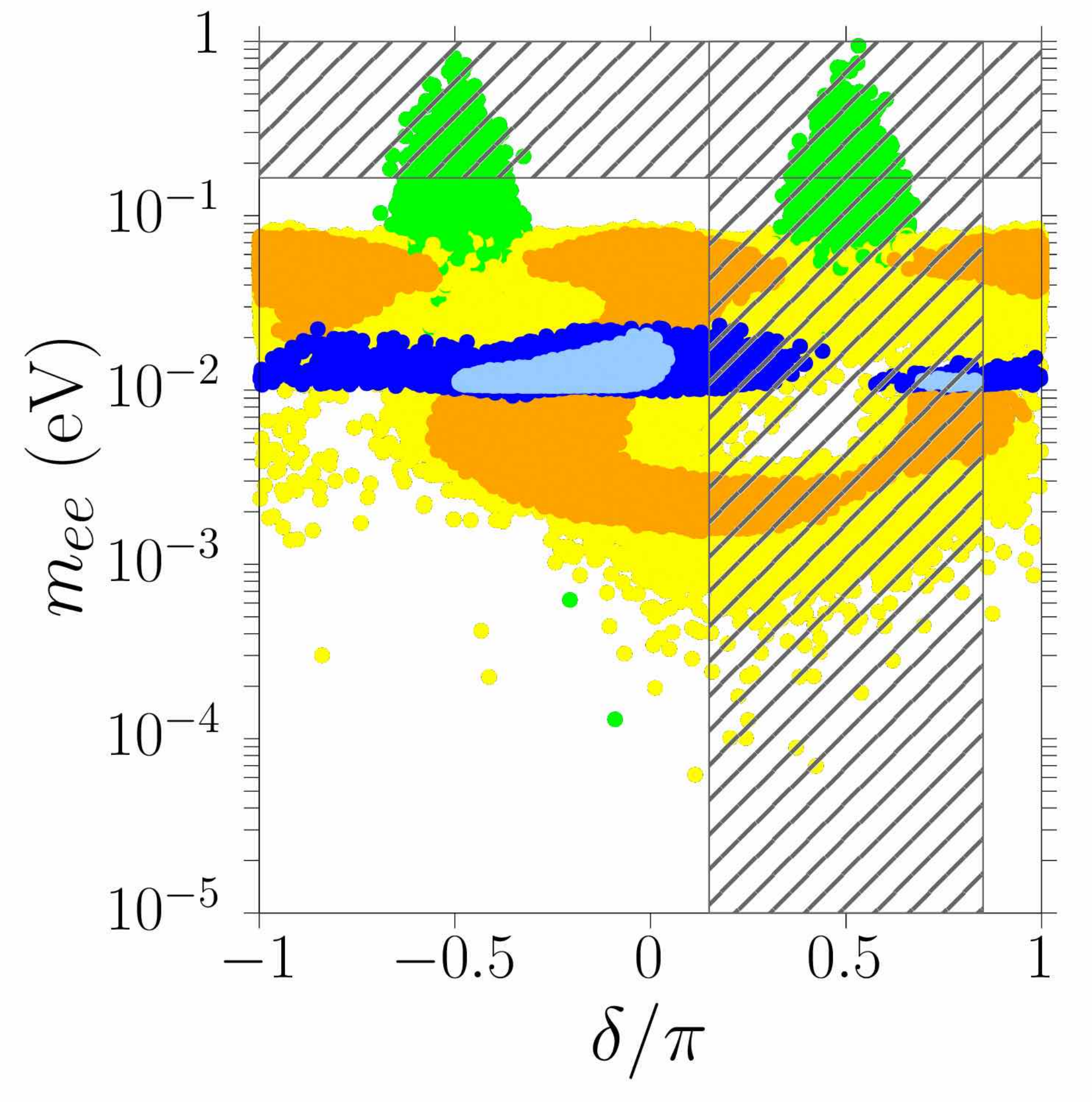} \hspace*{1mm}
\includegraphics[width=60mm]{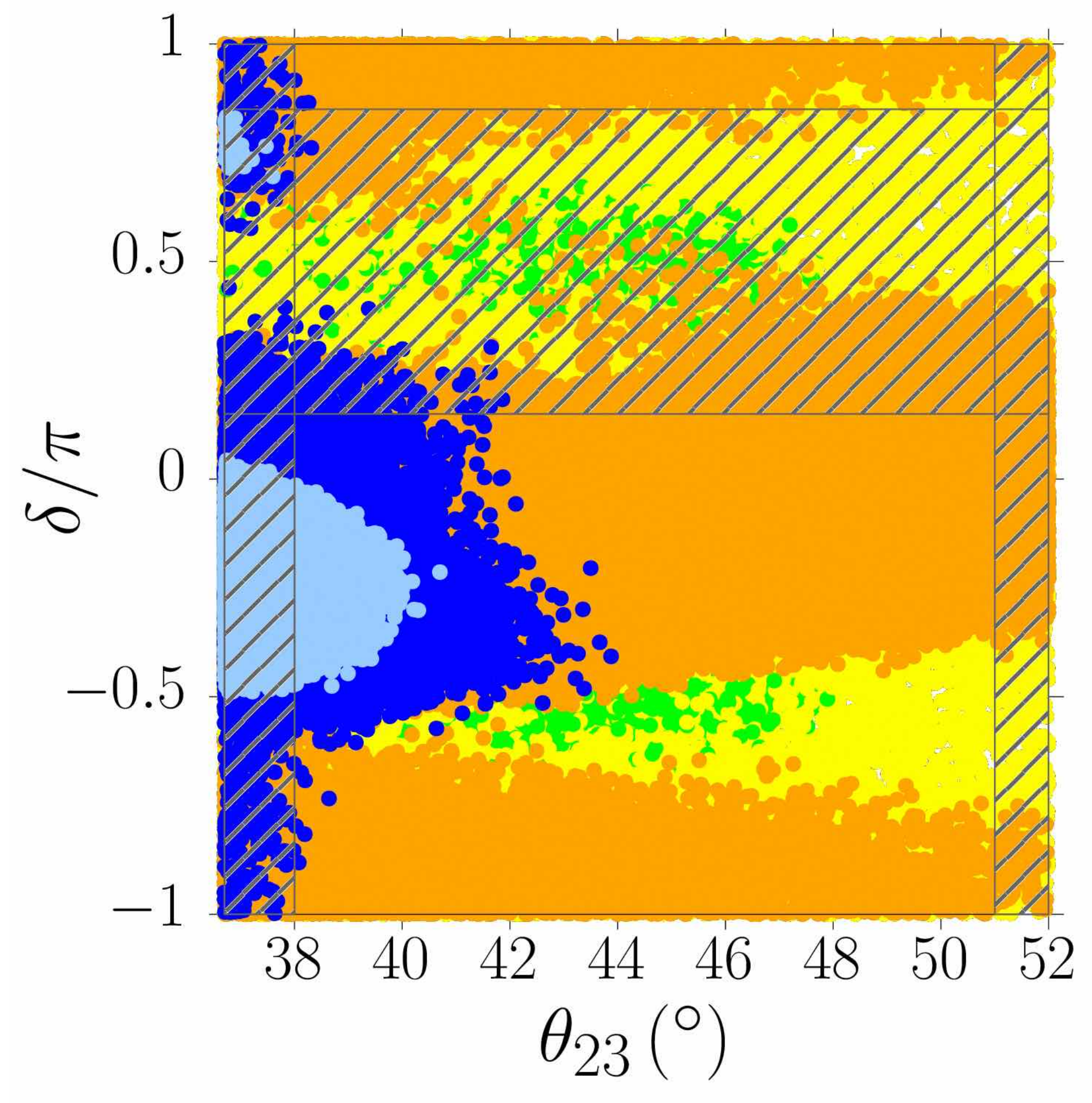}
\end{center}
\caption{Scatter plots of the solutions projected on the shown planes: $m_{ee}$ versus $\delta$ (left) and $\delta$ versus $\theta_{23}$ (right). The yellow points are obtained imposing just successful $SO(10)$-inspired solutions while the blue points are a subset imposing in addition the strong thermal leptogenesis condition (orange and light blue points are for $V_L=I$). Figure taken from Ref.~\cite{DiBari:2017uka}.}       
\end{figure}

It is possible to find very accurate expressions for all important quantities necessary to calculate the asymmetry in $SO(10)$-inspired leptogenesis. We refer the reader to Refs.~\cite{DiBari:2014eya, DiBari:2017uka} for a detailed discussion. Here, we just give some basic hints and results. The first step is that the Dirac mass matrix can be diagonalized by means of two unitary transformations $V_L$ and $U_R$, acting respectively on the left-handed and right-handed neutrino fields:
\begin{equation}
  m_D \ =\ V_L^{\dagger} \, D_{m_D} \, U_R \; ,
\end{equation}
where we have defined $D_{m_D} \equiv {\rm diag}(m_{D1},m_{D2},m_{D3})$. If one plugs this expression into the seesaw formula,~\eref{seesaw}, one finds $M^{-1} = U_R \, D_M^{-1} \,  U_R^{\mathsf{T}}$, where $M^{-1} \equiv D_{m_D}^{-1}\, V_L \, U_{\nu} \, D_m \, U_{\nu}^{\mathsf{T}} \, V_L^{\mathsf{T}} \, D_{m_D}^{-1}$ is the inverse of the Majorana mass matrix in the Yukawa basis (where $m_D$ is diagonal). Assuming $m_{D3} \gg m_{D2} \gg m_{D1}$, one can find accurate analytic expressions both for the RH-neutrino mixing matrix $U_R$ and for the RH-neutrino masses $M_i$. For example, for the RH-neutrino masses, one finds
\begin{equation}
M_1   \ \simeq \  \alpha_1^2 \, \frac{m^2_{\rm up}}{|(\widetilde{m}_\nu)_{11}|} \;, \;\;
M_2  \ \simeq  \  \alpha_2^2 \, \frac{m^2_{\rm charm}}{m_1 \, m_2 \, m_3 } \, \frac{|(\widetilde{m}_{\nu})_{11}|}{|(\widetilde{m}_{\nu}^{-1})_{33}|  } \;,  \;\;
M_3  \ \simeq \  \alpha_3^2\, {m^2_{\rm top}}\,|(\widetilde{m}_{\nu}^{-1})_{33}|  \;,
\end{equation}
where we have defined $(\alpha_1,\alpha_2,\alpha_3) \equiv (m_{D1}/m_{\rm up}, m_{D2}/m_{\rm charm}, m_{D3}/m_{\rm top})$ and $\widetilde{m}_{\nu} \equiv V_L\,m_{\nu}\,V_L^{\mathsf{T}}$. In this way, one arrives at a full analytic expression $\eta_B(m_{\nu};\alpha_i,V_L)$, allowing an analytic understanding of all constraints on low-energy neutrino parameters.


\subsubsection[Strong thermal SO(10)-inspired solution]{Strong thermal $SO(10)$-inspired solution}

As we discussed, when flavor effects are taken into account, there is only one scenario of (successful) leptogenesis allowing for independence of the initial conditions: the tau $N_2$-dominated scenario, where the asymmetry is produced by the $N_2$ decays in the tau flavor~\cite{Bertuzzo:2010et}. As we have seen, the conditions are quite special, since it is required that a large pre-existing asymmetry be washed out by the lightest RH neutrino in the electron and muon flavors. The next-to-lightest RH neutrinos both wash out a large pre-existing tau asymmetry  and also produce  the observed asymmetry in the same tau flavor, escaping the lightest RH-neutrino washout.

It is then highly non-trivial that this quite special set of conditions can be realised by a subset of the $SO(10)$-inspired solutions satisfying successful leptogenesis~\cite{DiBari:2013qja}. For this subset, the constraints are quite stringent and they pin down a well-defined solution: the strong thermal $SO(10)$-inspired solution. This is characterized by a non-vanishing reactor mixing angle, normally-ordered neutrino masses, an atmospheric mixing angle in the first octant and $\delta$ in the fourth quadrant  ($\sin\delta <0$ and $\cos\delta >0$).  In addition, the lightest neutrino mass has to be within a fairly narrow range of values about $m_1 \simeq 20\,{\rm meV}$, corresponding
to a sum of neutrino masses --- the quantity tested by cosmological observations --- $\sum_i m_i \simeq 95\,{\rm meV}$, implying a deviation from the normal-hierarchy prediction of $\sum_i m_i \simeq 60\,{\rm meV}$, detectable during the coming years. At the same time, the solution also predicts a $0\nu\beta\beta$ signal with $m_{ee} \simeq 0.8\, m_1 \simeq 16\,{\rm meV}$. In light of the latest experimental results discussed earlier, this solution is quite intriguing, since, in addition to relying on the same moderately strong washout as vanilla leptogenesis and due to the fact that both the solar and atmospheric scales are $\sim 10\,{\rm meV}$ --- the leptogenesis conspiracy~\cite{Blanchet:2008pw} --- it has also correctly predicted  a non-vanishing reactor mixing angle and is currently in very good agreement with the best-fit parameters from neutrino-mixing experiments. (To our knowledge, it is the only model that has truly predicted $\sin \delta < 0$.) Notice that the possibility to have a large pre-existing asymmetry prior to the onset of leptogenesis at the large reheat temperatures required is quite a plausible possibility, so that the assumption of strong thermal leptogenesis should be regarded as a reasonable setup. (In particular, one could have a traditional GUT baryogenesis followed by leptogenesis.)

It is also possible to consider a supersymmetric framework for $SO(10)$-inspired leptogenesis~\cite{DiBari:2015svd}. In this case, the most important modification to be taken into account is that the critical values for $M_1$, which set the transition from one flavor regime to another, are enhanced by a factor $1+\tan^2\beta$ and, for sufficiently large values of $\tan\beta$, the production might occur in a three-flavor regime rather than in a two-flavor regime. This typically goes in the direction of enhancing the final asymmetry, since the washout at the production is reduced.


\subsubsection{Realistic models}

A first example of realistic models satisfying $SO(10)$-inspired conditions and able to fit all lepton and quark mass and mixing parameters are, as one might expect, $SO(10)$ models. A specific example is given by renormalizable $SO(10)$ models for which the Higgs fields belong to 10-, 120-, 126-dim representations, yielding specific mass relations among the various fermion mass matrices.  Recently, reasonable fits have been obtained that typically point  to a compact RH-neutrino spectrum with all RH-neutrino masses falling in the two-flavor regime. This compact-spectrum solution implies, however, huge fine-tuned cancellations in the seesaw formula.  Even so, fits realising the $N_2$-dominated scenario have been obtained~\cite{Dueck:2013gca, Babu:2016bmy}, and, in this case, there is no fine-tuning in the seesaw formula. Note that $SO(10)$-inspired conditions can also be realised beyond $SO(10)$ models.
For example,  a Pati-Salam model combined with $A_4$ and $Z_5$ discrete symmetries has recently been proposed, satisfying $SO(10)$-inspired conditions and also successful $SO(10)$-inspired leptogenesis~\cite{DiBari:2015oca}. On the other hand, a realistic model realising strong thermal $SO(10)$-inspired leptogenesis has not yet been found.


\section{\label{sec:RL}Flavor and low-scale resonant leptogenesis}
\label{sec3}

\noindent When the mass splitting of two of the heavy neutrinos is small compared to their widths, self-energy effects on the CP asymmetry can dominate and the CP violation can be resonantly enhanced~\cite{Liu:1993tg,Flanz:1994yx,Flanz:1996fb,Covi:1996fm,Covi:1996wh,Pilaftsis:1997jf,Pilaftsis:1997dr,Buchmuller:1997yu} (see also~\cite{Kuzmin:1985mm}). This allows for the scale of successful leptogenesis to be lowered to energies in the $\mathrm{TeV}$ range~\cite{Pilaftsis:2005rv}, making \emph{resonant leptogenesis} (RL)~\cite{Pilaftsis:2003gt} directly testable at current and near-future experiments. A comprehensive discussion of RL is provided in the accompanying Chapter~\cite{leptogenesis:A03}, and we focus here only on the importance of flavor effects in these low-scale models.

The rate equations in the preceding section are covariant under flavor transformations of the SM lepton doublets. However, they are specifically written in the RH-neutrino mass eigenbasis. Therefore, it is natural to ask: is it possible to write rate equations that are \emph{fully} flavor-covariant, also maintaining flavor-covariance at each stage of the calculation?

This question, in addition to being of conceptual interest, has practical consequences for RL. As we will see below, amongst other things, flavor covariance requires us to take into account quantum coherences between different flavors; in the resonant regime, the RH neutrinos are quasi-degenerate and thus one can expect that their quantum coherences may play a significant role. Resonant leptogenesis allows the successful construction of low-scale models of leptogenesis and, at such low scales, one would naively expect to be in the fully-flavored regime discussed in \sref{sec_regimes} for the charged leptons, where their flavor decoherence has already taken place. However, when studying low-scale models of leptogenesis, one is particularly interested in their \emph{testability}, i.e.~in their observable effects at current and near-future experiments. As will be clear from the example discussed below, in low-scale models with observable signatures, at least some of the Yukawa couplings are sufficiently large that their effect will partially recreate coherences in the charged-lepton sector~\cite{Dev:2014laa, Dev:2015wpa}. Hence, a \emph{fully flavor-covariant} treatment~\cite{Dev:2014laa, Dev:2014wsa, Dev:2014tpa, Dev:2015dka, Dev:2015wpa}, which will describe coherences in both the charged-lepton and RH-neutrino sectors, is of particular and quantitative importance in \emph{low-scale testable models} of resonant leptogenesis.


\subsection{Flavor covariance}
\label{sec:3flavourcovariance}

The lepton-doublet and RH-neutrino field operators  $\ell_{\alpha}$ and $N_{Rk}$ transform under flavor rotations $U(\mathcal{N}_{\ell})\otimes U(\mathcal{N}_{N})$ as follows:\footnote{So as to avoid confusion, we do not suppress the $\dagger$ on Hermitian-conjugate fields as in Ref.~\cite{Dev:2014laa}.}
\begin{subequations}
\begin{gather}
  \ell_{\alpha} \ \to \ \ell'_{\alpha} \ = \ V_{\alpha}^{\phantom{\alpha}\beta}\ell_{\beta}\;,\qquad 
  \ell^{\dag\alpha} \ \to \ \ell^{\dag\prime\alpha} \ = \ V^{\alpha}_{\phantom{\alpha}\beta}\ell^{\dag \beta}\;, 
  \\
  N_{Rk} \ \to \ N'_{Rk} \ = \ U_{k}^{\phantom{k}l}N_{Rl},\qquad 
  N^{\dag k}_R \ \to \ 
  N^{\dag\prime k}_R \ =  \ U^{k}_{\phantom{k}l}N_R^{\dag l}\; ,
\end{gather}
\end{subequations}
where $V_{\alpha}^{\phantom{\alpha}\beta} \in U(\mathcal{N}_{\ell})$ and $U_{k}^{\phantom{k}l} \in U(\mathcal{N}_{N})$. Here and in the following, we adopt a flavor-covariant notation in which lower (upper) indices denote covariant (contravariant) transformation properties. In this notation, the relevant part of the Lagrangian in~\eref{eq:1_lagrangian_general} can be written as
\begin{equation}
  -\mathcal{L}_N  \  = \ \lambda_{\alpha}^{\phantom{\alpha}k}  \overline{\ell}^{\alpha} 
  \phi^c  N_{Rk} 
  + \frac{1}{2} \overline{N_{Rk}^{c}} [M_N]^{k l} 
  N_{Rl} + {\rm h.c.}\;,
  \label{3_Lagrangian}
\end{equation}
which is invariant under flavor transformations if the Yukawa couplings and Majorana mass matrix transform as spurions:
\begin{subequations}
\begin{gather}
  \lambda_{\alpha}^{\phantom{\alpha}k} \ \rightarrow \ \lambda_{\alpha}^{\prime\!\phantom{\alpha}k} \ = \ V_{\alpha}^{\phantom{\alpha}\beta}
  \;
  U^k_{\phantom{k} l} \; \lambda_{\beta}^{\phantom{\beta}l} \;, \\
  [M_N]^{kl} \ \rightarrow \ 
  [M'_N]^{kl} \ = \ U^k_{\phantom{k} m} \;
  U^l_{\phantom{l} n} \; [M_N]^{mn} \;.
\end{gather}
\end{subequations}
In order to maintain flavor covariance at all stages, the plane-wave decompositions of the field operators are written in a manifestly flavor-covariant way~\cite{Dev:2014laa}, e.g.
\begin{align}
  \ell_{\alpha}(x) \ & = \  \sum_{s\,=\,+,-} \int_{\mathbf{p}} \Big[\big(2E_{\ell}(\mathbf{p})\big)^{-1/2}\Big]_{\alpha}^{\phantom{\alpha}\beta}
  \notag \\ &\qquad \times \Big( \big[e^{-ip\cdot x}\big]_{\beta}^{\phantom{\beta}\gamma}\, [u(\mathbf{p},s)]_{\gamma}^{\phantom{\gamma}\delta}\,
    b_{\delta}(\mathbf{p},s)  \: +\: \big[e^{ip\cdot x}\big]_{\beta}^{\phantom{\beta}\gamma}\, [v(\mathbf{p},s)]_{\gamma}^{\phantom{\gamma}\delta}\,
    d_{\delta}^{\dagger}(\mathbf{p},s)
  \Big)\;,
\label{3_leptonfieldoperator}
\end{align}
where $[E_{\ell}^2(\mathbf{p})]_{\alpha}^{\phantom{\alpha}\beta}=\mathbf{p}^2\delta_{\alpha}^{\phantom{\alpha}\beta}+[M_{\ell}^{\dag}M_{\ell}]_{\alpha}^{\phantom{\alpha}\beta}$, with $M_{\ell}$ being the charged-lepton mass matrix, here generically taken as non-vanishing. We see that flavor covariance requires the Dirac four-spinors $ [u(\mathbf{p},s)]_{\gamma}^{\phantom{\gamma}\delta}$ and $[v(\mathbf{p},s)]_{\gamma}^{\phantom{\gamma}\delta}$ to transform as rank-$2$ tensors in flavor space, since they are solutions of the Dirac equation, which is matrix-valued in flavor space. 

Equation~\eqref{3_leptonfieldoperator} shows that the creation and annihilation operators for particles ($b^{\dag \alpha}$, $b_{\alpha}$), and anti-particles ($d^{\dag}_{\alpha}$, $d^{\alpha}$) need to transform in conjugate representations, in order to have flavor covariance. Therefore, relations such as the ordinary charge conjugation $C$ and the Majorana condition for the RH neutrinos, which relate particle and anti-particle operators, cannot be valid in an arbitrary flavor basis. Instead, one is forced to consider generalized $\rm C$ transformations, denoted $\widetilde{\rm C}$, which involve a unitary matrix $\mathcal{G}^{\alpha\beta}\equiv [V^*V^\dag]^{\alpha\beta}$, describing the rotations to and from the basis in which the ``standard'' $\rm C$-transformations are defined:\footnote{We emphasise that the $C$-transformations are defined only up to an arbitrary complex phase.}
\begin{equation}
 b_{\alpha}(\mathbf{p},s)^{\tilde{c}} \  \equiv \ \mathcal{G}^{\alpha\beta} \,
  b_{\beta}(\mathbf{p},s)^{c} \ = \  \mathcal{G}^{\alpha\beta} \,\mathcal{G}_{\beta\gamma}\,
  d^{\gamma}(\mathbf{p},s)
  = \  d^{\alpha}(\mathbf{p},s) \;.
\end{equation}
Analogously, the Majorana condition for the RH neutrinos involves a matrix $G^{k l}$, which can be taken equal to the identity in the mass eigenbasis. Notice also the order of flavor indices, dictated by flavor covariance, in the definition of the number densities:
\begin{equation}
  [n_{\ell}]_{\alpha}^{\phantom{\alpha}\beta} \  \sim \ \langle b^{\dag \beta} \,
  b_{\alpha} \rangle \;, \qquad [\bar{n}_{\ell}]_{\alpha}^{\phantom{\alpha}\beta} \  \sim \ \langle d^\dag_{\alpha} \,
  d^{\beta}\rangle \;,
\end{equation}
which implies that $n_{\ell}$ and $\bar{n}_{\ell}$ are $\widetilde{C}$-conjugate quantities: $n_{\ell}^{\tilde{c}} = \bar{n}_{\ell}^{\mathsf{T}}$, where $\mathsf{T}$ denotes the matrix transpose. Analogously, the RH-neutrino number densities are defined as
\begin{equation}
  [n_N]_{k}^{\phantom{k}l} \  \sim \ \langle a^{\dag l} \,
  a_k \rangle \;, \qquad [\bar{n}_N]_{k}^{\phantom{k}l} \  \sim \ G_{km} [n_N]_{n}^{\phantom{n}m} G^{nl}\;,
\end{equation}
and $n_N^{\tilde{c}} = \bar{n}_N^{\mathsf{T}}$. Thus, we can define number densities with definite $\widetilde{\rm C}{\rm P}$-transformation properties:
\begin{equation}
  \underline{n}_N\ = \ \frac{1}{2}\Big(n_N\:+\:\bar{n}_N\Big)\;,\qquad n_{\Delta N}\ =\ n_N\:-\:\bar{n}_N\;, \qquad n_{\Delta \ell}\ = \ n_{\ell}\:-\:\bar{n}_{\ell}\;.
\end{equation}
Notice that the $\widetilde{\rm C}{\rm P}$-odd $n_{\Delta N}$ is purely imaginary and off-diagonal in the RH-neutrino mass eigenbasis, i.e.~it encodes the CP-violating coherences present in the RH-neutrino sector. Instead, the $\widetilde{\rm C}{\rm P}$-even $\underline{n}_N$ describes the RH neutrino populations and $\widetilde{\rm C}{\rm P}$-even coherences, and $n_{\Delta \ell}$ is nothing other than the matrix of asymmetries in the LH charged leptons.


\subsection{Rate equations}
\label{subsec:rateequations}

The requirement of flavor covariance and the definite $\widetilde{\rm C}{\rm P}$-properties of the number densities introduced in~\sref{sec:3flavourcovariance} fix the form of the flavor-covariant generalization of the rate equations (cf.~Chapter~\cite{leptogenesis:A04}). For the moment, let us extract the $\widetilde{\rm C}{\rm P}$-even and -odd parts of the various rates as
\begin{equation}
  \gamma^X_Y \ \equiv \ \gamma(X \to Y) + \gamma(\bar{X} \to \bar{Y})\;, \qquad \delta\gamma^X_Y \ \equiv \ \gamma(X \to Y) - \gamma(\bar{X} \to \bar{Y})\;.
\end{equation}
We will discuss the physical issues related to $\widetilde{\rm C}{\rm P}$ violation in the rates later on. The Majorana nature of the RH neutrinos causes the appearance of real and imaginary parts of the rates in their rate equations that need to be defined conveniently in a covariant manner~\cite{Dev:2014laa}, and we will denote them here by a tilde.

With these considerations, the general form of the rate equations describing RH-neutrino oscillations, decays, inverse decays, $\Delta L=2$ scatterings and charged-lepton decoherence processes is~\cite{Dev:2014laa}:\\[0.25em]
\begin{subequations}
\begin{align}
 \frac{H_{N} \, s}{z}\,
   \frac{\mathrm{d}[\underline{Y}_N]_{k}^{\phantom{k}l}}{\mathrm{d}z} \   &= \ - \, i \, \frac{s}{2} \,
  \Big[\mathcal{E}_N,\, Y_{\Delta N}\Big]_k^{\phantom{k}l}
+ \, \Tdu{\big[\widetilde{\rm Re}
    (\gamma^{N}_{\ell \phi})\big]}{}{}{k}{l} \nonumber\\&\quad\;\;
  - \, \frac{1}{2 \, Y_N^{\rm eq}} \,
  \Big\{\underline{Y}_N, \, \widetilde{\rm Re}(\gamma^{N}_{\ell \phi})
  \Big\}_{k}^{\phantom{k}l} \;,
  \label{3_etanrateeq}\\[6pt]
  \frac{H_{N} \, s}{z}\,
  \frac{\mathrm{d}[Y_{\Delta N}]_{k}^{\phantom{k}l}}{\mathrm{d}z} \ 
  &= \ - \, 2 \, i \, s \,
  \Big[\mathcal{E}_N,\, \underline{Y}_N\Big]_k^{\phantom{k}l} \, + \, 2\, i\,  \Tdu{\big[\widetilde{\rm Im}
    (\delta \gamma^{N}_{\ell \phi})\big]}{}{}{k}{l} \notag\\
  &\quad\;\; - \, 
  \frac{i}{Y_N^{\rm eq}} \, \Big\{\underline{Y}_N, \,
  \widetilde{\rm Im}
  (\delta\gamma^{N}_{\ell \phi}) \Big\}_{k}^{\phantom{k}l}  - \, \frac{1}{2 \, Y_N^{\rm eq}}  \,
  \Big\{Y_{\Delta N}, \, \widetilde{\rm Re}(\gamma^{N}_{\ell \phi})
  \Big\}_{k}^{\phantom{k}l}
  \label{3_deltaetanrateeq}\;, \\[6pt]
 \frac{H_{N} \, s}{z}\, 
\frac{\mathrm{d}[Y_{\Delta \ell}]_{\alpha}^{\phantom{\alpha}\beta}}
  {\mathrm{d}z} \ 
  &= \ - \, \Tdu{[\delta \gamma^{N}_{\ell\phi}]}{\alpha}{\beta}{}{} \,
  +\, \frac{[\underline{Y}_N]_{l}^{\phantom{l}k}}
  {Y_N^{\rm eq}} \,
  \Tdu{[\delta \gamma^{N}_{\ell\phi}]}{\alpha}{\beta}{k}{l} 
   + \, \frac{[Y_{\Delta N}]_{l}^{\phantom{l} k}}{2\,Y_N^{\rm eq}}  \,
  \Tdu{[\gamma^{N}_{\ell \phi}]}{\alpha}{\beta}{k}{l} \notag\\ 
  &\quad\;\; - \frac{1}{3} \,
  \Big\{ Y_{\Delta \ell} , \,
  {\gamma}^{\ell\phi}_{\ell^{\tilde{c}} \phi^{\tilde{c}}} 
  + {\gamma}^{\ell\phi}_{\ell \phi}\Big\}_{\alpha}^{\phantom{\alpha} \beta}  
  \, - \, \frac{2}{3} \, \Tdu{[Y_{\Delta \ell}]}{\delta}{\epsilon}{}{} \,
  \Tdu{[{\gamma}^{\ell\phi}_{\ell^{\tilde{c}} \phi^{\tilde{c}}} - {\gamma}^{\ell\phi}_{\ell \phi}]}{\epsilon}{\delta}{\alpha}{\beta} 
 \notag\\[3pt]
  & \quad\;\; - \frac{2}{3} \, 
  \Big\{Y_{\Delta \ell}, \, 
  \gamma_{\rm dec } \Big\}_{\alpha}^{\phantom{\alpha}\beta} \,
  +\, [\delta \gamma_{\rm dec}^{\rm back}]_{\alpha}^{\phantom{\alpha}\beta}\;,
\end{align}
\end{subequations}
\\
\noindent where $z$ is defined in terms of the temperature $T$ and heavy-neutrino mass scale $M$ as $z\equiv M/T$ (see Chapter~\cite{leptogenesis:A04}). The generalized real and imaginary parts of an Hermitian matrix $A$ are defined via
\begin{subequations}
\begin{align}
  [\widetilde{\mathrm{Re}}(A)]_{\alpha}^{\phantom{\alpha}\beta}\ &\equiv\ \frac{1}{2}\Big(A_{\alpha}^{\phantom{\alpha}\beta}\:+\:G_{\alpha\lambda}A_{\mu}^{\phantom{\mu}\lambda}G^{\mu\beta}\Big)\;,\\
  [\widetilde{\mathrm{Im}}(A)]_{\alpha}^{\phantom{\alpha}\beta}\ &\equiv\ \frac{1}{2i}\Big(A_{\alpha}^{\phantom{\alpha}\beta}\:-\:G_{\alpha\lambda}A_{\mu}^{\phantom{\mu}\lambda}G^{\mu\beta}\Big)\;.
\end{align}
\end{subequations}
These rate equations have been written in terms of the yields (see Chapter~\cite{leptogenesis:A04})
\begin{equation}
  \underline{Y}_N(z)\ \equiv\ \frac{\underline{n}_N(z)}{s(z)}\;,\qquad Y_{\Delta N}(z)\ \equiv\ \frac{n_{\Delta N}(z)}{s(z)}\;,\qquad Y_{\Delta \ell}(z)\ \equiv\ \frac{n_{\Delta \ell}(z)}{s(z)}\;.
\end{equation}
While the form of the rate equations is essentially dictated by flavor covariance, it can be obtained explicitly by a semiclassical analysis~\cite{Dev:2014laa} and a field-theoretic Kadanoff-Baym treatment~\cite{Dev:2014wsa}.

\begin{figure}
\centering
\subfigure[][The in-medium inverse heavy-neutrino decay: $n_{\phi}\protect{[}n_{\ell}\protect{]}_{\beta}^{\protect{\phantom{\beta}}\alpha}\protect{[}\gamma(\ell\phi\to N)\protect{]}_{\alpha\protect{\phantom{\beta}}k}^{\protect{\phantom{\alpha}}\beta\protect{\phantom{k}}l}$.]{\parbox{\textwidth}{\centering \includegraphics[scale=0.7]{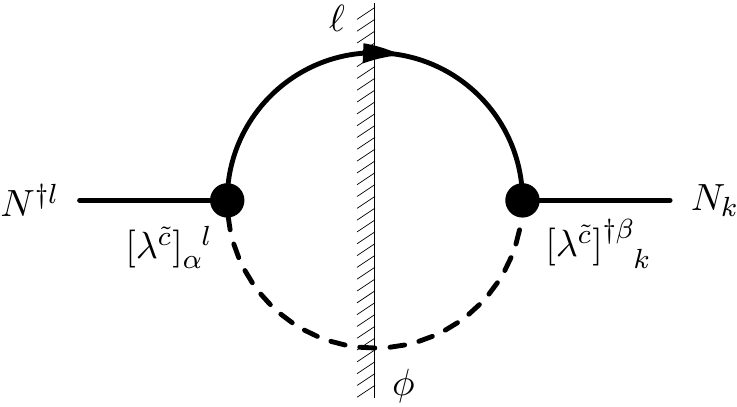}\vspace{0.5em}\\$\bigg\downarrow$\\\includegraphics[scale=0.7]{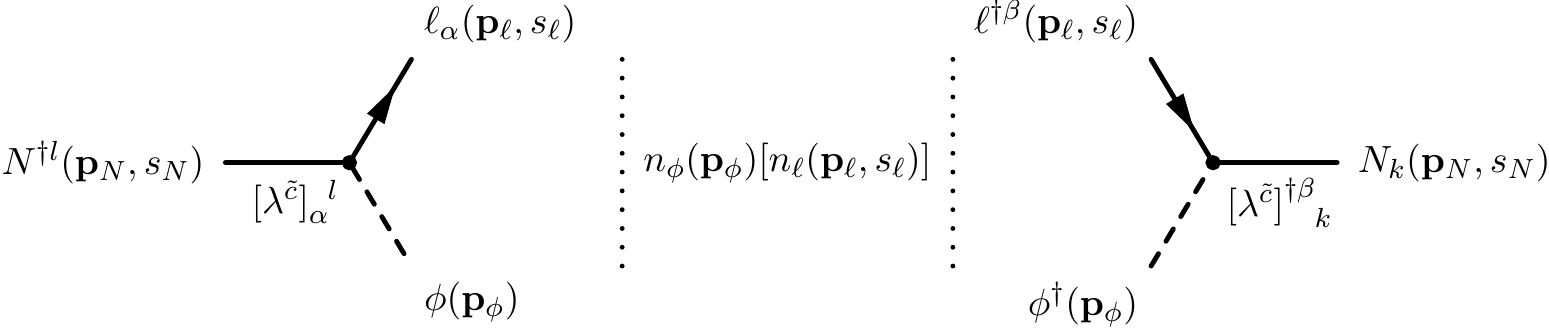}}}
\subfigure[][The in-medium inverse heavy-neutrino decay: $\bar{n}_{\phi}\protect{[}\bar{n}_{\ell}\protect{]}_{\beta}^{\protect{\phantom{\beta}}\alpha}\protect{[}\gamma(\ell^{\tilde{c}}\phi^{\tilde{c}}\to N)\protect{]}_{\alpha\protect{\phantom{\beta}}k}^{\protect{\phantom{\alpha}}\beta\protect{\phantom{k}}l}$.]{\parbox{\textwidth}{\centering \includegraphics[scale=0.7]{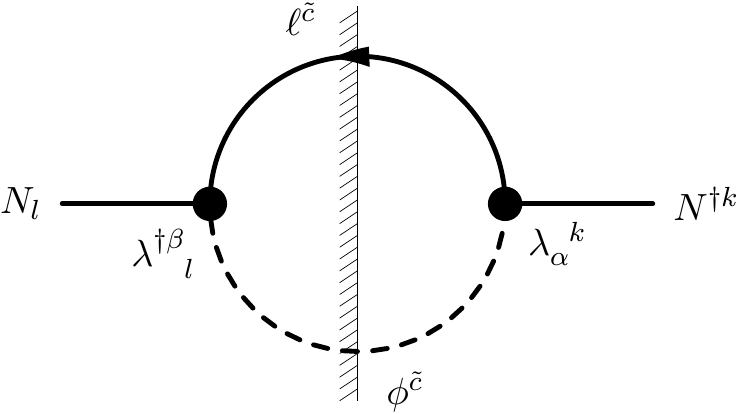}\vspace{0.5em}\\$\bigg\downarrow$\\\includegraphics[scale=0.7]{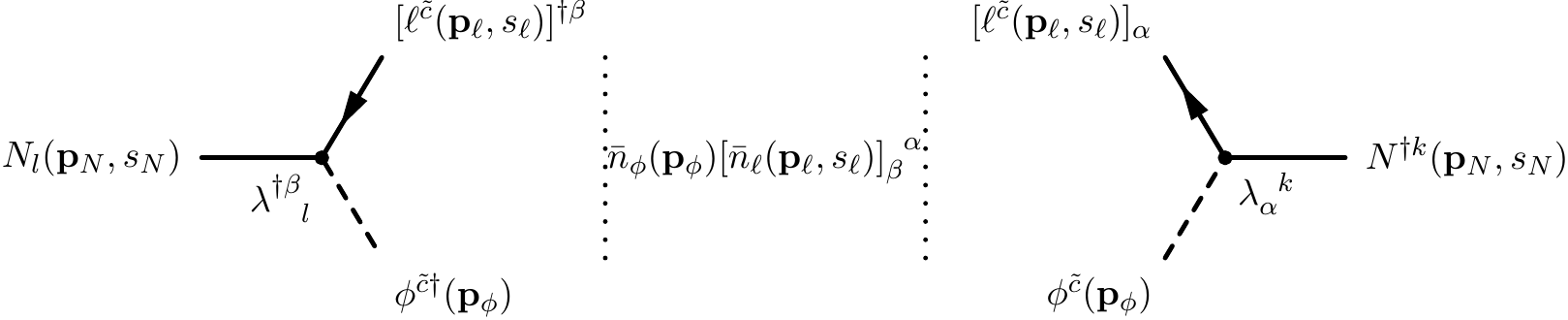}}}
\caption{\label{3_fig_cuts} Diagrammatic representation of the $2\to 1$ processes, illustrating the origin of the four-index rates from the unitarity cuts of the thermal heavy-neutrino self-energies~\cite{Dev:2014laa}. Notice that the shaded region of the cut appears to the left. Diagrams adapted from Ref.~\cite{Dev:2014laa}.}
\end{figure}

The necessary appearance of rates that carry high-rank structure in flavor space, e.g.~\smash{{\footnotesize $\Tdu{[\gamma^{N}_{\ell \phi}]}{\alpha}{\beta}{k}{l}$}}, can be understood in terms of partial cuts of the ``thermal'' self-energies (cf.~\sref{sec:methods_fieldtheory}) by means of a generalization of the optical theorem~\cite{Dev:2014laa}, where the cut is weighted by the matrix number density. For example, the inverse decay terms can be obtained directly from the cuts shown in~\fref{3_fig_cuts}, allowing us to extract the thermally-averaged rates (cf.~Chapter~\cite{leptogenesis:A04})
\begin{subequations}
\begin{gather}
  \Tdu{[\gamma(N\to\ell\phi)]}{\alpha}{\beta}{k}{l}  = \Tdu{[\gamma(\ell^{\tilde{c}}\phi^{\tilde{c}}\to N)]}{\alpha}{\beta}{k}{l} = \int_{N\ell\phi}g_{\ell}g_{\phi}(2p_N\cdot p_{\ell})\lambda^{\dag\beta}_{\phantom{\dag\beta}k}\lambda_{\alpha}^{\phantom{\alpha}l}\;,\\
  \Tdu{[\gamma(N\to\ell^{\tilde{c}}\phi^{\tilde{c}})]}{\alpha}{\beta}{k}{l} = \Tdu{[\gamma(\ell\phi\to N)]}{\alpha}{\beta}{k}{l} = \int_{N\ell\phi}g_{\ell}g_{\phi}(2p_N\cdot p_{\ell})[\lambda^{\tilde{c}}]^{\dag\beta}_{\phantom{\dag\beta}k}[\lambda^{\tilde{c}}]_{\alpha}^{\phantom{\alpha}l}\;,
\end{gather}
\end{subequations}
where the left-hand equalities follow from CPT, $g_{\ell}$ and $g_{\phi}$ are respectively the degeneracy factors of the internal degrees of freedom of the charged-lepton and Higgs doublets, and we employ the short-hand notation
\begin{align}
  \int_{N\ell\phi}\  &\equiv\  \int\!\frac{{\rm d}^3\mathbf{p}_N}{(2\pi)^32E_N(\mathbf{p}_N)}\,\frac{{\rm d}^3\mathbf{p}_{\ell}}{(2\pi)^32E_{\ell}(\mathbf{p}_{\ell})}\,\frac{{\rm d}^3\mathbf{p}_{\phi}}{(2\pi)^32E_{\phi}(\mathbf{p}_{\phi})}\nonumber\\&\qquad\times\:(2\pi)^4\delta^{(4)}(p_N-p_{\ell}-p_{\phi})e^{-p^0_N/T}
\end{align}
for the thermally-averaged phase-space integrals. In this way, we obtain
\begin{equation}
  \Tdu{[\gamma^{N}_{\ell \phi}]}{\alpha}{\beta}{k}{l}\ =\ \frac{M^4}{\pi^2z}\,\frac{\mathcal{K}_1(z)}{16\pi}\,\Big(\lambda^{\dag\beta}_{\phantom{\dag\beta}k}\lambda_{\alpha}^{\phantom{\alpha}l}\:+\:[\lambda^{\tilde{c}}]^{\dag\beta}_{\phantom{\dag\beta}k}[\lambda^{\tilde{c}}]_{\alpha}^{\phantom{\alpha}l}\Big)\;,
\end{equation}
where $\mathcal{K}_1(z)$ is the first-order modified Bessel function of the second kind.

In order to identify the physical origin of each of the terms in these rate equations, it is helpful to consider their flavor structure and, specifically, whether they arise from commutators or anti-commutators in flavor space.

The first term on each of the right-hand sides of~\eref{3_etanrateeq} and~\eref{3_deltaetanrateeq} originates from a commutator in flavor space. Working, for instance, in the mass eigenbasis, it is clear that these terms source CP asymmetry only when non-zero flavor correlations are encoded in the off-diagonal elements of the matrix number densities. Since these terms are non-zero only in the presence of such a misalignment, they predominantly capture the \emph{coherent oscillations} between heavy-neutrino flavors. These terms are of \emph{statistical} origin, and we emphasise that they would be absent in a flavor-diagonal treatment.

The remaining terms instead arise from anti-commutators in flavor space and persist in the flavor-diagonal limit. (The terms that do not explicitly carry braces began as anti-commutators involving the equilibrium number densities, which are taken to be diagonal in flavor space.) With the exception of the decoherence term, which will be described shortly, the anti-commutator structure predominantly captures the effect of \emph{mixing} between the heavy-neutrino flavors. The terms involving \smash{$\gamma^{N}_{\ell\phi}$} and \smash{$\delta\gamma^N_{\ell\phi}$} together describe decays and inverse decays. The terms involving \smash{$\gamma^{\ell\phi}_{\ell\phi}$} and $\gamma^{\ell\phi}_{\ell^{\tilde{c}}\phi^{\tilde{c}}}$ describe $\Delta L=2$ scatterings. In order to avoid double counting, the procedure of RIS subtraction~\cite{Kolb:1979qa} has been applied to these rate equations, as discussed in the accompanying Chapter~\cite{leptogenesis:A04}, with the necessary inclusion of thermal corrections~\cite{Dev:2014laa}.  Finally, the decoherence term \smash{$[\delta \gamma_{\rm dec}^{\rm back}]_{\alpha}^{\phantom{\alpha}\beta}$}~\cite{Dev:2014laa} (cf.~Ref.~\cite{Abada:2006fw}) accounts for processes mediated by the charged-lepton Yukawa couplings, which act in competition with the processes mediated by the heavy-neutrino Yukawa couplings. The former tend to decohere the charged leptons into their mass eigenbasis, whereas the latter tend to regenerate charged-lepton coherences.

The physically distinct sources of CP asymmetry from \emph{oscillations} and \emph{mixing} can also be isolated by considering the sequence of heavy-neutrino production, propagation and subsequent decay. The contribution from \emph{mixing} is associated with the heavy-neutrino production and decay processes, and the contribution from \emph{oscillations} is associated with the \emph{in-medium} propagation of the heavy neutrinos. The former is generated predominantly by the interference of the ($T=0$) one-loop and tree-level processes, capturing the usual $\varepsilon$- and $\varepsilon'$-type CP violation. The latter is contained in the thermal part of the intermediate heavy-neutrino propagator and is captured at leading order in the semi-classical rate equations by the presence of the commutator terms.

In the hierarchical limit, the source of CP asymmetry is dominated by \emph{mixing}. A semi-classical analysis of flavor-diagonal Boltzmann equations is then sufficient, and the source of asymmetry can be treated by means of effective or resummed Yukawa couplings (see~Ref.~\cite{Pilaftsis:2003gt}). In the quasi-degenerate limit, \emph{oscillations} become important, and we need also to keep track of the evolution of the off-diagonal flavor correlations, resulting in a non-vanishing contribution from the commutator term. Whilst it is clear that both \emph{mixing} and \emph{oscillations} contribute to the asymmetry in the quasi-degenerate regime, it remains an open question as to how to account consistently for both sources without under- or over-counting the final asymmetry. In semi-classical approaches, it has been claimed~\cite{Dev:2014laa} that both the commutator term and resummed Yukawa couplings should be included. This has also been argued in a field-theoretic approach~\cite{Dev:2014wsa} based on the interaction picture~\cite{Millington:2012pf,Millington:2013isa} (see also the discussion in Chapter~\cite{leptogenesis:A03}). Conversely, in other field-theoretic approaches, it has been claimed~\cite{Garbrecht:2011aw} that both sources are captured by the average mass shell approximation for the flavor-off-diagonal heavy-neutrino Wigner functions. The material difference amounts to a possible factor of 2 in the final asymmetry~\cite{Dev:2014wsa}. The main obstacle to resolving this debate is the technical difficulty of making direct comparisons between different approaches in the strong washout regime and in the presence of cosmological expansion.

A direct comparison was made in the weak washout regime and on a static and stationary background in Ref.~\cite{Kartavtsev:2015vto} (see also the discussion in Chapter~\cite{leptogenesis:A03} of this review). In this idealized setting, the sources of CP violation were studied in a field-theoretic approach, based on the Kadanoff-Baym formalism (in both interaction- and Heisenberg-picture descriptions), by analysing the effective shell structure of the would-be non-equilibrium heavy-neutrino propagators of a toy scalar model. Whilst both mixing and oscillation contributions can be identified --- the former living on the quasi-particle mass shells and the latter living on an intermediate average mass shell --- one also finds additional terms that can be interpreted as the \emph{destructive} interference between these contributions. In the hierarchical limit, the oscillation and interference terms are suppressed, such that the quasi-particle mass shells dominate and a flavor-diagonal semi-classical analysis with resummed Yukawa couplings is appropriate. In the fully degenerate limit, the destructive interference is complete (see also Ref.~\cite{Hohenegger:2014cpa}), and one finds zero asymmetry, as expected. In the problematic, quasi-degenerate limit, the degree of cancellation was shown~\cite{Kartavtsev:2015vto} to depend strongly on the distribution of particle number between the different flavors and is therefore model- and washout-dependent (i.e.~dependent upon the choice of initial conditions in the weak washout regime). If the asymmetry is distributed evenly between the different flavors (corresponding to symmetric initial conditions in the weak washout regime), the impact of the destructive interference is more severe, and there is a significant suppression of the \emph{mixing} source. If this result is extrapolated to the strong washout regime, the form of the CP source then agrees with the average mass shell approximation employed in Ref.~\cite{Garbrecht:2011aw}. Instead, if a particular diagonal element of the number density dominates (corresponding to asymmetric initial conditions in the weak washout regime), the interference does not significantly impact the magnitude of the mixing term. One then finds that both the mixing and oscillation sources contribute additively to the final asymmetry up to a maximum factor of 2 enhancement when compared with taking only one source into account.

We should, however, be careful in extrapolating the latter observations to the strong washout regime and an expanding background. Whilst it is the case that one diagonal element of the heavy-neutrino number densities dominates in the attractor limit of the scenario considered in Ref.~\cite{Dev:2014laa}, the behavior of the aforementioned destructive interference in the strong washout regime and the degree to which it is correctly captured remains an area of active discussion. In semi-classical approaches, the destructive interference is, at least in part, captured by ensuring that the regulator of the final asymmetry (obtained through consistent resummation of the effective Yukawa couplings) vanishes appropriately in the CP-conserving limit.


\subsection{Phenomenological aspects}

As already mentioned in the introduction, the flavor effects captured in the fully flavor-covariant treatment are both of qualitative and quantitative importance in testable leptogenesis models. In this section, we illustrate this with a minimal model of low-scale resonant $\tau$-genesis (RL$_\tau$) in which the lepton asymmetry is generated from and protected in a single lepton flavor $\ell=\tau$~\cite{Pilaftsis:2004xx, Deppisch:2010fr}. The Dirac Yukawa couplings involving electron and muon flavors in~\eref{3_Lagrangian} remain sizable, thus giving rise to potentially observable predictions for lepton number and flavor violation at both energy and intensity frontiers~\cite{Deppisch:2010fr, Dev:2014laa, Dev:2015wpa}.  

Within the minimal RL$_\ell$ setup, the heavy-neutrino sector possesses an $O({\cal N}_N)$ symmetry at some high energy scale $\mu_X$, i.e.~$M_N(\mu_X)=M\mathbb{I}$. The small mass splitting, as required for successful RL, can then be generated naturally at the phenomenologically relevant low-energy scale by renormalization group (RG) running effects induced by the Yukawa couplings $\lambda_\alpha^{\phantom{\alpha}k}$, i.e.~$M_N(M)=M\mathbb{I}+\Delta M_N^{\rm RG}$, where~\cite{Deppisch:2010fr}
\begin{align}
  \Delta M_N^{\rm RG} \ \simeq \ -\,\frac{M}{8\pi^2} \ln\left(\frac{\mu_X}{M}\right){\rm Re}[\lambda^\dag(\mu_X)\cdot\lambda(\mu_X)] \; .
\end{align}
However, it turns out that this minimal scenario is not viable due to a no-go theorem~\cite{Dev:2015wpa}, which ensures that the leptonic asymmetry vanishes identically at ${\cal O}(\lambda^4)$. To avoid this, we include a new source of flavor breaking $\Delta M_N$, which is not aligned with $\Delta M_N^{\rm RG}$. Thus, the relevant heavy-neutrino mass matrix for our case is given by
\begin{align}  
  M_N \ = \ M\mathbb{I}+\Delta M_N^{\rm RG}+\Delta M_N \; ,
\end{align}
which goes into the type I seesaw formula for the light neutrino mass matrix~\cite{Minkowski:1977sc, Mohapatra:1979ia, Yanagida:1979as, GellMann:1980vs, Glashow:1979nm}
\begin{align}
  M_\nu \ \simeq \ -\,\frac{v^2}{2}\,\lambda\cdot M_N^{-1}\cdot \lambda^{\sf T} \; .
  \label{3_lightmassmatrix}
\end{align} 

For the purpose of our illustration, we consider three RH neutrinos (i.e.~${\cal N}_N=3$) and the following diagonal form for $\Delta M_N$: 
\begin{align}
  \Delta M_N \ = \ {\rm diag}(\Delta M_1, \Delta M_2/2, -\,\Delta M_2/2) \; ,
\end{align}
where $\Delta M_2\neq \Delta M_1$ is needed to make the light neutrino mass matrix $M_\nu$ in~\eref{3_lightmassmatrix} rank-2, thus allowing us to fit successfully the low-energy neutrino oscillation data. 

As for the Yukawa coupling matrix $\lambda$,  we consider an $\mathrm{RL}_\tau$ model that
possesses a leptonic symmetry $U(1)_{\ell}$ and protects the lightness of the LH neutrino masses. In this scenario, the Yukawa
couplings $\lambda_{\alpha}^{\phantom{\alpha}k}$ have the following
structure~\cite{Pilaftsis:2004xx, Pilaftsis:2005rv}:
\begin{equation}
  \lambda \ = \begin{pmatrix}      0 & a \,e^{-i\pi/4} & a\,e^{i\pi/4}\cr
      0 & b\,e^{-i\pi/4} & b\,e^{i\pi/4}\cr
      0 & c\,e^{-i\pi/4} & c\,e^{i\pi/4}\end{pmatrix}
     \: + \: \delta \lambda \; .
  \label{3_Yukawastructure}
\end{equation}
In order to protect the $\tau$ asymmetry from excessive washout and simultaneously allow for large couplings in the electron and muon sectors so as to have experimentally observable effects, we take $|c| \ll |a|,|b| \approx
10^{-3}-10^{-2}$. The leptonic flavor-symmetry-breaking matrix is taken to be
\begin{equation}
  \delta \lambda \ = \ \begin{pmatrix}
      \varsigma_e & 0 & 0\cr
      \varsigma_\mu & 0 & 0\cr
      \varsigma_\tau & 0 & 0 \end{pmatrix}
 \; .
\end{equation}
To leading order in the symmetry-breaking parameters of $\Delta M_N$ and  $\delta \lambda$,  the tree-level light-neutrino mass matrix, given by~\eref{3_lightmassmatrix}, becomes
\begin{equation}
  M_\nu \ 
  \simeq \ \frac{v^2}{2M}\begin{pmatrix}
      \frac{\Delta M}{M} a^2 - \varsigma_e^2 &
     \frac{\Delta M}{M} ab - \varsigma_e\varsigma_\mu &
      - \varsigma_e\varsigma_\tau\cr
      \frac{\Delta M}{M} ab - \varsigma_e\varsigma_\mu &
      \frac{\Delta M}{M} b^2 - \varsigma_\mu^2 &
      - \varsigma_\mu\varsigma_\tau\cr
      - \varsigma_e\varsigma_\tau &
      - \varsigma_\mu\varsigma_\tau &
      - \varsigma_\tau^2 \end{pmatrix}\;,
\end{equation}
where $\Delta M = -i\Delta M_2$ and we have neglected subdominant terms $\frac{\Delta M}{M} \,c \times (a,b,c)$. Inverting this expression, we determine the following model parameters appearing in the Yukawa coupling matrix~\eqref{3_Yukawastructure}:
\begin{align}
  a^2 \ &= \ \frac{2M}{v^2}
  \left(M_{\nu,{11}}-\frac{M^2_{\nu,{13}}}{M_{\nu,{33}}}\right) \frac{M}{\Delta M}\; ,
\qquad
  b^2 \ = \ \frac{2M}{v^2}
  \left(M_{\nu,{22}}-\frac{M^2_{\nu,{23}}}{M_{\nu,{33}}}\right)\frac{M}{\Delta M}\; ,
  \nonumber \\[6pt]
  \varsigma_e^2 \  &= \ -\frac{2M}{v^2}\frac{M^2_{\nu,{13}}}{M_{\nu,{33}}} \; ,
\qquad
  \varsigma_\mu^2 \ = \ -\frac{2M}{v^2}\frac{M^2_{\nu,{23}}}{M_{\nu,{33}}}\; ,
\qquad
  \varsigma_\tau^2 \ = \ -\frac{2M}{v^2}M_{\nu,{33}}\; .
  \label{3_modelparameters}
\end{align}
Therefore, the Yukawa  coupling matrix in the RL$_\tau$ model can be fixed completely in terms of the heavy-neutrino mass scale $M$ and the input parameters $c$ and $\Delta M_{2}$, apart from the light-neutrino oscillation parameters, which determine the elements of $M_\nu$ from the diagonalization equation $M_\nu = U_{\nu}{\rm diag}(m_{\nu_1},m_{\nu 2},m_{\nu_3})U_{\nu}^{\sf T}$, where $U_{\nu}$ is the usual PMNS mixing matrix (see Eq.~\eqref{eq:PMNS}). 

\begin{table}[t!]
\tbl{The numerical values of the free parameters for three chosen benchmark points in our RL model. The parameters $a,b,\varsigma_{e,\mu,\tau}$ have been derived using~\eref{3_modelparameters}.}
{\begin{tabular}{c c c c}\hline
Input Parameter & BP1 & BP2 & BP3\\ \hline
$M$ & 400 GeV & 2000 GeV & 400 GeV \\ 
$\Delta M_1/M$ & $-5\times 10^{-5}$ & $-5\times 10^{-5}$ & $-5\times 10^{-5}$ \\
$\Delta M_2/M$ & $1.1\times 10^{-9}$ & $5\times 10^{-9}$ & $10^{-8}$ \\
$c$ & $2\times 10^{-7}$ & $2\times 10^{-7}$ & $2\times 10^{-7}$ \\
\hline
\end{tabular}\label{3_tab_benchmarks}}
\end{table}
\begin{figure}[t!]
\centering
\includegraphics[scale=0.35]{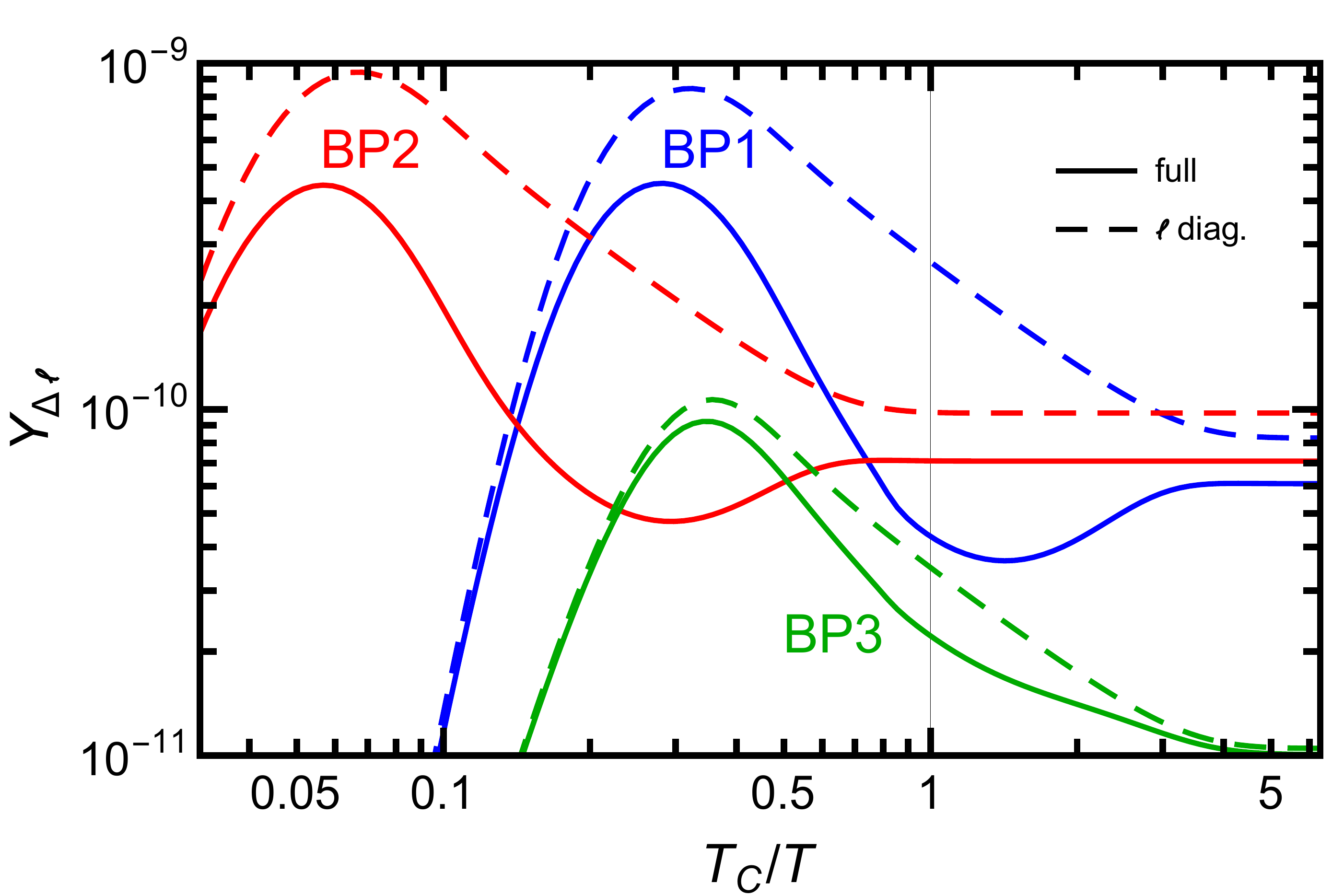}
\caption{Total lepton asymmetry $Y_{\Delta \ell}$ as a function of the inverse temperature, obtained using the fully flavor-covariant formalism (solid curves) for three benchmark points. For comparison, we also show the corresponding predictions as obtained using the Boltzmann equations diagonal in charged-lepton flavors (dashed curves), which overestimate the final asymmetry in all three cases. The vertical line shows the critical temperature $T_C$ beyond which the lepton asymmetry is frozen out due to the exponential suppression of the electroweak sphaleron transition rate.\label{3_fig_benchmarks}}
\end{figure}

For numerical purposes, we choose a normal hierarchy of light neutrino masses, with the lightest mass $m_{\nu_1}=0$, and use the best-fit values of the oscillation parameters (mass-squared differences and mixing angles) from a recent global fit~\cite{Capozzi:2013csa}. For illustration, we choose $\delta=0$ and $\phi_1=\pi,~\phi_2=\pi$ for the Dirac and Majorana phases, respectively.  To demonstrate the flavor dynamics of our RL$_\tau$ model, we select three benchmark points, as listed in~\tref{3_tab_benchmarks}. The results for the total lepton asymmetry in each case are shown in \fref{3_fig_benchmarks}. The ``bump'' in each case is due to an interplay between the heavy-neutrino coherence and charged-lepton decoherence effects~\cite{Dev:2014laa}. We find that the final lepton asymmetry obtained using the fully flavor-covariant treatment is smaller than that obtained from the solution of the Boltzmann equations diagonal in the charged-lepton flavor by up to a factor of 5. This clearly demonstrates the quantitative importance of the flavor effects captured by the flavor-covariant formalism.  

\begin{figure}[t!]
\centering
\includegraphics[scale=0.35]{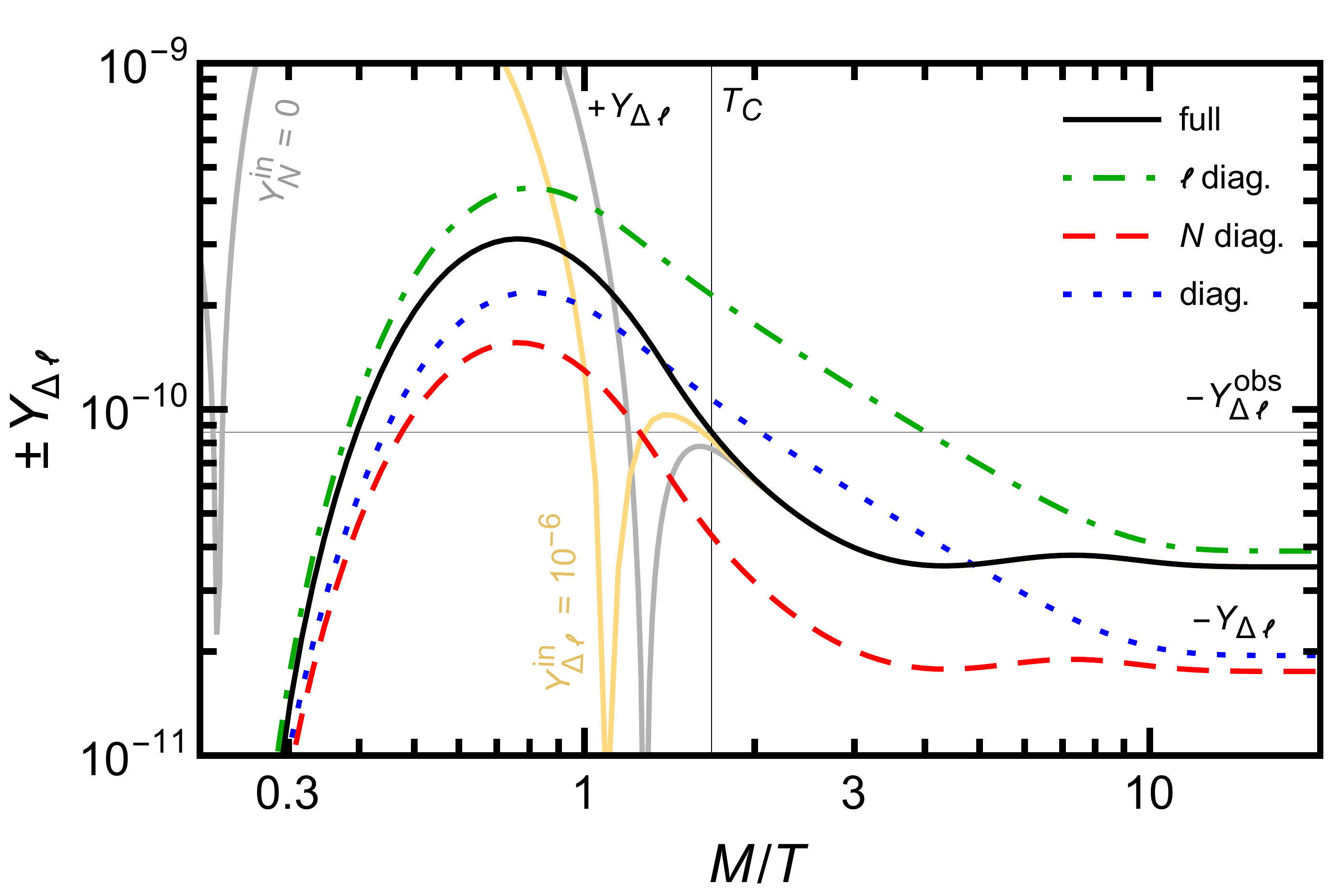}
\caption{Total lepton asymmetry  $Y_{\Delta \ell}$ as a function of the inverse temperature, obtained using the fully flavor-covariant formalism (black solid curve) versus that obtained using the Boltzmann equations diagonal in charged-lepton flavor (green dot-dashed), heavy-neutrino flavor (red dashed) and both (blue dotted). The yellow and grey solid curves show the total asymmetry in the flavor-covariant treatment for different initial conditions. The horizontal line corresponds to the lepton asymmetry that reproduces the observed baryon asymmetry. The vertical line shows the critical temperature $T_C$ beyond which the lepton asymmetry is frozen out due to the exponential suppression of the electroweak sphaleron transition rate.\label{3_fig_example}}
\end{figure}
\begin{table}[t!]
\tbl{The low-energy predictions for three chosen benchmark points in the RL model.}
{\begin{tabular}{c c c c c} \hline
Observable & BP1 & BP2 & BP3 & Current Upper Limit (90\% CL) \\ \hline
BR$(\mu\to e\gamma)$ & $3.9\times 10^{-13}$ & $1.2\times 10^{-15}$ & $4.7\times 10^{-15}$ & $4.2\times 10^{-13}$ [MEG]~\cite{TheMEG:2016wtm} \\
BR$(\tau\to \mu\gamma)$ & $3.2\times 10^{-23}$ & $1.7\times 10^{-25}$ & $7.0\times 10^{-24}$ & $4.4\times 10^{-8}$ [PDG]~\cite{Olive:2016xmw}\\
BR$(\tau\to e\gamma)$ & $1.2\times 10^{-23}$ & $6.5\times 10^{-26}$ & $2.6\times 10^{-24}$ & $3.3\times 10^{-8}$ [PDG]~\cite{Olive:2016xmw}\\ \hline
BR$(\mu\to 3e)$ & $1.9\times 10^{-14}$ & $1.5\times 10^{-16}$ & $2.3\times 10^{-16}$ & $1.0\times 10^{-12}$ [PDG]~\cite{Olive:2016xmw}\\ \hline
$R_{\mu-e}^{\rm Ti}$ & $5.9\times 10^{-13}$ &  $1.9\times 10^{-16}$&  $7.1\times 10^{-15}$ & $6.1\times 10^{-13}$ [SINDRUM II]~\cite{Kaulard:1998rb} \\
$R_{\mu-e}^{\rm Au}$ & $6.4\times 10^{-13}$ & $2.8\times 10^{-17}$ &  $7.1\times 10^{-15}$ & $7.0\times 10^{-13}$ [SINDRUM II]~\cite{Bertl:2006up} \\
$R_{\mu-e}^{\rm Pb}$ & $4.5\times 10^{-13}$ & $1.2\times 10^{-17}$ &  $7.1\times 10^{-15}$ & $4.6\times 10^{-11}$ [SINDRUM II]~\cite{Honecker:1996zf} \\ \hline
$\langle m_{\beta\beta}\rangle$ (meV) & $3.8\times 10^{-9}$ & $3.8\times 10^{-9}$ & $3.8\times 10^{-9}$ & $61-165$ [KamLAND-Zen]~\cite{KamLAND-Zen:2016pfg} \\ 
\hline
\end{tabular} \label{3_tab_predictions}}
\end{table}

The impact of flavor effects is further illustrated in~\fref{3_fig_example}. The solid curves show the total lepton asymmetry obtained from the fully flavor-covariant Boltzmann equations for {\it very} different initial conditions. It is reassuring to see that the final asymmetry is independent of any pre-existing initial abundance --- a hallmark of RL models~\cite{Pilaftsis:2005rv}. The dotted (blue), dashed (red) and dot-dashed (green) curves show the corresponding predictions from the solution of Boltzmann equations diagonal in both heavy-neutrino and charged-lepton flavors, only in the heavy-neutrino flavor, and only in the charged-lepton flavor, respectively. It is clear that none of the fully or partially diagonal rate equations are capable of capturing all flavor effects in a consistent manner, which necessitates the use of the flavor-covariant treatment. For this particular example, we have chosen $\delta=-\pi/2$, as mildly favored by the recent T2K data~\cite{Escudero:2016odp}, and $\phi_1=\pi,\phi_2=0$ for the PMNS CP phases in order to reproduce the observed baryon asymmetry in the flavor-covariant treatment. The other input parameters in this example are $M=250$ GeV, $\Delta M_1/M=-\,5\times 10^{-5}$, $\Delta M_2/M=1.5\times 10^{-9}$ and $c=2.8\times 10^{-7}$.  

As mentioned earlier, apart from explaining the matter-anti-matter asymmetry puzzle, the low-scale RL models offer the attractive possibility of being tested in various laboratory experiments at both energy and intensity frontiers. The benchmark scenarios shown in~\tref{3_tab_benchmarks}, having TeV-scale heavy neutrinos, can be probed at the LHC via multilepton final states~\cite{Deppisch:2015qwa}. Note that, due to the small mass splitting between the three heavy neutrinos, the same-sign dilepton signal at the LHC will be suppressed. However, the opposite-sign dilepton or trilepton signals can be useful in probing these scenarios. As for the low-energy probes at the intensity frontier, the model predictions for various low-energy observables are given in~\tref{3_tab_predictions}, along with the current experimental limits at 90\% C.L. For details of the theoretical calculations, see, e.g.,~Ref.~\cite{Dev:2014laa}. The $0\nu\beta\beta$ rate is suppressed in this case for the same reason as the suppression of the lepton number violating LHC signals, i.e.~due to the quasi-degeneracy of the heavy neutrinos. Even so, the $\mu\to e\gamma$  and $\mu-e$ conversion predictions are close to the current experimental bounds and could be tested in the near future by upcoming experiments, such as Mu2e~\cite{Bartoszek:2014mya} and PRISM/PRIME~\cite{Kuno:2005mm}. This is a characteristic feature of the RL$_\tau$ models being considered here, which have relatively large Yukawa couplings in the electron and muon sectors, thus giving rise to observable lepton flavor violating (LFV) effects. On the other hand, the Yukawa couplings in the tau sector are smaller, which suppresses the corresponding LFV effects. It is difficult to have any observable LFV effects in most of the other low-scale RL models~\cite{Heurtier:2016iac}, and this puts the RL$_\tau$ models discussed here on a unique footing.  


\section{\label{sec:typeII}Type II seesaw/scalar triplet leptogenesis}

Leptogenesis has mainly been studied in the framework of the type I seesaw mechanism, in which the source of the lepton asymmetry is the CP-violating decays of heavy Majorana neutrinos. Scalar triplet leptogenesis~\cite{Hambye:2003ka, Antusch:2004xy, Hambye:2005tk, Chun:2006sp, Hallgren:2007nq, Frigerio:2008ai, Felipe:2013kk, Sierra:2014tqa, Lavignac:2015gpa}, based on the type II seesaw mechanism~\cite{Magg:1980ut, Schechter:1980gr, Lazarides:1980nt, Mohapatra:1980yp}, has received much less attention in comparison. In particular, lepton flavor effects were included only recently in this scenario~\cite{Felipe:2013kk, Sierra:2014tqa, Lavignac:2015gpa}.


\subsection{The framework}
\label{4_sec_framework}


In spite of its simplicity, the type II seesaw mechanism is much less popular than its type I cousin, presumably because it is less easily implemented in GUTs. The only thing it requires is the addition to the SM of a massive scalar electroweak triplet, which couples to the LH leptons and to the Higgs doublet in the following way:
\begin{equation}
  \mathcal{L}_\Delta\ =\ -\,\frac{1}{2} \left( y_{\alpha\beta}\, \ell_\alpha^{\mathsf{T}} C i \sigma^2 \Delta \ell_\beta
    +\mu\, \phi^{\mathsf{T}} i \sigma^2 \Delta^\dagger \phi + \mbox{h.c.} \right) - M_\Delta^2\, \mbox{tr} (\Delta^\dagger \Delta)\; ,
\label{4_Lagrangian}
\end{equation}
where $C$ is the charge conjugation matrix defined by $C \gamma^{\mathsf{T}}_\mu C^{-1} = -\,\gamma_\mu$, and
\begin{equation}
  \Delta\ =\
    \left( \begin{array}{cc}  \Delta^+ / \sqrt{2} & \Delta^{++} \\
    \Delta^0 & - \Delta^+ / \sqrt{2}
    \end{array} \right) \;,  \qquad
  \Delta^\dagger\ =\
    \left( \begin{array}{cc}  \Delta^- / \sqrt{2} & \Delta^{0*} \\
    \Delta^{--} & - \Delta^- / \sqrt{2}
    \end{array} \right) \;.
\end{equation}

In~\eref{4_Lagrangian}, $\alpha$ and $\beta$ are lepton flavor indices, $y_{\alpha\beta}$ is a symmetric $3 \times 3$ matrix of complex dimensionless couplings, and $\mu$ is a complex mass parameter. Heavy scalar triplet exchange generates the neutrino mass matrix
\begin{equation}
  (M^\Delta_{\nu})_{\alpha\beta}\ =\ \frac{1}{4}\, \mu y_{\alpha\beta}\frac{v^2}{M_\Delta^2}\; ,
\label{4_mnu_Delta}
\end{equation}
where $v = \sqrt{2}\,\langle \phi^0 \rangle = 246\, \mbox{GeV}$ is the Higgs boson vacuum expectation value, providing the desired suppression of neutrino masses.

The Lagrangian in~\eref{4_Lagrangian} allows the scalar triplet to decay into a pair of anti-leptons or a pair of Higgs bosons, with respective tree-level decay rates and branching ratios
\begin{equation}
  \Gamma(\Delta \rightarrow \bar \ell \bar \ell)\ =\ \frac{\lambda^2_\ell}{32\pi}\, M_\Delta\, ,  \qquad
  \Gamma(\Delta \rightarrow \phi \phi)\ =\ \frac{\lambda^2_{\phi}}{32\pi}\, M_\Delta\; ,
\end{equation}
\begin{equation}
  B_\ell\ =\ \lambda_\ell^2 / (\lambda_\ell^2 + \lambda_{\phi}^2)\;,  \qquad \qquad
    B_{\phi}\ =\ \lambda_{\phi}^2 / (\lambda_\ell^2 + \lambda_{\phi}^2)\; ,
\end{equation}
where we have introduced the notations
\begin{equation}
  \lambda_\ell\, \equiv\, \sqrt{\mathrm{tr}(yy^\dagger)}\;,  \qquad \qquad  \lambda_{\phi}\ \equiv\ |\mu| / M_\Delta\;.
\end{equation}
This minimal setup is, however, not enough for leptogenesis: to generate an asymmetry between triplet and anti-triplet decays, another heavy state must be added to the model that couples to the lepton and Higgs doublets. Examples of such states are additional scalar triplets, which induce a CP asymmetry in $\Delta / \bar \Delta$ decays through self-energy corrections, or right-handed neutrinos, which give rise to vertex corrections. If the additional particles are significantly heavier than the scalar triplet, they are not present in the thermal bath at the time of leptogenesis, and one can parametrize their effects~\cite{Hambye:2005tk} by the effective dimension-5 operators\footnote{In full generality, one should also consider the effective dimension-6 operators
\begin{equation*}
  -\frac{1}{4}\, \frac{\eta_{\alpha\beta\gamma\delta}}{\Lambda^2}
    \left( \ell_\alpha^{\mathsf{T}} C i \sigma^2 \vec{\sigma} \ell_\beta \right)\! \cdot\!
    \left( \bar{\ell}_\gamma \vec{\sigma} i \sigma^2 C \bar{\ell}_\delta^{\mathsf{T}} \right)\;,
\label{4_4lepton_operator}
\end{equation*}
which arise at tree level if the heavier particles are scalar triplets and at the one-loop level if they are right-handed neutrinos. These operators, which contribute to the flavor-dependent CP  asymmetries $\epsilon_{\alpha\beta}$
but not to the total CP asymmetry $\epsilon_\Delta \equiv \sum_{\alpha, \beta} \epsilon_{\alpha \beta}$, play a crucial role in the scenario of ``purely flavored leptogenesis,'' discussed in Refs.~\cite{Felipe:2013kk, Sierra:2014tqa}.
Given that they are suppressed by an additional power of $\Lambda$ and possibly also by a loop factor, their effects are typically subdominant in less specific scenarios, and we will omit them in the following.}
\begin{equation}
  \frac{1}{4}\, \frac{\kappa_{\alpha\beta}}{\Lambda}\,
    (\ell^{\mathsf{T}}_\alpha i \sigma^2 \phi)\, C\, (\phi^{\mathsf{T}} i\sigma^2 \ell_\beta)\: +\: \mbox{h.c.}\; ,
\label{4_Weinberg_operator}
\end{equation}
which are suppressed by $\Lambda \gg M_\Delta$. These operators induce a new contribution to neutrino masses proportional to $\kappa_{\alpha\beta} / \Lambda$, so that the total neutrino mass matrix can be written
\begin{equation}
  M_\nu\ =\ M_{\nu}^{\Delta} + M_{\nu}^H\; ,   \quad
  (M_{\nu}^\Delta)_{\alpha\beta} \ =\  \frac{\lambda_{\phi} y_{\alpha\beta}}{4 M_\Delta}\, v^2\; ,   \quad
  (M_{\nu}^H)_{\alpha\beta} \ =\  \frac{\kappa_{\alpha\beta}}{4 \Lambda}\, v^2\; .
\label{4_mnu}
\end{equation}
The CP asymmetries between triplet and anti-triplet decays arise from the interference between a tree-level diagram and a one-loop diagram with insertion of the operators in~\eref{4_Weinberg_operator}. They are given
by~\cite{Hambye:2005tk, Lavignac:2015gpa}
\begin{equation}
  \epsilon_{\phi}\ \equiv\ 2\ \frac{\Gamma(\Delta \rightarrow \phi\phi)-\Gamma(\bar \Delta \rightarrow \bar{\phi}\bar{\phi})}
    {\Gamma_\Delta+\Gamma_{\bar \Delta}}\ =\
  \frac{1}{2\pi}\frac{M_\Delta}{v^2}\sqrt{B_\ell B_{\phi}}\
 \frac{\mbox{Im} \left[ \mbox{tr} (M^{\Delta\dagger}_\nu M^H_{\nu}) \right]}{\bar{M}_\nu^\Delta}\; ,
\label{4_epsilon_Delta}
\end{equation}
\begin{align}
  \epsilon_{\alpha\beta}\ & \equiv\ 
    \frac{\Gamma(\bar \Delta\rightarrow \ell_\alpha \ell_\beta)-\Gamma(\Delta\rightarrow \bar \ell_\alpha \bar \ell_\beta)}
    {\Gamma_\Delta+\Gamma_{\bar \Delta}}\ \left( 1 + \delta_{\alpha \beta} \right)  \nonumber \\
  & =\, \frac{1}{2\pi}\frac{M_\Delta}{v^2}\sqrt{B_\ell B_{\phi}}\
    \frac{\mbox{Im}\left[(M^{\Delta*}_\nu)_{\alpha\beta} (M^H_\nu)_{\alpha\beta}\right]}{\bar{M}_\nu^\Delta}\; ,
\label{4_epsilon_alphabeta}
\end{align}
where $(M^\Delta_\nu)_{\alpha\beta}$ and $(M^H_\nu)_{\alpha\beta}$ are defined in~\eref{4_mnu}, $\Gamma_\Delta = \Gamma_{\bar \Delta}$ is the total triplet decay rate, and
\begin{equation}
  \bar{M}_\nu^\Delta\ \equiv\ \sqrt{\mathrm{tr}(M_\nu^{\Delta\dagger} M_{\nu}^\Delta)}\; .
\end{equation}
Unitarity and CPT invariance ensure that the CP asymmetry in decays into Higgs bosons $\epsilon_{\phi}$ is equal to the total CP asymmetry in leptonic decays $\sum_{\alpha, \beta} \epsilon_{\alpha \beta}$.

The first quantitative study of scalar triplet leptogenesis, in which flavor effects were omitted, was performed in Ref.~\cite{Hambye:2005tk}. Flavor effects were discussed in a flavor non-covariant approach in Refs.~\cite{Felipe:2013kk, Sierra:2014tqa}, and spectator processes were included in Ref.~\cite{Sierra:2014tqa}. Flavor-covariant Boltzmann equations were first presented in Ref.~\cite{Lavignac:2015gpa}.


\subsection{Flavor-covariant Boltzmann equations}
\label{4_sec_covariant_BEs}


In order to describe flavor effects in a covariant way, we introduce, as was done for the type I seesaw case in Ref.~\cite{Barbieri:1999ma}, a $3 \times 3$ matrix in lepton flavor space~\cite{Dolgov:1980cq, Stodolsky:1986dx, Raffelt:1992uj, Sigl:1992fn} --- the matrix of flavor asymmetries $[Y_{\Delta \ell}]_{\alpha\beta}$. The diagonal entries of this matrix are the asymmetries $Y_{\Delta \ell_\alpha} \equiv (n_{\ell_\alpha} - \bar{n}_{\ell_{\alpha}})/s$ stored
in the lepton doublets $\ell_\alpha$, while its off-diagonal entries encode the quantum correlations between the different flavor asymmetries. Explicitly, one first defines the phase-space distribution functions $f_{\ell\alpha\beta}(\mathbf{p})$ and $\bar{f}_{\ell\alpha\beta}(\mathbf{p})$ as matrices in flavor space by~\cite{Sigl:1992fn}
\begin{subequations}
\begin{align}
  \langle b_\alpha^\dagger(\mathbf{p}) b_\beta(\mathbf{p}') \rangle\ =\
    (2\pi)^3 \delta^{(3)}(\mathbf{p}-\mathbf{p}') f_{\ell\alpha\beta}(\mathbf{p})\; ,
\label{4_rho}  \\
  \langle d_\beta^\dagger(\mathbf{p}) d_\alpha(\mathbf{p}') \rangle\ =\
    (2\pi)^3 \delta^{(3)}(\mathbf{p}-\mathbf{p}') \bar{f}_{\ell\alpha\beta}(\mathbf{p})\; ,
\label{4_rhobar}
\end{align}
\end{subequations}
where $b_\alpha^\dagger$ (resp. $d_\alpha^\dagger$) is the operator that creates a lepton (anti-lepton) doublet of flavor $\alpha$ (the opposite order of the flavor indices $\alpha$ and $\beta$
in~\eref{4_rho} and~\eref{4_rhobar} is required by flavor covariance). The matrix of flavor asymmetries is then given by
\begin{equation}
  [Y_{\Delta \ell}]_{\alpha\beta}\ \equiv\ \frac{n_{\ell\alpha\beta} - \bar n_{\ell\alpha\beta}}{s}\;,
\label{4_Deltal_def}
\end{equation}
where the (matrix) number densities $n_{\ell\alpha\beta}$ and $\bar n_{\ell\alpha\beta}$ are obtained by integrating $f_{\ell\alpha\beta}(\mathbf{p})$ and $\bar{f}_{\ell\alpha\beta}(\mathbf{p})$ over phase space (with a factor $g_\ell = 2$ due to the $SU(2)_L$ degeneracy):
\begin{equation}
  n_{\ell\alpha\beta}\
    =\ 2 \int\! \frac{{\rm d}^3\mathbf{p}}{(2\pi)^3}\;f_{\ell\alpha\beta}(\mathbf{p})\; ,   \qquad
  \bar n_{\ell\alpha\beta}\
    =\ 2 \int\! \frac{{\rm d}^3\mathbf{p}}{(2\pi)^3}\;\bar{f}_{\ell\alpha\beta}(\mathbf{p})\; .
\end{equation}
With this definition, the matrix of flavor asymmetries transforms as $Y_{\Delta \ell} \to U^*Y_{\Delta \ell} U^{\mathsf{T}}$ under flavor rotations $\ell \to U \ell$, where $U$ is a $3 \times 3$ unitary matrix. We also need to define asymmetries for the Higgs doublet and scalar triplet:
\begin{equation}
  Y_{\Delta\chi}\ \equiv\ \frac{n_\chi - \bar{n}_{\chi}}{s}\;,  \qquad \qquad  \chi\ =\ \phi, \Delta\; ,
\label{4_Deltaphi_def}
\end{equation}
where $n_\chi$ and $\bar{n}_{\chi}$ are the number densities of the scalars $\chi$ and of their anti-particles:
\begin{equation}
  n_\chi\
    =\ g_\chi \int\! \frac{{\rm d}^3\mathbf{p}}{(2\pi)^3}\; f_\chi (\mathbf{p})\; ,   \qquad
  \bar{n}_{\chi}\
    =\ g_\chi \int\! \frac{{\rm d}^3\mathbf{p}}{(2\pi)^3}\; f_{\bar \chi} (\mathbf{p})\; ,
\end{equation}
with $g_\chi = 2$ for Higgs doublets and $g_\chi = 3$ for scalar triplets.

The time evolution of the matrix of flavor asymmetries is governed by a flavor-covariant Boltzmann equation of the form
\begin{equation}
  sHz\,\frac{{\rm d} [Y_{\Delta \ell}]_{\alpha\beta}}{{\rm d}z}\ =\
    \left(\frac{Y_{\Delta}+\bar{Y}_{\Delta}}{Y_{\Delta}^{\rm eq}+\bar{Y}_{\Delta}^{\rm eq}}-1\right)\! \gamma_D\, \mathcal{E}_{\alpha\beta}
    - \mathcal{W}_{\alpha\beta}\;,
\label{4_covariant_BE}
\end{equation}
where the first term on the right-hand side is the source term proportional to the CP-asymmetry matrix $\mathcal{E}_{\alpha\beta}$, and the second term is the washout term. In the parenthesis, $Y_\Delta \equiv n_\Delta / s$ and $\bar Y_\Delta \equiv \bar n_\Delta / s$ are the triplet and anti-triplet yields, respectively, and \smash{$Y^\text{eq}_\Delta$} and \smash{$\bar Y^\text{eq}_\Delta$} are their equilibrium values. Flavor covariance requires that, under rotations $\ell \to U \ell$, the matrices $\mathcal{E}$ and $\mathcal{W}$ transform in the same way as $Y_{\Delta \ell}$, namely as $\mathcal{E} \to U^* \mathcal{E} U^{\mathsf{T}}$ and $\mathcal{W} \to U^* \mathcal{W} U^{\mathsf{T}}$.

\begin{figure}[t!]
\centerline{\includegraphics[scale=0.1]{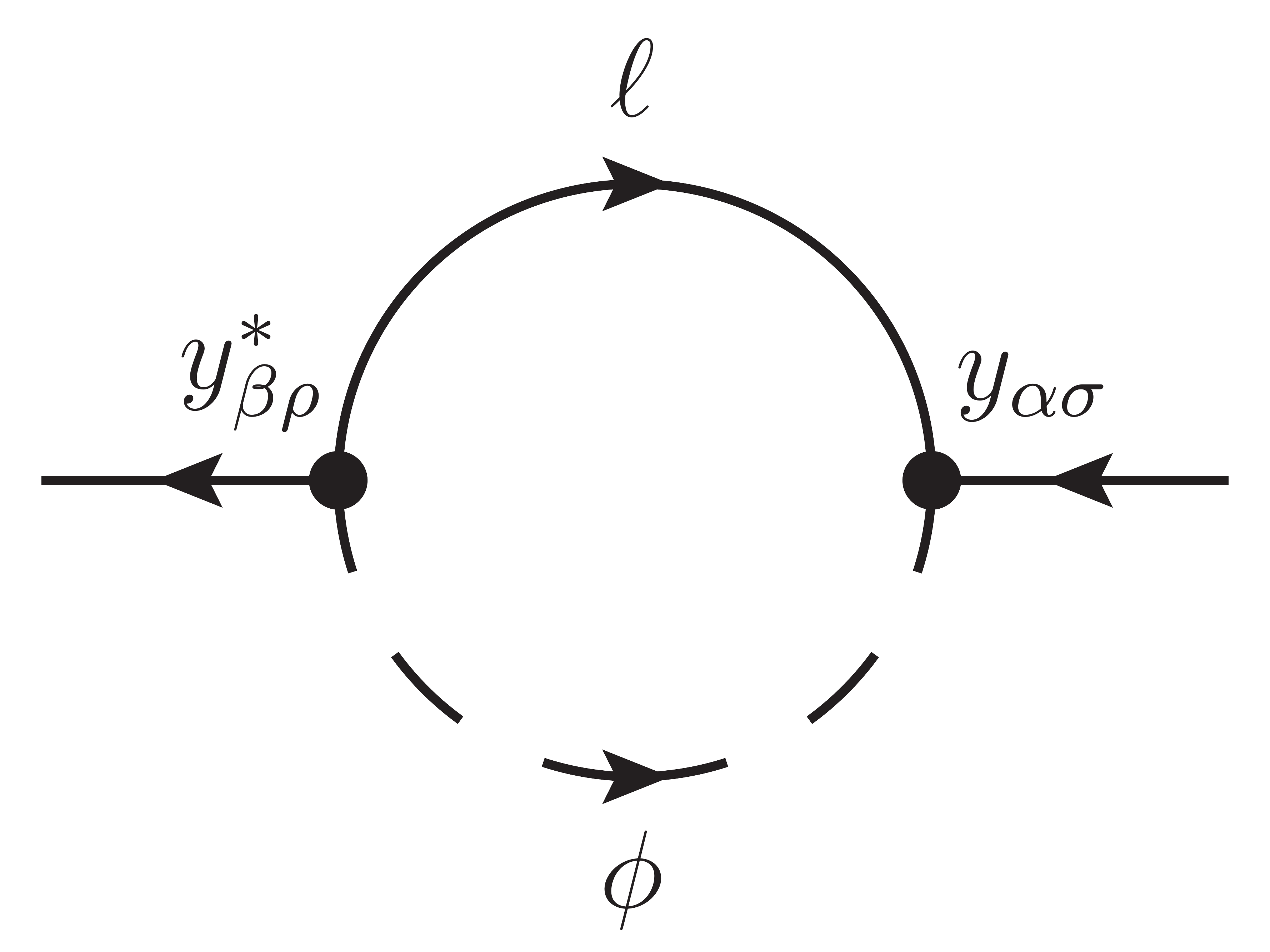}\qquad\quad
\includegraphics[scale=0.1]{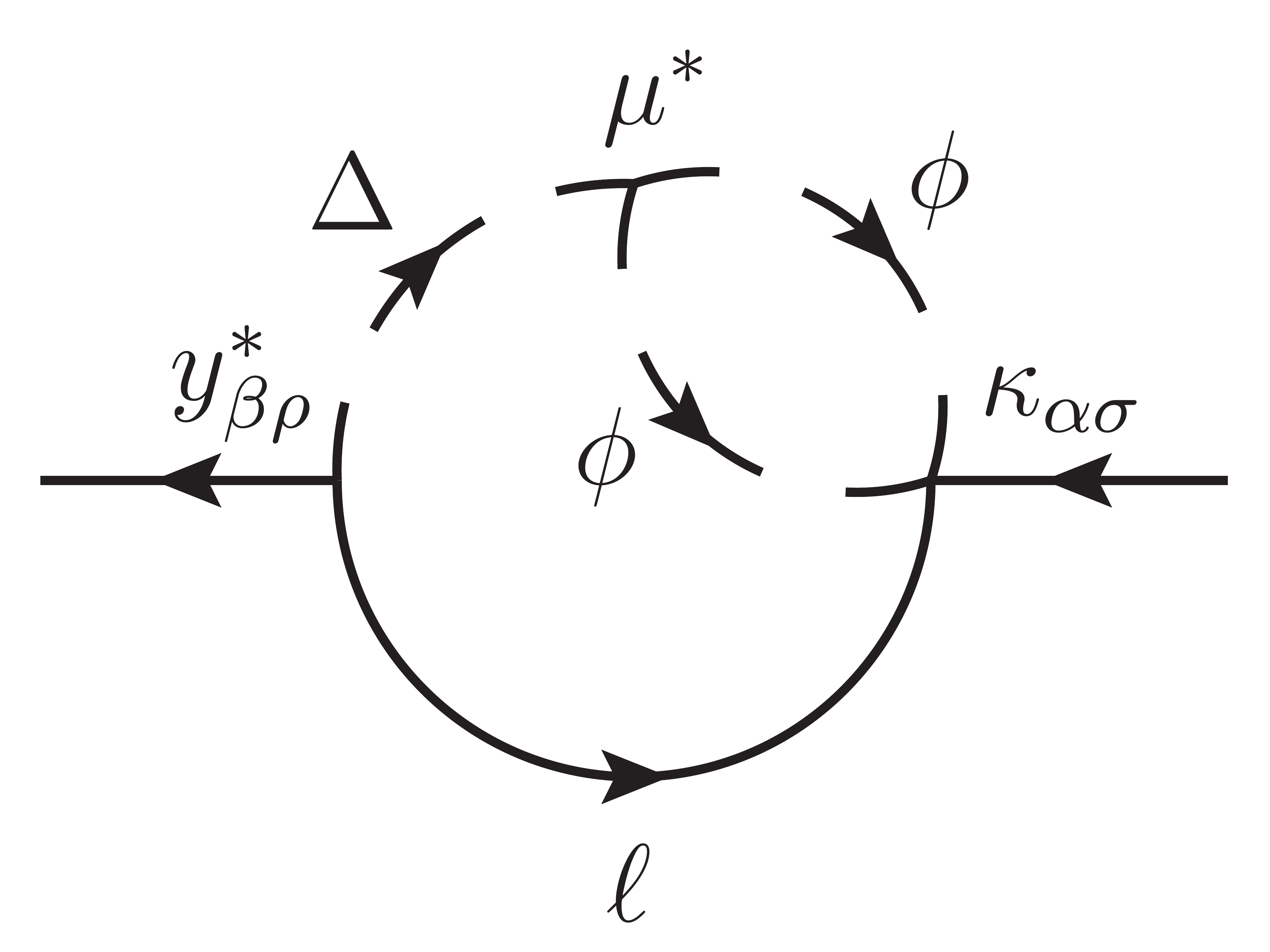}\quad
\includegraphics[scale=0.1]{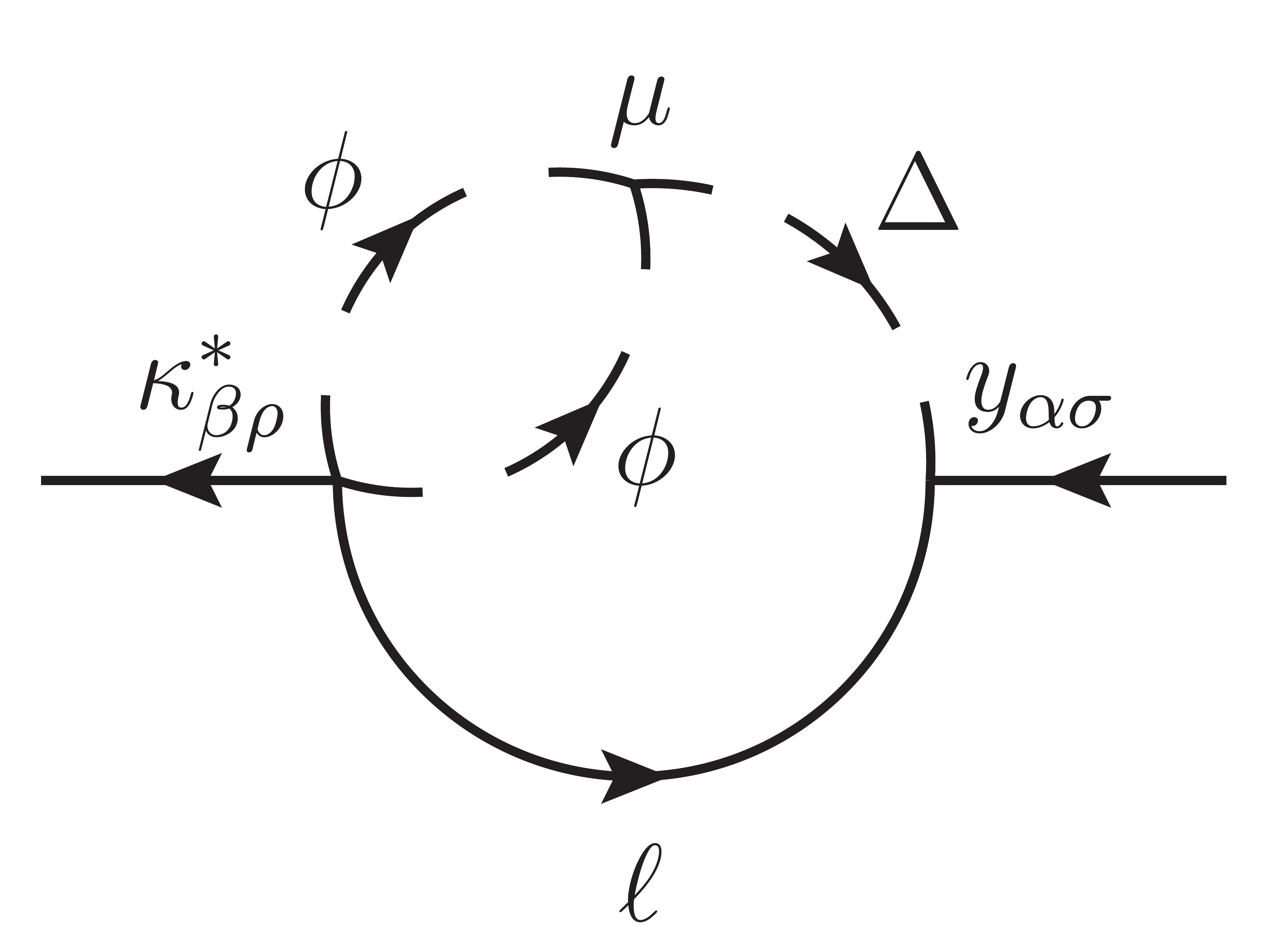}}
\hskip 1.8cm $(a)$ \hskip 6.1cm $(b)$
\vskip .1cm
\caption{$(a)$ One-loop contribution to the lepton doublet self-energy $\Sigma_{\ell\beta\alpha}$. $(b)$ Two-loop contributions to the lepton doublet self-energy
giving rise to the CP asymmetry $\mathcal{E}_{\alpha\beta}$.}
\label{4_fig_self-energy}
\end{figure}

The Boltzmann equation,~\eref{4_covariant_BE}, can be derived using the CTP formalism~\cite{Schwinger:1960qe, Keldysh:1964ud, Bakshi:1962dv, Bakshi:1963bn} (see also \sref{sec:methods_fieldtheory}), in a similar way to the flavored quantum Boltzmann equations of type I seesaw leptogenesis~\cite{Buchmuller:2000nd, DeSimone:2007gkc, Garny:2009rv, Garny:2009qn, Cirigliano:2009yt, Beneke:2010wd, Beneke:2010dz, Anisimov:2010dk, Dev:2014wsa}. In the CTP approach, particle densities are replaced by Green's functions defined on a closed path in the complex time plane going from an initial instant $t=0$ to $t=+\infty$ and back. Starting from the Schwinger-Dyson equations satisfied by the lepton-doublet Green's functions, one arrives, after some manipulations, at the quantum Boltzmann equation (see Ref.~\cite{Lavignac:2015gpa} for details)
\begin{align}
  sHz\, \frac{{\rm d}[Y_{\Delta \ell}]_{\alpha\beta}}{{\rm d}z}\, =\, & - \int\! {\rm d}^3\mathbf{w} \int_0^t {\rm d}t_w\;
    \mathrm{tr} \left[ \Sigma^>_{\ell\beta\gamma} (x,w) S^<_{\ell\gamma\alpha} (w,x)
    - \Sigma^<_{\ell\beta\gamma} (x,w) S^>_{\ell\gamma\alpha} (w,x) \right.  \nonumber \\
  & \left. -\, S^>_{\ell\beta\gamma} (x,w) \Sigma^<_{\ell\gamma\alpha} (w,x)
    + S^<_{\ell\beta\gamma} (x,w) \Sigma^>_{\ell\gamma\alpha} (w,x) \right]\;,
\label{4_QBE_Deltal}
\end{align}
where $S^<_{\ell\alpha\beta} (x,y)$ and $S^>_{\ell\alpha\beta} (x,y)$ are lepton-doublet Green's functions path-ordered along the closed time contour, and $\Sigma^<_{\ell\alpha\beta} (x,y)$ and $\Sigma^>_{\ell\alpha\beta} (x,y)$ are self-energies. The expansion of the Universe has been taken into account by making the replacement $\frac{{\rm d}}{{\rm d}t} \to sHz \frac{{\rm d}}{{\rm d}z}$ on the left-hand side of~\eref{4_QBE_Deltal}. Since we are not interested in quantum effects, we take the classical limit of~\eref{4_QBE_Deltal} by extending the time integral to infinity, which amounts to keeping only the contribution of on-shell intermediate states in the self-energy functions.
In this way, we obtain the semi-classical, flavor-covariant Boltzmann equation
\begin{equation}
  sHz\,\frac{{\rm d} [Y_{\Delta \ell}]_{\alpha\beta}}{{\rm d}z}\ =\
    \left(\frac{Y_{\Delta}+\bar{Y}_{\Delta}}{Y_{\Delta}^{\rm eq}+\bar{Y}^{\rm eq}_{\Delta}}-1\right)\! \gamma_D\, \mathcal{E}_{\alpha\beta}
    - \mathcal{W}^D_{\alpha\beta} - \mathcal{W}^{\ell \phi}_{\alpha\beta}
    - \mathcal{W}^{4\ell}_{\alpha\beta} - \mathcal{W}^{\ell\Delta}_{\alpha\beta}\;,
\label{4_BE_Delta_l}
\end{equation}
in which the terms $\mathcal{W}^D_{\alpha\beta}$, $\mathcal{W}^{\ell \phi}_{\alpha\beta}$, $\mathcal{W}^{4\ell}_{\alpha\beta}$ and $\mathcal{W}^{\ell\Delta}_{\alpha\beta}$ correspond to different washout processes, to be specified below.
The source term of~\eref{4_BE_Delta_l} arises from the two-loop self-energy diagrams of~\fref{4_fig_self-energy}, which provide the flavor-covariant CP-asymmetry matrix
\begin{equation}
  \mathcal{E}_{\alpha\beta}\, =\, \frac{1}{4\pi i}\frac{M_\Delta}{v^2}\sqrt{B_\ell B_\phi}\
    \frac{(M^H_\nu M_\nu^{\Delta\dagger} - M_\nu^\Delta M^{H\dagger}_\nu)_{\alpha\beta}}{\bar{M}_\nu^\Delta}\;.
\label{4_E_alphabeta}
\end{equation}
It is straightforward to check that the trace of this matrix is equal to the total CP asymmetry between triplet and anti-triplet decays: $\mathrm{tr}\, \mathcal{E} = \sum_{\alpha, \beta} \epsilon_{\alpha \beta} = \epsilon_\Delta$.

The washout term $\mathcal{W}^D_{\alpha\beta}$ is associated with triplet and anti-triplet inverse decays. It arises from the one-loop contribution to the lepton doublet self-energy, shown in~\fref{4_fig_self-energy}, and is given by
\begin{align}
  \mathcal{W}^D_{\alpha\beta}\ & =\ \frac{2 B_\ell}{\mathrm{tr}(yy^\dagger)}
    \bigg[ (yy^\dagger)_{\alpha\beta}\, \frac{Y_{\Delta_{\Delta}}}{Y_{\Delta}^{\rm eq}+\bar{Y}_{\Delta}^{\rm eq}}\nonumber\\&\qquad +\:\frac{1}{4 Y_\ell^{\text{eq}}}
    \left( 2 y [Y_{\Delta \ell}]^{\mathsf{T}} y^\dagger + yy^\dagger Y_{\Delta  \ell} + Y_{\Delta \ell} yy^\dagger \right)_{\alpha\beta}\bigg]\! \gamma_D\;.
\label{4_W_D}
\end{align}
In~\eref{4_W_D}, $Y_{\Delta_{\Delta}} \equiv (n_\Delta - \bar{n}_{\Delta}) / s$ is the triplet asymmetry, $Y_\ell^{\text{eq}} \equiv n_\ell^{\text{eq}} / s$ and $\gamma_D$ is the total, thermally-averaged decay rate
of triplets and anti-triplets:
\begin{align}
  \gamma_D\, =\, & \int\! \frac{{\rm d}^3\mathbf{p}}{(2\pi)^32 \omega_{\mathbf{p}}} \int\! \frac{{\rm d}^3\mathbf{k}}{(2\pi)^32 \omega_{\mathbf{k}}}
    \int\! \frac{{\rm d}^3\mathbf{q}}{(2\pi)^32 \omega_{\mathbf{q}}}\; 3 \left( \lambda_\ell^2 + \lambda_\phi^2 \right) (k\cdot q)  \nonumber \\
  & \times (2\pi)^4 \delta^{(4)} (p-k-q) \left\lbrace f_\Delta^{\mathrm{eq}} (\mathbf{p})
    + \bar{f}_\Delta^{\mathrm{eq}} (\mathbf{p}) \right\rbrace\;.
\end{align}

The other washout terms are associated with $2 \to 2$ scattering processes and originate from two-loop contributions to the lepton doublet self-energy. $\mathcal{W}^{ \ell \phi}_{\alpha\beta}$ accounts for the washout of the flavor
asymmetries by the $\Delta L =2$ scatterings $\ell_\gamma \ell_\delta \leftrightarrow \bar \phi \bar \phi$ and $\ell_\gamma \phi \leftrightarrow \bar \ell_\delta \bar \phi$, and is given by
\begin{align}
  \mathcal{W}^{ \ell \phi}_{\alpha\beta}\, =\ & 2 \left\lbrace \frac{1}{\mathrm{tr}(yy^\dagger)}
    \left[ \frac{ \left( 2y [Y_{\Delta\ell}]^{\mathsf{T}} y^\dagger + yy^\dagger Y_{\Delta\ell}
    + Y_{\Delta\ell} yy^\dagger \right)_{\alpha\beta} }{4Y_\ell^\text{eq}}\,
    + \frac{Y_{\Delta \phi}}{Y_\phi^{\text{eq}}}\, (yy^\dagger)_{\alpha\beta} \right] \right.\!
    \gamma_{\ell \phi}^\Delta  \nonumber \\
  & + \frac{1}{\mbox{Re} \left[ \mathrm{tr}(y\kappa^\dagger) \right]}
    \left[ \frac{ \left( 2 y [Y_{\Delta\ell}]^{\mathsf{T}} \kappa^\dagger + y \kappa^\dagger Y_{\Delta\ell}
    + Y_{\Delta\ell} y \kappa^\dagger \right)_{\alpha\beta} }{4Y_\ell^\text{eq}}\,
    + \frac{Y_{\Delta \phi}}{Y_\phi^{\text{eq}}}\, (y \kappa^\dagger)_{\alpha\beta} \right]\!
    \gamma_{\ell \phi}^\mathcal{I}  \nonumber \\
  & + \frac{1}{\mbox{Re} \left[ \mathrm{tr}(y\kappa^\dagger) \right]}
    \left[ \frac{ \left( 2 \kappa [Y_{\Delta\ell}]^{\mathsf{T}} y^\dagger + \kappa y^\dagger Y_{\Delta\ell}
    + Y_{\Delta\ell} \kappa y^\dagger \right)_{\alpha\beta}}{4Y_\ell^\text{eq}}\,
    + \frac{Y_{\Delta \phi}}{Y_\phi^{\text{eq}}}\, (\kappa y^\dagger)_{\alpha\beta} \right]\!
    \gamma_{\ell \phi}^\mathcal{I}  \nonumber \\
  & \left. + \frac{1}{\mathrm{tr}(\kappa\kappa^\dagger)}
    \left[ \frac{ \left( 2 \kappa [Y_{\Delta\ell}]^{\mathsf{T}} \kappa^\dagger + \kappa \kappa^\dagger Y_{\Delta\ell}
    + Y_{\Delta\ell} \kappa \kappa^\dagger \right)_{\alpha\beta}}{4Y_\ell^\text{eq}}\,
    + \frac{Y_{\Delta \phi}}{Y_\phi^{\text{eq}}}\, (\kappa\kappa^\dagger)_{\alpha\beta} \right]\!
    \gamma_{\ell \phi}^H \right\rbrace\;,
\label{4_W_lh}
\end{align}
in which $\gamma_{\ell \phi}^\Delta$ and $\gamma_{\ell \phi}^H$ are respectively the contributions of scalar-triplet exchange and of the $d=5$ operators in~\eref{4_Weinberg_operator} to the rate of $\Delta L =2$ scatterings $\gamma_{\ell \phi}$, and $\gamma_{\ell \phi}^\mathcal{I}$ is the interference term (more precisely, $\gamma_{\ell \phi}=\gamma_{\ell \phi}^\Delta+2\gamma_{\ell \phi}^\mathcal{I}+\gamma_{\ell \phi}^H$). The remaining washout terms $\mathcal{W}^{4\ell}_{\alpha\beta}$ and $\mathcal{W}^{\ell\Delta}_{\alpha\beta}$ are associated with $\Delta L = 0$ scatterings. Even though they do not violate lepton number, they modify the dynamics of leptogenesis
by redistributing the lepton asymmetry among the different flavors, thus affecting the value of the final $B-L$ asymmetry. For the washout term due to the lepton-lepton scatterings $\ell_\gamma\ell_\delta\leftrightarrow\ell_\rho\ell_\sigma$ and $\ell_\gamma\bar \ell_\rho\leftrightarrow\bar \ell_\delta\ell_\sigma$, one obtains
\begin{equation}
  \mathcal{W}^{4\ell}_{\alpha\beta}\, =\, \frac{2}{\lambda^4_\ell}
    \left[ \lambda^2_\ell\, \frac{ \left( 2 y [Y_{\Delta\ell}]^{\mathsf{T}} y^\dagger + yy^\dagger Y_{\Delta\ell}
    + Y_{\Delta\ell} yy^\dagger \right)_{\alpha\beta}}{4Y_\ell^\text{eq}}\,
    - \frac{\mathrm{tr}(Y_{\Delta\ell} yy^\dagger)}{Y_\ell^{\text{eq}}}\, (yy^\dagger)_{\alpha\beta} \right]\! \gamma_{4\ell}\;,
\label{4_W_4l}
\end{equation}
while for the lepton-triplet scatterings $\ell_\gamma\Delta\leftrightarrow\ell_\delta\Delta$, $\ell_\gamma\bar \Delta\leftrightarrow\ell_\delta\bar \Delta$ and $\ell_\gamma\bar \ell_\delta\leftrightarrow\Delta\bar \Delta$:
\begin{equation}
  \mathcal{W}^{\ell\Delta}_{\alpha\beta}\, =\, \frac{1}{\mathrm{tr}(yy^\dagger yy^\dagger)\, 2 Y_\ell^\text{eq}}\,
    \left( yy^\dagger yy^\dagger Y_{\Delta\ell} - 2 yy^\dagger Y_{\Delta\ell} yy^\dagger
    + Y_{\Delta\ell} yy^\dagger yy^\dagger \right)_{\alpha\beta} \gamma_{\ell\Delta}\;.
\label{4_W_ellDelta}
\end{equation}
The scattering rates $\gamma_{4\ell}$, $\gamma_{\ell\Delta}$ and the contributions $\gamma_{\ell \phi}^\Delta$, $\gamma_{\ell \phi}^\mathcal{I}$ and $\gamma_{\ell \phi}^H$ to $\gamma_{\ell \phi}$ are computed with the appropriate subtraction of on-shell intermediate states when necessary (see the discussion in Chapter~\cite{leptogenesis:A04}). Their expressions can be found in Ref.~\cite{Lavignac:2015gpa}.

Since the couplings $y_{\alpha \beta}$ and $\kappa_{\alpha \beta}$ transform as $(y, \kappa) \to U^* (y, \kappa)\, U^\dagger$ under flavor rotations $\ell \to U \ell$, one immediately sees from~\eref{4_E_alphabeta}, \eref{4_W_D}, \eref{4_W_lh}, \eref{4_W_4l} and~\eref{4_W_ellDelta} that the CP-asymmetry matrix $\mathcal{E}$ and the various washout terms transform as $(\mathcal{E}, \mathcal{W}) \to U^* (\mathcal{E}, \mathcal{W})\, U^{\mathsf{T}}$, as required by flavor covariance.

In order to have a closed set of Boltzmann equations, one must supplement \eref{4_BE_Delta_l} with equations for $Y_{\Delta}+\bar{Y}_{\Delta}$ and $Y_{\Delta_{\Delta}}$ (an equation for $Y_{\Delta \phi}$ is not needed, as $Y_{\Delta \phi}$ can be expressed as a function\footnote{For instance, in the limit where all spectator processes (electroweak and QCD sphalerons, Standard Model Yukawa couplings) are neglected, which has been implicitly considered so far, one has $Y_{\Delta \phi} \ =\  \mathrm{tr} Y_{\Delta\ell} \:-\: 2 Y_{\Delta_{\Delta}}$ from hypercharge and baryon number conservation.} of $Y_{\Delta_{\Delta}}$ and $[Y_{\Delta\ell}]_{\alpha\beta}$):
\begin{subequations}
\begin{align}
  sHz\,\frac{{\rm d}\big(Y_{\Delta}+\bar{Y}_{\Delta}\big)}{{\rm d}z}\ & =\ -\:\bigg[\frac{Y_{\Delta}+\bar{Y}_{\Delta}}{Y_{\Delta}^\text{eq}+\bar{Y}^{\rm eq}_{\Delta}}\:-\:1\bigg]\gamma_D\:
    -\:2\bigg[ \bigg(\frac{Y_{\Delta}+\bar{Y}_{\Delta}}{Y_{\Delta}^\text{eq}+\bar{Y}^{\rm eq}_{\Delta}} \bigg)^{\! 2}\: -\: 1\bigg] \gamma_A\; ,
\label{4_BE_Sigma_Delta}  \\
  sHz\,\frac{{\rm d}Y_{\Delta_{\Delta}}}{{\rm d}z}\ & =\ -\:\frac{1}{2}\big[\mathrm{tr}(\mathcal{W}^D)-W^D_\phi\big]\;,
\label{4_BE_Delta_Delta}
\end{align}
\end{subequations}
where the first and second terms in~\eref{4_BE_Sigma_Delta} are due to triplet/anti-triplet decays and to triplet-anti-triplet annihilations, respectively, and the term $W^D_\phi$ in~\eref{4_BE_Delta_Delta} is associated with the decays $\Delta \to \phi \phi$, $\bar \Delta \to \bar \phi \bar \phi$ and with their inverse decays:
\begin{equation}
  W^D_\phi\, =\, 2B_\phi\left(\frac{Y_{\Delta \phi}}{Y_\phi^\text{eq}}-\frac{Y_{\Delta_{\Delta}}}{Y_{\Delta}^\text{eq}+\bar{Y}_{\Delta}^\text{eq}}\right)\! \gamma_D\;.
\label{4_W^D_phi}
\end{equation}
Using~\eref{4_W_D} and~\eref{4_W^D_phi}, the Boltzmann equation for $Y_{\Delta_{\Delta}}$ can be rewritten as
\begin{equation}
  sHz\,\frac{{\rm d}Y_{\Delta_{\Delta}}}{{\rm d}z}\ =\ -\left(\frac{Y_{\Delta_{\Delta}}}{Y_{\Delta}^\text{eq}+\bar{Y}_{\Delta}^\text{eq}}
    +B_\ell\frac{\mathrm{tr}(yy^\dagger Y_{\Delta \ell})}{\lambda_\ell^2 Y_\ell^\text{eq}}
    -B_\phi\frac{Y_{\Delta \phi}}{Y_\phi^\text{eq}}\right)\! \gamma_D\;.
\label{4_BE_Delta_Delta_bis}
\end{equation}
%


\subsection{Flavor regimes and spectator processes}
\label{4_sec_regimes}

In deriving the flavor-covariant Boltzmann equation,~\eref{4_BE_Delta_l}, we assumed that the quantum correlations between the different lepton flavors are not affected by charged-lepton Yukawa interactions, which, strictly speaking, is true only above $T = 10^{12}\, \mbox{GeV}$ (see~\sref{sec_regimes}). At lower temperatures, the scatterings induced by charged-lepton Yukawa couplings can no longer be neglected, and their effects must be taken into account by appropriate terms on the right-hand side of~\eref{4_BE_Delta_l}. Alternatively, one can neglect the quantum correlations between lepton flavors that these processes, when they are sufficiently fast, tend to destroy.
For instance, below $T = 10^{12}\, \mbox{GeV}$, the tau Yukawa coupling is in equilibrium and drives the $(e,\tau)$ and $(\mu,\tau)$ entries of $Y_{\Delta\ell}$ to zero. The relevant dynamical variables in the temperature range
$10^{9}\, \mbox{GeV} < T < 10^{12}\, \mbox{GeV}$ are therefore $Y_{\Delta\ell_{\tau}}$ (the asymmetry stored in the tau lepton doublet) and the $2 \times 2$ matrix $[Y_{\Delta \ell}^0]_{\alpha \beta}$ (the flavor asymmetries stored in $\ell_e$ and $\ell_\mu$ and their quantum correlations). Accordingly, \eref{4_BE_Delta_l} must be replaced by two separate Boltzmann equations for $Y_{\Delta \ell_{\tau}}$ and $[Y_{\Delta \ell}^0]_{\alpha \beta}$, the second one being covariant with respect to rotations in the ($\ell_e$, $\ell_\mu$) flavor space. Below $T = 10^9\, \mbox{GeV}$, the muon Yukawa coupling also enters equilibrium and destroys the correlations between the $e$ and $\mu$ flavors. The Boltzmann equation,~\eref{4_BE_Delta_l}, then reduces to three equations for the flavor asymmetries $Y_{\Delta \ell_{\alpha}}$ ($\alpha = e, \mu, \tau$).

Finally, the effect of spectator processes~\cite{Buchmuller:2001sr, Nardi:2005hs}, which affect the dynamics of leptogenesis even though they do not violate lepton number, must be taken into account~\cite{Sierra:2014tqa}.
Working in the usual approximation that, in a given temperature range, each of these reactions is either negligible or in equilibrium, one obtains relations among the various particle asymmetries in the plasma. Using these relations, one can write the Boltzmann equations solely in terms of asymmetries that are conserved by all spectator processes relevant in the temperature range considered. These asymmetries are $Y_{\Delta_{\Delta}}$, the $3 \times 3$ and $2 \times 2$ flavor-covariant matrices
\begin{equation}
  Y_{\Delta_{\alpha\beta}}\ \equiv\ \frac 1 3 \,Y_{\Delta B}\, \delta_{\alpha\beta} - [Y_{\Delta\ell}]_{\alpha\beta}
    \quad\  \text{and}\  \quad
    Y^0_{\Delta_{\alpha\beta}}\, \equiv\, \frac 1 3 Y_{\Delta B}\, \delta_{\alpha\beta} - [Y^0_{\Delta \ell}]_{\alpha\beta}
\end{equation}
(relevant in the temperature regimes $T > 10^{12}\, \mbox{GeV}$ and $10^9\, \mbox{GeV} < T < 10^{12}\, \mbox{GeV}$, respectively), which are conserved by all spectator processes except charged-lepton Yukawa interactions,
and
\begin{equation}
  Y_{\Delta_\alpha}\ \equiv\ Y_{\Delta B/3-L_{\alpha}}\ =\ \frac 1 3\,Y_{\Delta B} - Y_{\Delta\ell_{\alpha}}-Y_{\Delta e_{R\alpha}}\;,
\end{equation}
which are preserved by all SM interactions. In addition to $Y_{\Delta}+\bar{Y}_{\Delta}$ and $Y_{\Delta_{\Delta}}$, the dynamical variables appearing in the Boltzmann equations (after making use of the equilibrium relations)
are $Y_{\Delta_{\alpha\beta}}$ above $T = 10^{12}\, \mbox{GeV}$, $(Y^0_{\Delta_{\alpha\beta}}, Y_{\Delta_\tau})$ between $T = 10^9\, \mbox{GeV}$ and $T = 10^{12}\, \mbox{GeV}$, and $(Y_{\Delta_e}, Y_{\Delta_\mu}, Y_{\Delta_\tau})$ below $T = 10^9\, \mbox{GeV}$.

The expressions for the Boltzmann equations valid in each temperature regime, with proper inclusion of the spectator processes, can be found in Ref.~\cite{Lavignac:2015gpa}.


\subsection{The relevance of flavor effects}
\label{4_sec_covariance}

A remarkable property of scalar triplet leptogenesis, as opposed to leptogenesis in the type I seesaw framework, is that lepton flavor effects are relevant in all temperature regimes. In particular, there is no well-defined single-flavor approximation in scalar triplet leptogenesis. The basic reason for this is that the scalar triplet couples to a pair of leptons rather than to a specific combination of lepton flavors. By contrast, in the leptogenesis scenario with right-handed neutrinos, the couplings of the lightest singlet neutrino $N_1$ can be written as
\begin{equation}
  - \sum_\alpha \lambda_{\alpha 1} \bar \ell_\alpha \phi^c N_1 \:+\: \mbox{h.c.}\ =\
    -\:\lambda_{N_1} \bar \ell_{N_1\,}\! \phi^c N_1 \:+\: \mbox{h.c.}\; ,
\end{equation}
where $\ell_{N_1} \equiv \sum_\alpha \lambda^*_{\alpha 1} \ell_\alpha / \lambda_{N_1}$ and $\lambda_{N_1} \equiv \sqrt{\sum_\alpha |\lambda_{\alpha 1}|^2}\, $. Assuming hierarchical right-handed neutrinos, so that the heavier singlet neutrinos $N_2$ and $N_3$ are not present in the plasma when $N_1$ starts to decay (and neglecting the $\Delta L=2$ scattering processes mediated by $N_2$ and $N_3$), the coherence of $\ell_{N_1}$ is preserved
as long as the scatterings induced by the charged-lepton Yukawa couplings remain out of equilibrium, i.e.~in  the temperature regime $T > 10^{12}\, \mbox{GeV}$. Leptogenesis can then be described in terms of a single lepton flavor\footnote{An exception to this statement is when the lepton asymmetries generated in $N_2$ and $N_3$ decays have not been completely washed out before the out-of-equilibrium decays of $N_1$ start to occur.} --- hence the name {\it single-flavor approximation}. This can be understood in more technical terms by going to the flavor basis $(\ell_{N_1}, \ell_{\perp 1}, \ell_{\perp 2})$, where $\ell_{\perp 1}$ and $\ell_{\perp 2}$ are two directions perpendicular to $\ell_{N_1}$ in flavor space. When the charged-lepton Yukawa couplings and the washout terms mediated by $N_2$ and $N_3$ are switched off, the Boltzmann equation for $[Y_{\Delta\ell}]_{11} \equiv Y_{\Delta\ell_{N_1}}$ becomes independent of the other entries of the matrix $[Y_{\Delta\ell}]_{\alpha\beta}$, and the source terms for $Y_{\Delta\ell_{\perp1}}$ and $Y_{\Delta\ell_{\perp 2}}$ vanish. Analogously, in the temperature regime $10^9\, \mbox{GeV} < T < 10^{12}\, \mbox{GeV}$, where the tau Yukawa coupling is in equilibrium but the muon and electron ones are not, leptogenesis can be described in terms of the flavor asymmetries $Y_{\Delta\ell_{\tau}}$ and $Y_{\Delta\ell_{0}}$ (where $\ell_0 \propto \lambda^*_{e 1} \ell_e + \lambda^*_{\mu 1}\ell_\mu$), provided that $N_2$ and $N_3$ play a negligible role in the generation and washout of the lepton asymmetry.

In scalar triplet leptogenesis, one may formally define a single-flavor approximation by making the substitutions $[Y_{\Delta\ell}]_{\alpha\beta} \to Y_{\Delta\ell}$, $y_{\alpha\beta} \to \lambda_\ell$, $\kappa_{\alpha\beta} \to \lambda_\kappa \equiv \sqrt{\mbox{tr}(\kappa \kappa^\dagger)}$ and $\mathcal{E}_{\alpha\beta} \to \epsilon_\Delta$ in~\eref{4_BE_Delta_l} and~\eref{4_BE_Delta_Delta_bis}, but the resulting Boltzmann equations\footnote{These equations are the ones that were derived and used in the first quantitative study of scalar triplet leptogenesis~\cite{Hambye:2005tk}, which did not include flavor effects.} cannot be obtained as limits of the flavor-covariant ones. As a consequence, neglecting flavor effects in scalar triplet leptogenesis does not, in general, provide a good approximation to the flavor-covariant computation, even above $T = 10^{12}\, \mbox{GeV}$.
This is a clear difference with the standard leptogenesis scenario with hierarchical right-handed neutrinos. The analogue of the single-flavor approximation of the type I seesaw case is in fact a ``three-flavor approximation'' in which flavor effects still play a prominent role. Namely, in the basis where the triplet couplings to leptons are flavor diagonal, the Boltzmann equations for the diagonal entries of the matrix $[Y_{\Delta\ell}]_{\alpha\beta}$ become independent
of the off-diagonal ones when the contribution of the dimension-5 operators in~\eref{4_Weinberg_operator} to the $\Delta L = 2$ scatterings in~\eref{4_W_lh} vanishes. Equation~\eqref{4_BE_Delta_l} may then be replaced by three Boltzmann equations for the flavor asymmetries $Y_{\Delta\ell_1}$, $Y_{\Delta\ell_2}$ and $Y_{\Delta\ell_3}$, where the $\ell_i$ define the basis of flavor space in which the couplings $y_{\alpha \beta}$ are diagonal. It should be clear that the three-flavor approximation is valid only in this particular basis; in any other basis, the diagonal and off-diagonal entries of $[Y_{\Delta\ell}]_{\alpha\beta}$ are coupled. Furthermore, the flavor-covariant Boltzmann equations must be used as soon as the contribution of the operators in~\eref{4_Weinberg_operator} to $\Delta L = 2$ scatterings is sizable. Finally, between $T = 10^9\, \mbox{GeV}$ and $T = 10^{12}\, \mbox{GeV}$, there is no flavor basis in which~\eref{4_BE_Delta_l} can be substituted for Boltzmann equations for ``diagonal'' flavor asymmetries, even when $\Delta L = 2$ scatterings are negligible. The use of the flavor-covariant formalism involving the $2 \times 2$ matrix $[Y^0_{\Delta\ell}]_{\alpha\beta}$ is therefore unavoidable in this regime.


\subsection{Quantitative impact of flavor effects}
\label{4_sec_pheno}

Let us now illustrate the relevance of flavor effects by means of some numerical examples. Given the large number of parameters involved, we shall concentrate on two suitably chosen Ans\"atze. We can take as independent parameters the scalar triplet mass $M_\Delta$ and its couplings to Higgs doublets ($\lambda_\phi$) and to lepton doublets ($y_{\alpha \beta}$). Once values for these parameters and for the neutrino parameters are chosen (including the yet unknown mass ordering, lightest neutrino mass and phases of the PMNS matrix), the coefficients $\kappa_{\alpha \beta} / \Lambda$ of the effective dimension-5 operators in~\eref{4_Weinberg_operator} are completely fixed by the neutrino mass formula in~\eref{4_mnu}. For definiteness, we work in the charged-lepton mass eigenbasis, in which the neutrino mass matrix takes the form $M_\nu = U_{\nu}^*\, \mbox{diag}(m_1, m_2, m_3)\, U_{\nu}^\dagger$, where the $m_i$ ($i = 1, 2, 3$) are the neutrino masses and $U_\nu$ is the PMNS matrix. For the mixing angles and squared mass differences, we take values within $1 \sigma$ of the best fit to global neutrino data of Ref.~\cite{Gonzalez-Garcia:2014bfa}. Finally, we set all phases of the PMNS matrix to zero, assume a normal mass ordering and take the lightest neutrino mass to be $m_1 = 10^{-3}\, \mbox{eV}$ at the triplet mass scale. 

For the triplet parameters, we choose the following Ans\"atze, defined in terms of the triplet contribution to the neutrino mass matrix $m_\Delta$:
\begin{itemize}
\item {\bf Ansatz 1:} $M_\nu^\Delta = i M_\nu$ \
\item {\bf Ansatz 2:} $M_\nu^\Delta = i \bar M_\nu\, U_\nu^* \left( \begin{array}{ccc} 0.949 & 0 & 0 \\
  0 & 0.048 & 0 \\
  0 & 0 & 0.312
  \end{array} \right) U_\nu^\dagger\, $, \\ \\
where $\bar M_\nu \equiv \sqrt{\mathrm{tr} (M^\dagger_\nu M_\nu)} = \sqrt{\sum_i m^2_i}$.
\end{itemize}
Both Ans\"atze are characterized by $\bar M_\nu^\Delta = \bar M_\nu$. Since $[M_\nu^\Delta]_{\alpha \beta} = \lambda_\phi y_{\alpha \beta}\, v^2 / (4 M_\Delta)$, the hierarchical structure of the triplet couplings to leptons $y_{\alpha \beta}$ is completely determined in each case, while two parameters, which can be chosen to be $\lambda_\ell$ and $M_\Delta$, remain free. In Ansatz~1, the triplet couplings to leptons are proportional to the entries of the neutrino mass matrix, while, in Ansatz 2, the hierarchical structures of $y_{\alpha \beta}$ and $[M_\nu]_{\alpha \beta}$ are very different. Ansatz~1 also has the property of maximizing the total CP asymmetry $\epsilon_\Delta$.

\begin{figure}[!t]
\centering
\includegraphics[scale=0.45]{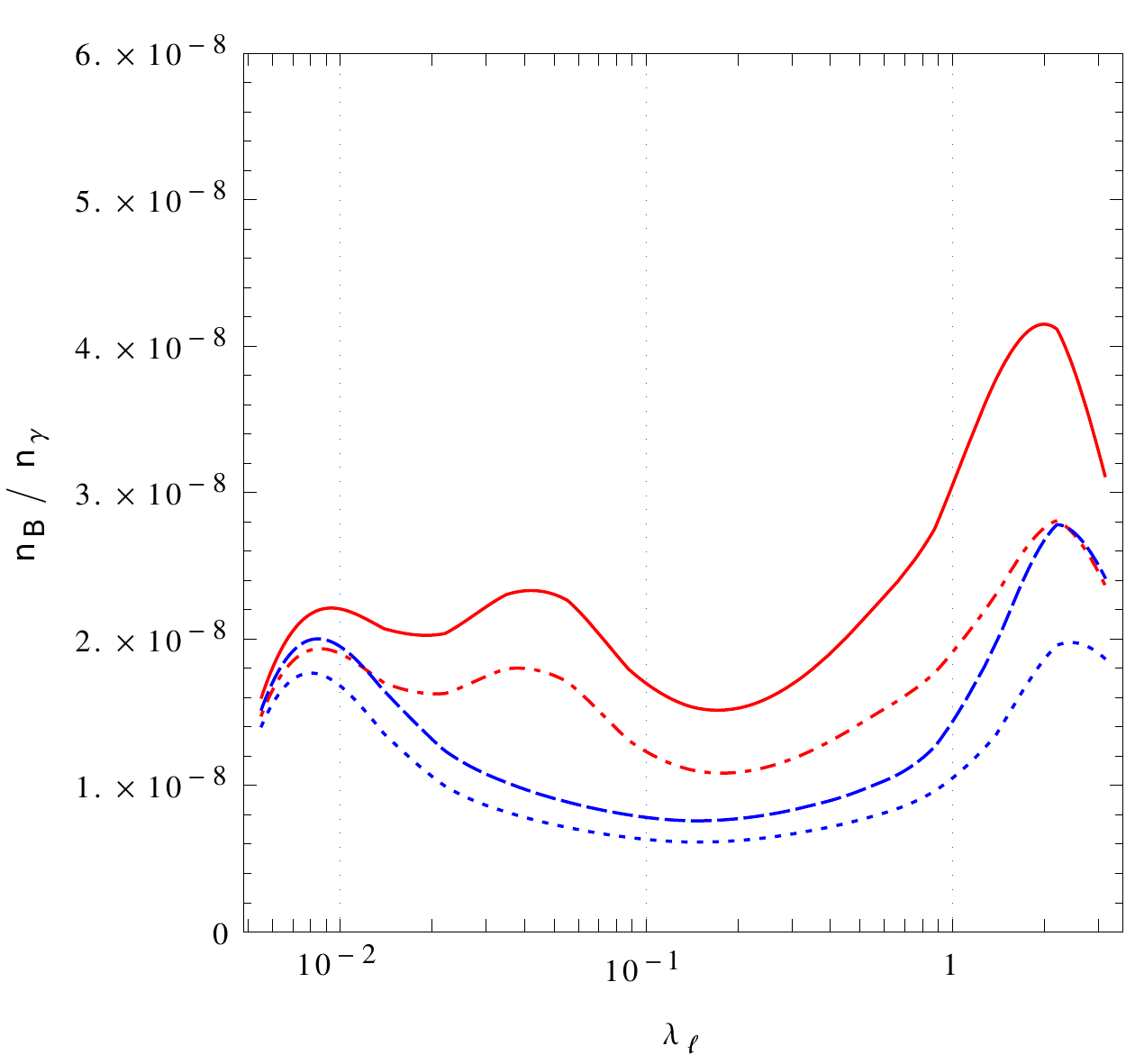} \qquad
\includegraphics[scale=0.45]{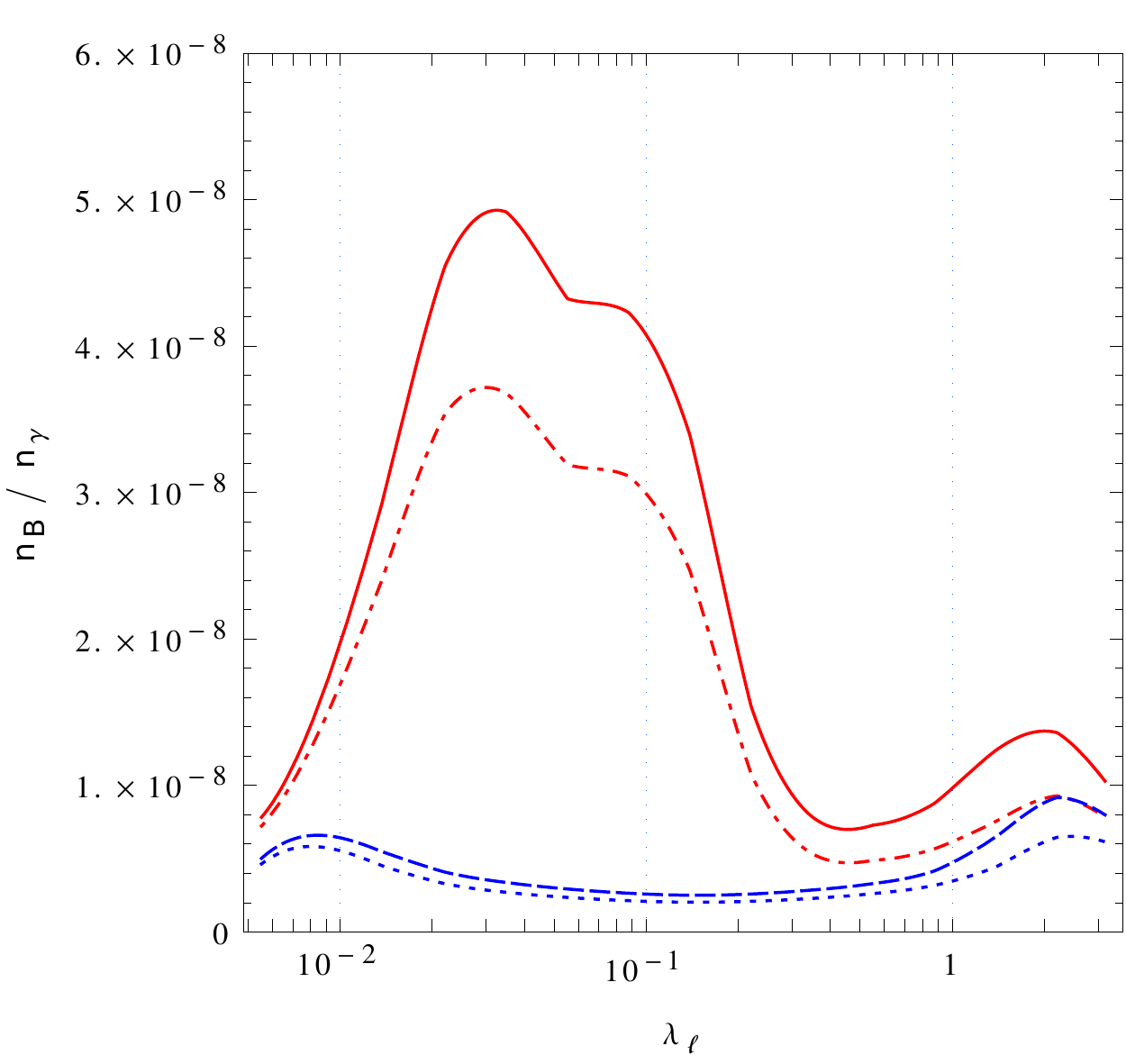}
\caption{Baryon-to-photon ratio $n_B/n_\gamma$ as a function of $\lambda_\ell$ for $M_\Delta=5\times10^{12}\, \mbox{GeV}$, assuming Ansatz~1 (left panel) or Ansatz~2 (right panel). The red lines show the result of the flavor-covariant computation involving the $3 \times 3$ matrix $[Y_{\Delta \ell}]_{\alpha\beta}$, with (solid red line) or without (dashed-dotted red line) spectator processes taken into account, while the blue lines correspond to the result of the single-flavor approximation, including spectator processes (blue dashed line) or not (blue dotted line). The branching ratios $B_\ell$ and $B_\phi$ are equal for $\lambda_\ell \simeq 0.15$. Figure taken from Ref.~\cite{Lavignac:2015gpa}.}
\label{4_fig_comp1}
\end{figure}

Figure~\ref{4_fig_comp1} shows the impact of lepton flavor effects and spectator processes on the generated baryon-to-photon ratio for Ansatz~1 (left panel) and Ansatz~2 (right panel). The triplet mass has been chosen to be $M_\Delta = 5 \times 10^{12}\, \mbox{GeV}$, so that most of the $B-L$ asymmetry is produced at $T > 10^{12}\, \mbox{GeV}$. The flavor-covariant computation involving the $3 \times 3$ matrix of flavor asymmetries
$[Y_{\Delta \ell}]_{\alpha \beta}$ is compared with the single-flavor approximation, with and without spectator processes. Flavor effects are sizable for practically all parameter values and typically lead to an enhancement of the generated baryon asymmetry by a factor of order one (up to an order of magnitude for Ansatz~2 with $\lambda_\ell \sim 0.03$). However, for small values of $\lambda_\ell$ (corresponding to $B_\ell \ll B_\phi$), the difference between
the flavor-covariant computation and the single flavor approximation is much less significant. This can easily be understood by noting that, in this limit, the washout of the flavored lepton asymmetries, which is mainly due to the inverse decays $\ell_\alpha \ell_\beta \to \bar \Delta$ and $\bar \ell_\alpha \bar \ell_\beta \to \Delta$, becomes less important. Neglecting all washout terms in the Boltzmann equation,~\eref{4_BE_Delta_l}, and taking the trace over lepton flavors, one obtains
\begin{align}
  &sHz\,\frac{{\rm d} [Y_{\Delta\ell}]_{\alpha\beta}}{{\rm d}z}\, =
    \bigg(\frac{Y_{\Delta}+\bar{Y}_{\Delta}}{Y_\Delta^\text{eq}+\bar{Y}^{\rm eq}_{\Delta}}\:-\:1\bigg) \gamma_D\, \mathcal{E}_{\alpha\beta}\;,
    \nonumber\\&\qquad \qquad \Longrightarrow \qquad
  sHz\,\frac{{\rm d} Y_{\Delta\ell}}{{\rm d}z}\, =
    \bigg(\frac{Y_{\Delta}+\bar{Y}_{\Delta}}{Y_\Delta^\text{eq}+\bar{Y}^{\rm eq}_{\Delta}}\:-\:1\bigg) \gamma_D \epsilon_\Delta\;,
\end{align}
which is the equation of the single-flavor approximation in the same limit. Flavor effects also tend to become relatively less important in the opposite limit $\lambda_\ell \gg 1$ (corresponding to $B_\ell \gg B_\phi$), because
the lepton flavor asymmetries are more efficiently washed out than for smaller values of $\lambda_\ell$, and the asymmetry generated in the Higgs sector becomes the dominant source of the final baryon-to-photon ratio~\cite{Hambye:2005tk, Lavignac:2015gpa}.

\begin{figure}[!t]
\centering
\includegraphics[scale=0.45]{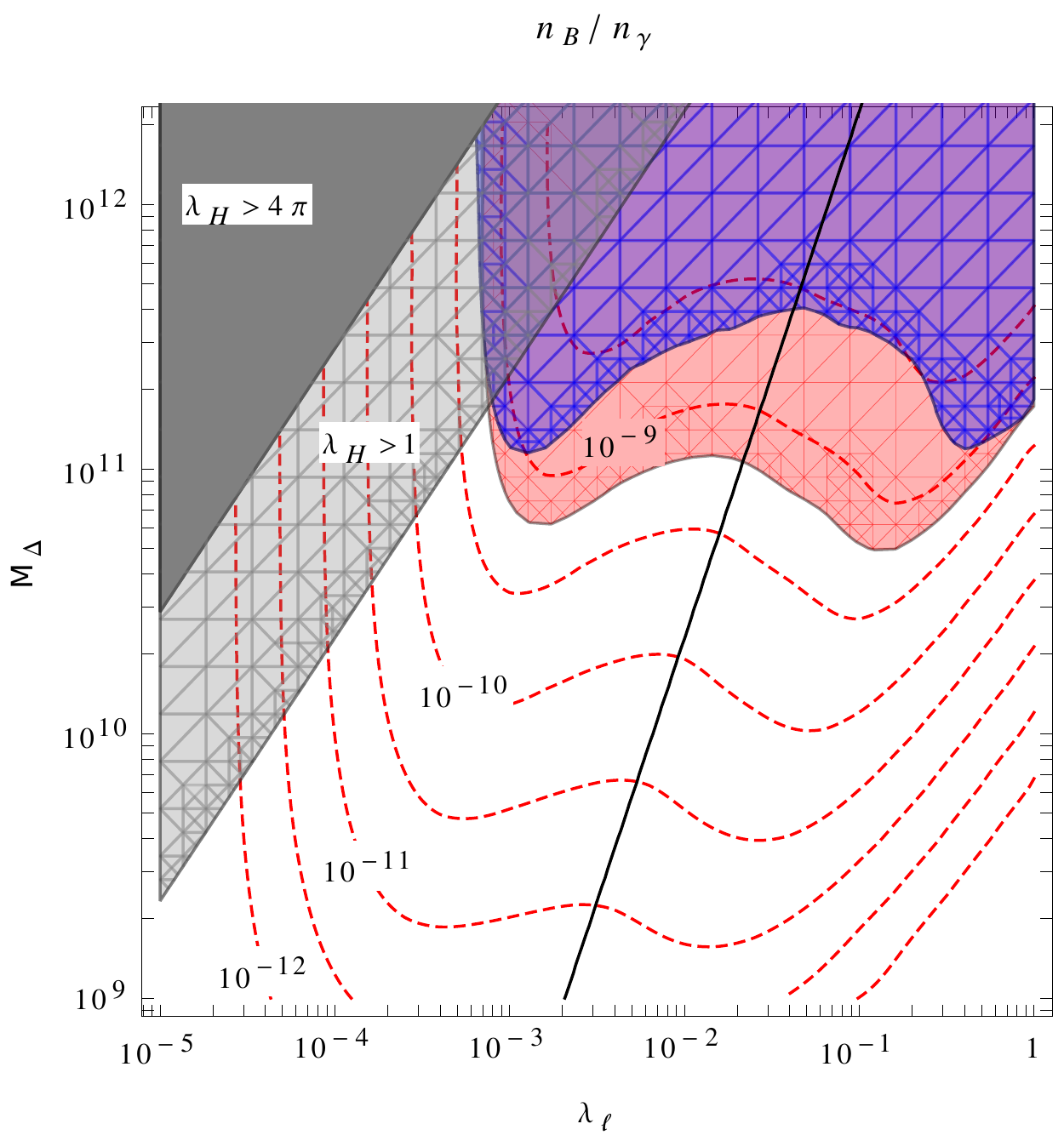} \qquad
\includegraphics[scale=0.45]{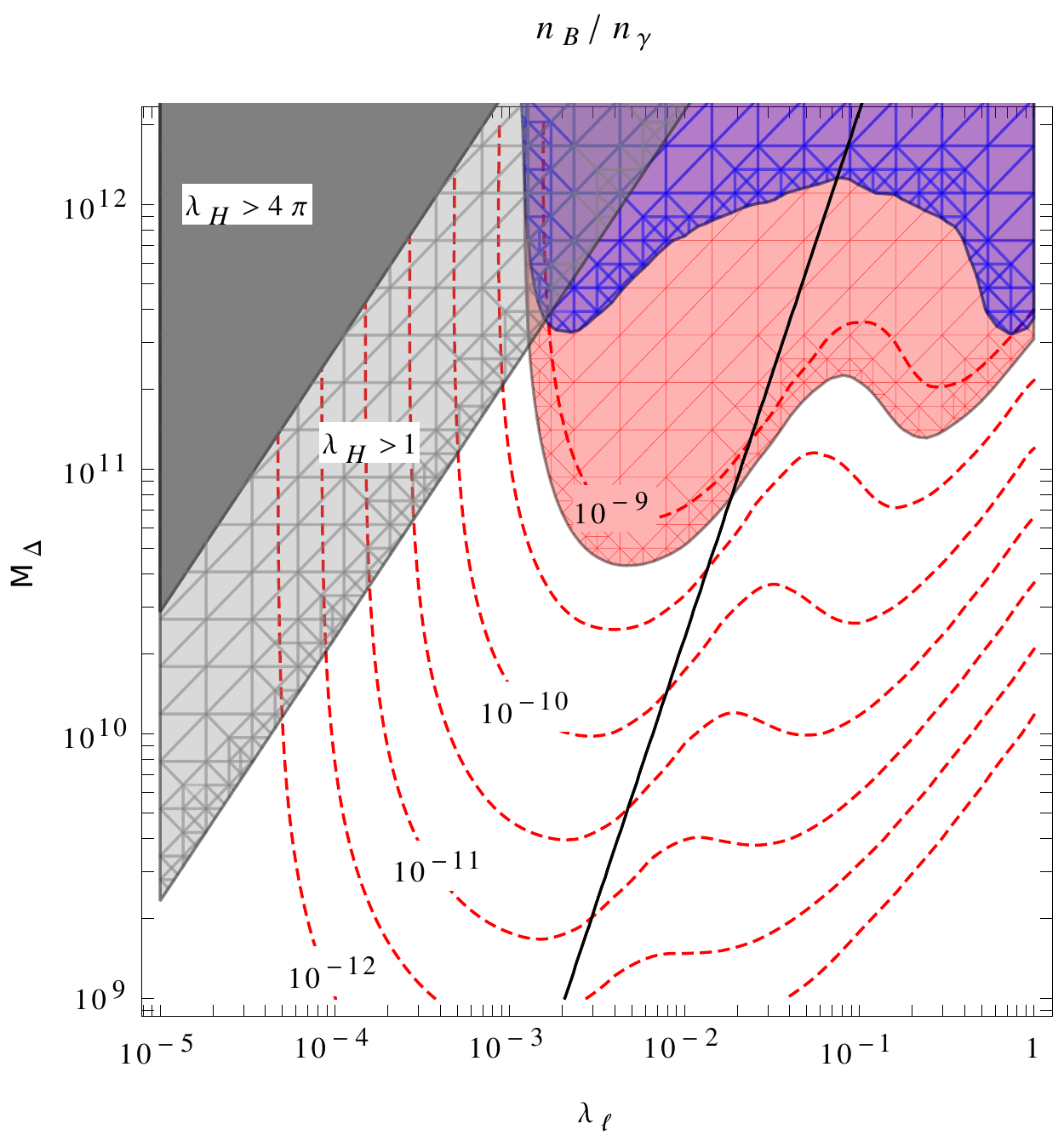}
\caption{Isocurves of the baryon-to-photon ratio $n_B/n_\gamma$ in the $(\lambda_\ell, M_\Delta)$ plane, obtained performing the flavor-covariant computation including spectator processes, assuming Ansatz~1 (left panel) or Ansatz~2 (right panel). The shaded, colored areas correspond to the regions of the parameter space where the observed baryon asymmetry can be reproduced in the flavor-covariant computation (light red shading) or in the single-flavor approximation neglecting spectator processes (dark blue shading). The solid black line corresponds to $B_\ell = B_\phi$. Also shown are the regions where $\lambda_\phi$ is greater than $1$ or $4\pi$. Figure taken from Ref.~\cite{Lavignac:2015gpa}.}
\label{4_fig_M_Delta_dependence}
\end{figure}

Figure~\ref{4_fig_M_Delta_dependence} shows the dependence of the generated baryon asymmetry on $\lambda_\ell$ and $M_\Delta$ for Ansatz~1 (left panel) and Ansatz~2 (right panel). The isocurves of the baryon-to-photon ratio correspond to the flavor-covariant computation including spectator processes. The comparison of the two shaded, colored areas shows that the inclusion of flavor effects significantly enlarges the region of parameter space where successful scalar triplet leptogenesis is possible. For the Ans\"atze considered, the observed baryon-to-photon ratio can be reproduced for triplet masses as low as $4.4 \times 10^{10}\, \mbox{GeV}$, to be compared with $1.2\times10^{11}\, \mbox{GeV}$ in the approximation where flavor effects and spectator processes are neglected. These values are not absolute lower bounds, as different assumptions about the triplet parameters can lead to successful leptogenesis for lower triplet masses (for instance, Ref.~\cite{Hambye:2005tk} found a lower bound $M_\Delta > 2.8 \times 10^{10}\, \mbox{GeV}$ for $\bar M_\nu^\Delta = 0.001\, \mbox{eV} \ll \bar M_\nu$ in the single-flavor approximation).


\section{Importance of flavor in other models}
\label{sec:other}

Before concluding this chapter, we remark on the importance of flavor effects in other models of leptogenesis. We focus, in particular, on those models detailed in the other chapters of this review, and cross references are included where appropriate.

\paragraph{ARS mechanism.} If the sterile-neutrino Yukawa couplings are sufficiently small, successful leptogenesis can be achieved within type I seesaw scenarios at scales as low as $M \sim 1\,$--$\,100\ {\rm GeV}$, whilst at the same time satisfying the observational and experimental constraints on the SM neutrino masses. The smallness of these Yukawa couplings delays the thermalization of the sterile states, such that at least one of them can still be out of equilibrium at the onset of the electroweak phase transition. Their CP-violating oscillations are then able to distribute lepton asymmetry unevenly amongst the different flavors. These individual asymmetries can then be communicated to the charged leptons by any of the sterile neutrinos that are in equilibrium and reprocessed into baryon asymmetry by sphaleron processes. The resulting baryon asymmetry is protected from the eventual equilibration of the sterile states, since this occurs after the sphaleron processes have switched off. This scenario of baryogenesis via leptogenesis is known as the ARS mechanism, after Akhmedov, Rubakov and Smirnov~\cite{Akhmedov:1998qx} (see also Ref.~\cite{Asaka:2005pn}). In contrast to the scenarios described in the rest of this chapter, the ARS mechanism does not rely on the Majorana nature of the sterile neutrinos, and it therefore allows for successful leptogenesis also for Dirac-type neutrinos. Even if Majorana masses are present, the lepton number violating processes that they mediate are suppressed in the regime $T\gg M$ relevant to the ARS mechanism. With the exception of contributions to the asymmetry from thermally-induced $L$- and CP-violating decays of the Higgs doublet~\cite{Hambye:2016sby, Hambye:2017elz}, ARS leptogenesis is therefore a purely flavored scenario, and further discussions can be found in the dedicated Chapter~\cite{leptogenesis:A02} along with an overview of its experimental signatures in Chapter~\cite{leptogenesis:A05}.

\paragraph{Extended low-scale type II and type III leptogenesis.} The resonant enhancement of CP violation in type II (scalar triplet) and type III (fermion triplet) seesaw scenarios can be implemented through the addition of new scalars and fermions. Further discussions and references can be found in the discussions in Sec. 4.2 of Chapter~\cite{leptogenesis:A05}. 

\paragraph{Left-right symmetric models.} Further discussions of the embeddings of low-scale resonant scenarios in left-right-symmetric~\cite{Mohapatra:1974hk, Mohapatra:1974gc, Senjanovic:1975rk} extensions of the SM gauge groups ($SU(2)_L\times SU(2)_R\times U(1)_{B-L}$) can be found in Sec. 5.2 of Chapter~\cite{leptogenesis:A05}.

\paragraph{Type I soft leptogenesis.} Soft SUSY breaking terms can give rise to additional sources of CP violation, allowing leptogenesis to be realised in supersymmetric type I seesaw scenarios at temperatures $T\lesssim 10^9\ {\rm GeV}$ lower than the bound from gravitino over-production. Further details of type I soft leptogenesis and the importance of lepton flavor effects are discussed in Sec. 6.1 of Chapter~\cite{leptogenesis:A05}.

\paragraph{Flavor symmetries.} In order to predict the mixing angles and phases of the PMNS matrix, one can assume that the three generations of SM leptons form a triplet of a flavor symmetry group $G_f$, which may be taken together with a CP symmetry that acts non-trivially in flavor space.  A comprehensive discussion of flavor symmetries and their implications for leptogenesis can be found in Chapter~\cite{leptogenesis:A06}.

\section{Conclusions}
\label{sec:conclusions}

In this chapter, we have highlighted the potential importance of accounting fully for flavor effects in order to obtain accurate estimates of the final lepton (and therefore baryon) asymmetry in scenarios of leptogenesis. Flavor correlations in the heavy-neutrino sector contribute to the source of the CP asymmetry, and flavor correlations in the charged-lepton sector are important for determining the washout of the lepton asymmetry.  The effect on the latter can even allow for successful leptogenesis when total lepton number is conserved (or the violation of total lepton number is suppressed). 

In the case of thermal leptogenesis based on the type I seesaw scenario, we have seen that the region of parameter space where the next-to-lightest RH neutrino dominates the production of the asymmetry is enhanced when charged-lepton flavor effects are taken into account. Moreover, once these effects are accounted for, only one scenario of thermal leptogenesis can successfully generate the observed asymmetry whilst remaining independent of the initial conditions: the tau $N_2$-dominated scenario, wherein the asymmetry is mostly produced by decays of the next-to-lightest heavy neutrino via the tau channel. In these flavored regimes, the evolution of the individual flavor asymmetries can be coupled by spectator effects, and this can expand and open up viable regions of parameter space for $N_2$-dominated scenarios.

In resonant leptogenesis, we have seen that coherences in the charged-lepton and heavy-neutrino sectors play significant and opposing roles in determining the final asymmetry.  This is because, for the quasi-degenerate heavy-neutrino mass spectra relevant to these scenarios, flavor oscillations also contribute to the source of the CP asymmetry in addition to the flavor mixing that dominates for hierarchical mass spectra. Treating only coherences in the heavy-neutrino flavors but neglecting coherences amongst the charged-lepton flavors can overestimate the asymmetry by as much as a factor of 5.  Doing the opposite, i.e.~treating only coherences in the charged-lepton flavors but neglecting coherences amongst the heavy-neutrino flavors, can instead underestimate the asymmetry by as much as a factor of 2.  This motivates the use of fully flavor-covariant approaches that are able to yield rate equations for the matrices of charged-lepton and heavy-neutrino number densities. Such approaches can be realised both in semi-classical and field-theoretic descriptions of leptogenesis, and we have briefly reviewed these complementary methodologies.

Furthermore, for models of leptogenesis embedded in the type II seesaw scenario, we have seen that charged-lepton flavor effects are relevant in all temperature regimes, since the scalar triplet couples to a pair of lepton doublets. A flavor-covariant treatment then shows that accounting fully for these effects typically leads to an order-one enhancement of the asymmetry compared to a single-flavor approximation, where the latter may be justified for small triplet-lepton couplings.

Aside from having an important impact on the final asymmetry, flavor effects are also relevant to the testability of leptogenesis.  Specifically, when flavor effects cannot be neglected, leptogenesis becomes sensitive to the phases of the PMNS matrix. Moreover, in low-scale resonant scenarios, some of the Yukawa couplings remain sizable, allowing such models to be directly testable in current and near-future experiments, including the LHC, as well as low-energy experiments looking for lepton flavor and lepton number violation.


\newpage

\section*{Acknowledgments}

We thank Emiliano Molinaro and Serguey Petcov for helpful comments. PDB acknowledges financial support from the STFC Consolidated Grant ST/L000296/1. The work of SL has been supported in part by the European Union Horizon 2020 Research and Innovation Programme under the Marie Sk\l odowska-Curie Grant Agreements No.~690575 and No.~674896. The work of PM is supported by STFC Grant No. ST/L000393/1 and a Leverhulme Trust Research Leadership Award. The work of DT is supported by a ULB postdoctoral fellowship and the Belgian Federal Science Policy (IAP P7/37). We gratefully acknowledge the hospitality of the Munich Institute for Astro- and Particle Physics (MIAPP) of the DFG cluster of excellence ``Origin and Structure of the Universe'', where this work has been initiated.


\bibliography{A01_flavour}{}

\begin{thebibliography}{149}
\providecommand{\natexlab}[1]{#1}
\providecommand{\url}[1]{\texttt{#1}}
\expandafter\ifx\csname urlstyle\endcsname\relax
  \providecommand{\doi}[1]{doi: #1}\else
  \providecommand{\doi}{doi: \begingroup \urlstyle{rm}\Url}\fi

\bibitem{Abada:2006fw}
A.~Abada, S.~Davidson, F.-X. Josse-Michaux, M.~Losada, and A.~Riotto, {Flavor
  issues in leptogenesis}, \emph{JCAP}. {\bf 0604}, \penalty0 004,  (2006).

\bibitem{Nardi:2006fx}
E.~Nardi, Y.~Nir, E.~Roulet, and J.~Racker, {The Importance of flavor in
  leptogenesis}, \emph{JHEP}. {\bf 01}, \penalty0 164,  (2006).

\bibitem{Abada:2006ea}
A.~Abada, S.~Davidson, A.~Ibarra, F.~X. Josse-Michaux, M.~Losada, and
  A.~Riotto, {Flavour matters in leptogenesis}, \emph{JHEP}. {\bf 09},
  \penalty0 010,  (2006).

\bibitem{Blanchet:2006be}
S.~Blanchet and P.~Di~Bari, {Flavor effects on leptogenesis predictions},
  \emph{JCAP}. {\bf 0703}, \penalty0 018,  (2007).

\bibitem{Pascoli:2006ie}
S.~Pascoli, S.~T. Petcov, and A.~Riotto, {Connecting low energy leptonic $CP$
  violation to leptogenesis}, \emph{Phys. Rev.} {\bf D75}, \penalty0 083511,
  (2007).

\bibitem{DeSimone:2006nrs}
A.~De~Simone and A.~Riotto, {On the impact of flavour oscillations in
  leptogenesis}, \emph{JCAP}. {\bf 0702}, \penalty0 005,  (2007).

\bibitem{Fukugita:1986hr}
M.~Fukugita and T.~Yanagida, {Baryogenesis without grand unification},
  \emph{Phys. Lett.} {\bf B174}, \penalty0 45--47,  (1986).

\bibitem{Pilaftsis:1998pd}
A.~Pilaftsis, {Heavy Majorana neutrinos and baryogenesis}, \emph{Int. J. Mod.
  Phys.} {\bf A14}, \penalty0 1811--1858,  (1999).

\bibitem{Davidson:2008bu}
S.~Davidson, E.~Nardi, and Y.~Nir, {Leptogenesis}, \emph{Phys. Rept.} {\bf
  466}, \penalty0 105--177,  (2008).

\bibitem{Blanchet:2012bk}
S.~Blanchet and P.~Di~Bari, {The minimal scenario of leptogenesis}, \emph{New
  J. Phys.} {\bf 14}, \penalty0 125012,  (2012).

\bibitem{Fong:2013wr}
C.~S. Fong, E.~Nardi, and A.~Riotto, {Leptogenesis in the Universe}, \emph{Adv.
  High Energy Phys.} {\bf 2012}, \penalty0 158303,  (2012).

\bibitem{leptogenesis:A03}
P.~S.~B. Dev, M.~Garny, J.~Klaric, P.~Millington, and D.~Teresi, {Resonant
  enhancement in leptogenesis}, \emph{Int.~J.~Mod.~Phys.} {\bf A33}, \penalty0
  1842003,  (2018).

\bibitem{Akhmedov:1998qx}
E.~{\relax Kh}. Akhmedov, V.~A. Rubakov, and A.~{\relax Yu}. Smirnov,
  {Baryogenesis via Neutrino Oscillations}, \emph{Phys. Rev. Lett.} {\bf 81},
  \penalty0 1359--1362,  (1998).

\bibitem{leptogenesis:A02}
M.~Drewes, B.~Garbrecht, P.~Hern\'{a}ndez, M.~Kekic, J.~Lopez-Pavon, J.~Racker,
  N.~Rius, J.~Salvado, and D.~Teresi, {ARS leptogenesis},
  \emph{Int.~J.~Mod.~Phys.} {\bf A33}, \penalty0 1842002,  (2018).

\bibitem{Baym:1961zz}
G.~Baym and L.~P. Kadanoff, {Conservation Laws and Correlation Functions},
  \emph{Phys. Rev.} {\bf 124}, \penalty0 287--299,  (1961).

\bibitem{KadanoffBaym}
G.~Baym and L.~P. Kadanoff, \emph{{Quantum Statistical Mechanics}}. (Benjamin,
  New York, 1962).

\bibitem{Schwinger:1960qe}
J.~S. Schwinger, {Brownian Motion of a Quantum Oscillator}, \emph{J. Math.
  Phys.} {\bf 2}, \penalty0 407--432,  (1961).

\bibitem{Keldysh:1964ud}
L.~V. Keldysh, {Diagram Technique for Nonequilibrium Processes}, \emph{Zh.
  Eksp. Teor. Fiz.} {\bf 47}, \penalty0 1515--1527,  (1964).
\newblock {[Sov. Phys. JETP {\bf20}, 1018, (1965)]}.

\bibitem{Dolgov:1980cq}
A.~D. Dolgov, {Neutrinos in the early Universe}, \emph{Sov. J. Nucl. Phys.}
  {\bf 33}, \penalty0 700--706,  (1981).
\newblock [Yad.~Fiz.~{\bf 33}, 1309, (1981)].

\bibitem{Stodolsky:1986dx}
L.~Stodolsky, {Treatment of neutrino oscillations in a thermal environment},
  \emph{Phys. Rev.} {\bf D36}, \penalty0 2273,  (1987).

\bibitem{Raffelt:1992uj}
G.~Raffelt, G.~Sigl, and L.~Stodolsky, {Non-Abelian Boltzmann equation for
  mixing and decoherence}, \emph{Phys. Rev. Lett.} {\bf 70}, \penalty0
  2363--2366,  (1993).
\newblock {[Erratum: Phys. Rev. Lett.{\bf 98}, 069902, (2007)]}.

\bibitem{Sigl:1992fn}
G.~Sigl and G.~Raffelt, {General kinetic description of relativistic mixed
  neutrinos}, \emph{Nucl. Phys.} {\bf B406}, \penalty0 423--451,  (1993).

\bibitem{Garbrecht:2013urw}
B.~Garbrecht, F.~Glowna, and P.~Schwaller, {Scattering rates for leptogenesis:
  Damping of lepton flavour coherence and production of singlet neutrinos},
  \emph{Nucl. Phys.} {\bf B877}, \penalty0 1--35,  (2013).

\bibitem{Garbrecht:2014kda}
B.~Garbrecht and P.~Schwaller, {Spectator effects during leptogenesis in the
  strong washout regime}, \emph{JCAP}. {\bf 1410}\penalty0 (10), \penalty0 012,
   (2014).

\bibitem{Kolb:1979qa}
E.~W. Kolb and S.~Wolfram, {Baryon number generation in the early Universe},
  \emph{Nucl. Phys.} {\bf B172}, \penalty0 224,  (1980).
\newblock {[Erratum: Nucl.~Phys.~{\bf B195}, 542, (1982)]}.

\bibitem{Luty:1992un}
M.~A. Luty, {Baryogenesis via leptogenesis}, \emph{Phys. Rev.} {\bf D45},
  \penalty0 455--465,  (1992).

\bibitem{leptogenesis:A04}
S.~Biondini, D.~B{\"o}deker, N.~Brambilla, M.~Garny, J.~Ghiglieri,
  A.~Hohenegger, M.~Laine, S.~Mendizabal, P.~Millington, A.~Salvio, and
  A.~Vairo, {Status of rates and rate equations for thermal leptogenesis},
  \emph{Int.~J.~Mod.~Phys.} {\bf A33}, \penalty0 1842004,  (2018).

\bibitem{Dev:2014laa}
P.~S. Bhupal~Dev, P.~Millington, A.~Pilaftsis, and D.~Teresi, {Flavour
  covariant transport equations: An application to resonant leptogenesis},
  \emph{Nucl. Phys.} {\bf B886}, \penalty0 569--664,  (2014).

\bibitem{Weisskopf:1930au}
V.~Weisskopf and E.~P. Wigner, {Berechnung der nat\"{u}rlichen Linienbreite auf
  Grund der Diracschen Lichttheorie (Calculation of the natural brightness of
  spectral lines on the basis of Dirac's theory)}, \emph{Z. Phys.} {\bf 63},
  \penalty0 54--73,  (1930).

\bibitem{Calzetta:1986cq}
E.~Calzetta and B.~L. Hu, {Nonequilibrium quantum fields: Closed-time-path
  effective action, Wigner function and Boltzmann equation}, \emph{Phys. Rev.}
  {\bf D37}, \penalty0 2878,  (1988).

\bibitem{Lee:2004we}
C.~Lee, V.~Cirigliano, and M.~J. Ramsey-Musolf, {Resonant relaxation in
  electroweak baryogenesis}, \emph{Phys. Rev.} {\bf D71}, \penalty0 075010,
  (2005).

\bibitem{Millington:2012pf}
P.~Millington and A.~Pilaftsis, {Perturbative nonequilibrium thermal field
  theory}, \emph{Phys. Rev.} {\bf D88}\penalty0 (8), \penalty0 085009,  (2013).

\bibitem{Millington:2013isa}
P.~Millington and A.~Pilaftsis, {Perturbative non-equilibrium thermal field
  theory to all orders in gradient expansion}, \emph{Phys. Lett.} {\bf B724},
  \penalty0 56--62,  (2013).

\bibitem{Buchmuller:2000nd}
W.~Buchm{\"u}ller and S.~Fredenhagen, {Quantum mechanics of baryogenesis},
  \emph{Phys. Lett.} {\bf B483}, \penalty0 217--224,  (2000).

\bibitem{DeSimone:2007gkc}
A.~De~Simone and A.~Riotto, {Quantum Boltzmann equations and leptogenesis},
  \emph{JCAP}. {\bf 0708}, \penalty0 002,  (2007).

\bibitem{Garny:2009rv}
M.~Garny, A.~Hohenegger, A.~Kartavtsev, and M.~Lindner, {Systematic approach to
  leptogenesis in nonequilibrium QFT: Vertex contribution to the $CP$-violating
  parameter}, \emph{Phys. Rev.} {\bf D80}, \penalty0 125027,  (2009).

\bibitem{Garny:2009qn}
M.~Garny, A.~Hohenegger, A.~Kartavtsev, and M.~Lindner, {Systematic approach to
  leptogenesis in nonequilibrium QFT: Self-energy contribution to the
  $CP$-violating parameter}, \emph{Phys. Rev.} {\bf D81}, \penalty0 085027,
  (2010).

\bibitem{Beneke:2010wd}
M.~Beneke, B.~Garbrecht, M.~Herranen, and P.~Schwaller, {Finite number density
  corrections to leptogenesis}, \emph{Nucl. Phys.} {\bf B838}, \penalty0 1--27,
   (2010).

\bibitem{Anisimov:2010aq}
A.~Anisimov, W.~Buchm{\"{u}}ller, M.~Drewes, and S.~Mendizabal, {Leptogenesis
  from Quantum Interference in a Thermal Bath}, \emph{Phys. Rev. Lett.} {\bf
  104}, \penalty0 121102,  (2010).

\bibitem{Anisimov:2010dk}
A.~Anisimov, W.~Buchm{\"{u}}�ller, M.~Drewes, and S.~Mendizabal, {Quantum
  leptogenesis I}, \emph{Annals Phys.} {\bf 326}, \penalty0 1998--2038,
  (2011).
\newblock {[Erratum: Annals Phys. {\bf 338}, 376, (2011)]}.

\bibitem{Beneke:2010dz}
M.~Beneke, B.~Garbrecht, C.~Fidler, M.~Herranen, and P.~Schwaller, {Flavoured
  leptogenesis in the CTP formalism}, \emph{Nucl. Phys.} {\bf B843}, \penalty0
  177--212,  (2011).

\bibitem{Garbrecht:2011xw}
B.~Garbrecht and M.~Garny, {Finite width in out-of-equilibrium propagators and
  kinetic theory}, \emph{Annals Phys.} {\bf 327}, \penalty0 914--934,  (2012).

\bibitem{Fidler:2011yq}
C.~Fidler, M.~Herranen, K.~Kainulainen, and P.~M. Rahkila, {Flavoured quantum
  Boltzmann equations from cQPA}, \emph{JHEP}. {\bf 02}, \penalty0 065,
  (2012).

\bibitem{Cline:1993bd}
J.~M. Cline, K.~Kainulainen, and K.~A. Olive, {Protecting the primordial baryon
  asymmetry from erasure by sphalerons}, \emph{Phys. Rev.} {\bf D49}, \penalty0
  6394--6409,  (1994).

\bibitem{Capozzi:2017ipn}
F.~Capozzi, E.~Di~Valentino, E.~Lisi, A.~Marrone, A.~Melchiorri, and
  A.~Palazzo, {Global constraints on absolute neutrino masses and their
  ordering}, \emph{Phys. Rev.} {\bf D95}\penalty0 (9), \penalty0 096014,
  (2017).

\bibitem{leptogenesis:A06}
C.~Hagedorn, R.~N. Mohapatra, E.~Molinaro, C.~C. Nishi, and S.~T. Petcov, {CP
  violation in the lepton sector and implications for leptogenesis},
  \emph{Int.~J.~Mod.~Phys.} {\bf A33}, \penalty0 1842006,  (2018).

\bibitem{Esteban:2016qun}
I.~Esteban, M.~C. Gonzalez-Garcia, M.~Maltoni, I.~Martinez-Soler, and
  T.~Schwetz, {Updated fit to three neutrino mixing: exploring the
  accelerator-reactor complementarity}, \emph{JHEP}. {\bf 01}, \penalty0 087,
  (2017).

\bibitem{KamLAND-Zen:2016pfg}
A.~Gando et~al., {Search for Majorana Neutrinos Near the Inverted Mass
  Hierarchy Region with KamLAND-Zen}, \emph{Phys. Rev. Lett.} {\bf
  117}\penalty0 (8), \penalty0 082503,  (2016).
\newblock [Addendum: Phys.~Rev.~Lett.~{\bf 117}, 109903, (2016)].

\bibitem{Agostini:2017dxu}
M.~Agostini et~al.
\newblock {Searching for neutrinoless double beta decay with GERDA}.
\newblock In \emph{{15th International Conference on Topics in Astroparticle
  and Underground Physics (TAUP 2017), 24--28 July, 2017 Sudbury, Ontario,
  Canada}},  (2017).

\bibitem{Albert:2017owj}
J.~B. Albert et~al., {Search for Neutrinoless Double-Beta Decay with the
  Upgraded EXO-200 Detector}, {\ttfamily arXiv:1707.08707},  (2017).

\bibitem{Aalseth:2017btx}
C.~E. Aalseth et~al., {Search for Zero-Neutrino Double Beta Decay in 76Ge with
  the Majorana Demonstrator}, {\ttfamily arXiv:1710.11608},  (2017).

\bibitem{Aghanim:2016yuo}
N.~Aghanim et~al., {Planck intermediate results. XLVI. Reduction of large-scale
  systematic effects in HFI polarization maps and estimation of the
  reionization optical depth}, \emph{Astron. Astrophys.} {\bf 596}, \penalty0
  A107,  (2016).

\bibitem{Buchmuller:2004nz}
W.~Buchm{\"{u}}ller, P.~Di~Bari, and M.~Pl{\"{u}}macher, {Leptogenesis for
  pedestrians}, \emph{Annals Phys.} {\bf 315}, \penalty0 305--351,  (2005).

\bibitem{Covi:1996wh}
L.~Covi, E.~Roulet, and F.~Vissani, {CP violating decays in leptogenesis
  scenarios}, \emph{Phys. Lett.} {\bf B384}, \penalty0 169--174,  (1996).

\bibitem{Blanchet:2006ch}
S.~Blanchet, P.~Di~Bari, and G.~G. Raffelt, {Quantum Zeno effect and the impact
  of flavor in leptogenesis}, \emph{JCAP}. {\bf 0703}, \penalty0 012,  (2007).

\bibitem{Kuzmin:1985mm}
V.~A. Kuzmin, V.~A. Rubakov, and M.~E. Shaposhnikov, {On the anomalous
  electroweak baryon-number non-conservation in the early universe},
  \emph{Phys. Lett.} {\bf 155B}, \penalty0 36,  (1985).

\bibitem{Ade:2015xua}
P.~A.~R. Ade et~al., {Planck 2015 results. XIII. Cosmological parameters},
  \emph{Astron. Astrophys.} {\bf 594}, \penalty0 A13,  (2016).

\bibitem{Davidson:2002qv}
S.~Davidson and A.~Ibarra, {A lower bound on the right-handed neutrino mass
  from leptogenesis}, \emph{Phys. Lett.} {\bf B535}, \penalty0 25--32,  (2002).

\bibitem{Buchmuller:2002rq}
W.~Buchm{\"{u}}ller, P.~Di~Bari, and M.~Pl{\"{u}}macher, {Cosmic microwave
  background, matter-antimatter asymmetry and neutrino masses}, \emph{Nucl.
  Phys.} {\bf B643}, \penalty0 367--390,  (2002).
\newblock {[Erratum: Nucl.~Phys.~B{\bf 793}, 362, (2002)]}.

\bibitem{Khlopov:1984pf}
M.~{\relax Yu}. Khlopov and A.~D. Linde, {Is it easy to save the gravitino?},
  \emph{Phys. Lett.} {\bf 138B}, \penalty0 265--268,  (1984).

\bibitem{Ellis:1984eq}
J.~R. Ellis, J.~E. Kim, and D.~V. Nanopoulos, {Cosmological gravitino
  regeneration and decay}, \emph{Phys. Lett.} {\bf 145B}, \penalty0 181--186,
  (1984).

\bibitem{Kawasaki:2008qe}
M.~Kawasaki, K.~Kohri, T.~Moroi, and A.~Yotsuyanagi, {Big-bang nucleosynthesis
  and gravitinos}, \emph{Phys. Rev.} {\bf D78}, \penalty0 065011,  (2008).

\bibitem{Hambye:2003rt}
T.~Hambye, Y.~Lin, A.~Notari, M.~Papucci, and A.~Strumia, {Constraints on
  neutrino masses from leptogenesis models}, \emph{Nucl. Phys.} {\bf B695},
  \penalty0 169--191,  (2004).

\bibitem{Blanchet:2008pw}
S.~Blanchet and P.~Di~Bari, {New aspects of leptogenesis bounds}, \emph{Nucl.
  Phys.} {\bf B807}, \penalty0 155--187,  (2009).

\bibitem{Barbieri:1999ma}
R.~Barbieri, P.~Creminelli, A.~Strumia, and N.~Tetradis, {Baryogenesis through
  leptogenesis}, \emph{Nucl. Phys.} {\bf B575}, \penalty0 61--77,  (2000).

\bibitem{Pascoli:2006ci}
S.~Pascoli, S.~T. Petcov, and A.~Riotto, {Leptogenesis and low energy
  CP-violation in neutrino physics}, \emph{Nucl. Phys.} {\bf B774}, \penalty0
  1--52,  (2007).

\bibitem{Anisimov:2007mw}
A.~Anisimov, S.~Blanchet, and P.~Di~Bari, {Viability of Dirac phase
  leptogenesis}, \emph{JCAP}. {\bf 0804}, \penalty0 033,  (2008).

\bibitem{Molinaro:2007uv}
E.~Molinaro, S.~T. Petcov, T.~Shindou, and Y.~Takanishi, {Effects of lightest
  neutrino mass in leptogenesis}, \emph{Nucl. Phys.} {\bf B797}, \penalty0
  93--116,  (2008).

\bibitem{Molinaro:2008rg}
E.~Molinaro and S.~T. Petcov, {The interplay between the {``}low{''} and
  {``}high{''} energy CP-violation in leptogenesis}, \emph{Eur. Phys. J.} {\bf
  C61}, \penalty0 93--109,  (2009).

\bibitem{Molinaro:2008cw}
E.~Molinaro and S.~T. Petcov, {A case of subdominant/suppressed {``}high
  energy{''} contribution to the baryon asymmetry of the Universe in flavoured
  leptogenesis}, \emph{Phys. Lett.} {\bf B671}, \penalty0 60--65,  (2009).

\bibitem{Blanchet:2011xq}
S.~Blanchet, P.~Di~Bari, D.~A. Jones, and L.~Marzola, {Leptogenesis with heavy
  neutrino flavours: from density matrix to Boltzmann equations}, \emph{JCAP}.
  {\bf 1301}, \penalty0 041,  (2013).

\bibitem{Buchmuller:2001sr}
W.~Buchm{\"u}ller and M.~Pl{\"u}macher, {Spectator processes and baryogenesis},
  \emph{Phys. Lett.} {\bf B511}, \penalty0 74--76,  (2001).

\bibitem{Nardi:2005hs}
E.~Nardi, Y.~Nir, J.~Racker, and E.~Roulet, {On Higgs and sphaleron effects
  during the leptogenesis era}, \emph{JHEP}. {\bf 01}, \penalty0 068,  (2006).

\bibitem{JosseMichaux:2007zj}
F.~X. Josse-Michaux and A.~Abada, {Study of flavour dependencies in
  leptogenesis}, \emph{JCAP}. {\bf 0710}, \penalty0 009,  (2007).

\bibitem{DiBari:2005st}
P.~Di~Bari, {See-saw geometry and leptogenesis}, \emph{Nucl. Phys.} {\bf B727},
  \penalty0 318--354,  (2005).

\bibitem{DiBari:2015svd}
P.~Di~Bari and M.~Re~Fiorentin, {Supersymmetric $SO(10)$-inspired leptogenesis
  and a new $N_2$-dominated scenario}, \emph{JCAP}. {\bf 1603}\penalty0 (03),
  \penalty0 039,  (2016).

\bibitem{Vives:2009zz}
O.~Vives.
\newblock {Flavoured leptogenesis: A successful thermal leptogenesis with $N_1$
  mass below $10^8$ GeV}.
\newblock In eds. J.~Bernab{\'{e}}u, F.~J. Botella, N.~E. Mavromatos, and V.~A.
  Mitsou, \emph{{Proceedings of DISCRETE '08: Symposium on the prospects in the
  physics of discrete symmetries, 11--16 December 2008, Valencia, Spain, J.
  Phys.: Conf. Ser.}}, vol. 171, p. 012076,  (2009).

\bibitem{Engelhard:2006yg}
G.~Engelhard, Y.~Grossman, E.~Nardi, and Y.~Nir, {Importance of the Heavier
  Singlet Neutrinos in Leptogenesis}, \emph{Phys. Rev. Lett.} {\bf 99},
  \penalty0 081802,  (2007).

\bibitem{Bertuzzo:2010et}
E.~Bertuzzo, P.~Di~Bari, and L.~Marzola, {The problem of the initial conditions
  in flavoured leptogenesis and the tauon $N_2$-dominated scenario},
  \emph{Nucl. Phys.} {\bf B849}, \penalty0 521--548,  (2011).

\bibitem{DiBari:2014eqa}
P.~Di~Bari, S.~King, and M.~Re~Fiorentin, {Strong thermal leptogenesis and the
  absolute neutrino mass scale}, \emph{JCAP}. {\bf 1403}, \penalty0 050,
  (2014).

\bibitem{Antusch:2010ms}
S.~Antusch, P.~Di~Bari, D.~A. Jones, and S.~F. King, {A fuller flavour
  treatment of $N_2$-dominated leptogenesis}, \emph{Nucl. Phys.} {\bf B856},
  \penalty0 180--209,  (2012).

\bibitem{Casas:2001sr}
J.~A. Casas and A.~Ibarra, {Oscillating neutrinos and $\mu\to e,\gamma$},
  \emph{Nucl. Phys.} {\bf B618}, \penalty0 171--204,  (2001).

\bibitem{Buchmuller:2002jk}
W.~Buchm{\"{u}}ller, P.~Di~Bari, and M.~Pl{\"{u}}macher, {A bound on neutrino
  masses from baryogenesis}, \emph{Phys. Lett.} {\bf B547}, \penalty0 128--132,
   (2002).

\bibitem{DiBari:2008mp}
P.~Di~Bari and A.~Riotto, {Successful type I leptogenesis with
  $SO(10)$-inspired mass relations}, \emph{Phys. Lett.} {\bf B671}, \penalty0
  462--469,  (2009).

\bibitem{DiBari:2010ux}
P.~Di~Bari and A.~Riotto, {Testing SO(10)-inspired leptogenesis with low energy
  neutrino experiments}, \emph{JCAP}. {\bf 1104}, \penalty0 037,  (2011).

\bibitem{DiBari:2017uka}
P.~Di~Bari and M.~Re~Fiorentin, {A full analytic solution of $SO(10)$-inspired
  leptogenesis}, \emph{JHEP}. {\bf 10}, \penalty0 029,  (2017).

\bibitem{DiBari:2014eya}
P.~Di~Bari, L.~Marzola, and M.~Re~Fiorentin, {Decrypting $SO(10)$-inspired
  leptogenesis}, \emph{Nucl. Phys.} {\bf B893}, \penalty0 122--157,  (2015).

\bibitem{DiBari:2013qja}
P.~Di~Bari and L.~Marzola, {$SO(10)$-inspired solution to the problem of the
  initial conditions in leptogenesis}, \emph{Nucl. Phys.} {\bf B877}, \penalty0
  719--751,  (2013).

\bibitem{Dueck:2013gca}
A.~Dueck and W.~Rodejohann, {Fits to SO(10) grand unified models}, \emph{JHEP}.
  {\bf 09}, \penalty0 024,  (2013).

\bibitem{Babu:2016bmy}
K.~S. Babu, B.~Bajc, and S.~Saad, {Yukawa sector of minimal SO(10)
  unification}, \emph{JHEP}. {\bf 02}, \penalty0 136,  (2017).

\bibitem{DiBari:2015oca}
P.~Di~Bari and S.~F. King, {Successful $N_2$ leptogenesis with flavour coupling
  effects in realistic unified models}, \emph{JCAP}. {\bf 1510}\penalty0 (10),
  \penalty0 008,  (2015).

\bibitem{Liu:1993tg}
J.~Liu and G.~Segr{\`e}, {Reexamination of generation of baryon and lepton
  number asymmetries by heavy particle decay}, \emph{Phys. Rev.} {\bf D48},
  \penalty0 4609--4612,  (1993).

\bibitem{Flanz:1994yx}
M.~Flanz, E.~A. Paschos, and U.~Sarkar, {Baryogenesis from a lepton asymmetric
  universe}, \emph{Phys. Lett.} {\bf B345}, \penalty0 248--252,  (1995).
\newblock [Errata: \emph{Phys.\ Lett.}\ {\bf B382} (1996) 447 and \emph{Phys.\
  Lett.}\ {\bf B384} (1996) 487].

\bibitem{Flanz:1996fb}
M.~Flanz, E.~A. Paschos, U.~Sarkar, and J.~Weiss, {Baryogenesis through mixing
  of heavy Majorana neutrinos}, \emph{Phys. Lett.} {\bf B389}, \penalty0
  693--699,  (1996).

\bibitem{Covi:1996fm}
L.~Covi and E.~Roulet, {Baryogenesis from mixed particle decays}, \emph{Phys.
  Lett.} {\bf B399}, \penalty0 113--118,  (1997).

\bibitem{Pilaftsis:1997jf}
A.~Pilaftsis, {CP violation and baryogenesis due to heavy Majorana neutrinos},
  \emph{Phys. Rev.} {\bf D56}, \penalty0 5431--5451,  (1997).

\bibitem{Pilaftsis:1997dr}
A.~Pilaftsis, {Resonant CP violation induced by particle mixing in transition
  amplitudes}, \emph{Nucl. Phys.} {\bf B504}, \penalty0 61--107,  (1997).

\bibitem{Buchmuller:1997yu}
W.~Buchm{\"u}ller and M.~Pl{\"u}macher, {CP asymmetry in Majorana neutrino
  decays}, \emph{Phys. Lett.} {\bf B431}, \penalty0 354--362,  (1998).

\bibitem{Pilaftsis:2005rv}
A.~Pilaftsis and T.~E.~J. Underwood, {Electroweak-scale resonant leptogenesis},
  \emph{Phys. Rev.} {\bf D72}, \penalty0 113001,  (2005).

\bibitem{Pilaftsis:2003gt}
A.~Pilaftsis and T.~E.~J. Underwood, {Resonant leptogenesis}, \emph{Nucl.
  Phys.} {\bf B692}, \penalty0 303--345,  (2004).

\bibitem{Dev:2015wpa}
P.~S.~B. Dev, P.~Millington, A.~Pilaftsis, and D.~Teresi, {Corrigendum to
  ``Flavour covariant transport equations: An application to resonant
  leptogenesis''}, \emph{Nucl. Phys.} {\bf B897}, \penalty0 749--756,  (2015).

\bibitem{Dev:2014wsa}
P.~S. Bhupal~Dev, P.~Millington, A.~Pilaftsis, and D.~Teresi, {Kadanoff-Baym
  approach to flavour mixing and oscillations in resonant leptogenesis},
  \emph{Nucl. Phys.} {\bf B891}, \penalty0 128--158,  (2015).

\bibitem{Dev:2014tpa}
P.~S. Bhupal~Dev, P.~Millington, A.~Pilaftsis, and D.~Teresi.
\newblock {Flavour Covariant Formalism for Resonant Leptogenesis}.
\newblock In eds. M.~Aguilar-Ben{\'{i}}tez, J.~Fuster,
  S.~Mart{\'{i}}-Garc{\'{i}}a, and A.~Santamar{\'{i}}a, \emph{{Proceedings of
  the 37th International Conference on High Energy Physics (ICHEP), 2--9 July
  2014, Valencia, Spain, Nucl. Part. Phys. Proc.}}, vol. 273-275, pp. 268--274,
   (2016).

\bibitem{Dev:2015dka}
P.~S. Bhupal~Dev, P.~Millington, A.~Pilaftsis, and D.~Teresi.
\newblock {Flavour effects in Resonant Leptogenesis from semi-classical and
  Kadanoff-Baym approaches}.
\newblock In eds. A.~Di~Domenico, N.~E. Mavromatos, V.~A. Mitsou, and D.~P.
  Skliros, \emph{{Proceedings of the 4th Symposium on Prospects in the Physics
  of Discrete Symmetries (DISCRETE2014), 2--6 December 2014, London, UK, J.
  Phys.: Conf. Ser.}}, vol. 631, p. 012087,  (2015).

\bibitem{Garbrecht:2011aw}
B.~Garbrecht and M.~Herranen, {Effective theory of Resonant Leptogenesis in the
  Closed-Time-Path approach}, \emph{Nucl. Phys.} {\bf B861}, \penalty0 17--52,
  (2012).

\bibitem{Kartavtsev:2015vto}
A.~Kartavtsev, P.~Millington, and H.~Vogel, {Lepton asymmetry from mixing and
  oscillations}, \emph{JHEP}. {\bf 06}, \penalty0 066,  (2016).

\bibitem{Hohenegger:2014cpa}
A.~Hohenegger and A.~Kartavtsev, {Leptogenesis in crossing and runaway
  regimes}, \emph{JHEP}. {\bf 07}, \penalty0 130,  (2014).

\bibitem{Pilaftsis:2004xx}
A.~Pilaftsis, {Resonant $\tau$ Leptogenesis with Observable Lepton Number
  Violation}, \emph{Phys. Rev. Lett.} {\bf 95}, \penalty0 081602,  (2005).

\bibitem{Deppisch:2010fr}
F.~F. Deppisch and A.~Pilaftsis, {Lepton flavor violation and $\theta_{13}$ in
  minimal resonant leptogenesis}, \emph{Phys. Rev.} {\bf D83}, \penalty0
  076007,  (2011).

\bibitem{Minkowski:1977sc}
P.~Minkowski, {$\mu \to e\gamma$ at a rate of one out of $10^{9}$ muon
  decays?}, \emph{Phys. Lett.} {\bf B67}, \penalty0 421--428,  (1977).

\bibitem{Mohapatra:1979ia}
R.~N. Mohapatra and G.~Senjanovi\'c, {Neutrino Mass and Spontaneous Parity
  Nonconservation}, \emph{Phys. Rev. Lett.} {\bf 44}, \penalty0 912,  (1980).

\bibitem{Yanagida:1979as}
T.~Yanagida.
\newblock {Horizontal symmetry and masses of neutrinos}.
\newblock In eds. O.~Sawada and A.~Sugamoto, \emph{{Proceedings of the Workshop
  on Unified Theories and Baryon Number in the Universe, 13--14 February 1979
  National Laboratory for High Energy Physics, Tsukuba, Japan}}, pp. 95--99,
  (1979).

\bibitem{GellMann:1980vs}
M.~Gell-Mann, P.~Ramond, and R.~Slansky.
\newblock {Complex spinors and unified theories}.
\newblock In eds. P.~van Nieuwenhuizen and D.~Z. Freedman, \emph{{Proceedings
  of Supergravity, 27--28 September 1979, Stony Brook, New York}}, pp.
  315--321,  (1979).

\bibitem{Glashow:1979nm}
S.~L. Glashow.
\newblock {The Future of Elementary Particle Physics}.
\newblock In eds. M.~L\'evy, J.-L. Basdevant, D.~Speiser, J.~Weyers,
  R.~Gastmans, and M.~Jacob, \emph{{Proceedings of the Carg\`ese Summer
  Institute: Quarks and Leptons, 9--29 July 1979, Carg\`ese, France, NATO Sci.
  Ser. B}}, vol.~61, p. 687,  (1980).

\bibitem{Capozzi:2013csa}
F.~Capozzi, G.~L. Fogli, E.~Lisi, A.~Marrone, D.~Montanino, and A.~Palazzo,
  {Status of three-neutrino oscillation parameters, circa 2013}, \emph{Phys.
  Rev.} {\bf D89}, \penalty0 093018,  (2014).

\bibitem{TheMEG:2016wtm}
A.~M. Baldini et~al., {Search for the lepton flavour violating decay $\mu
  ^+\rightarrow \mathrm {e}^+ \gamma $ with the full dataset of the MEG
  experiment}, \emph{Eur. Phys. J.} {\bf C76}\penalty0 (8), \penalty0 434,
  (2016).

\bibitem{Olive:2016xmw}
C.~Patrignani et~al., {Review of Particle Physics}, \emph{Chin. Phys.} {\bf
  C40}\penalty0 (10), \penalty0 100001,  (2016).

\bibitem{Kaulard:1998rb}
J.~Kaulard et~al., {Improved limit on the branching ratio of $\mu\to e$
  conversion on titanium}, \emph{Phys. Lett.} {\bf B422}, \penalty0 334--338,
  (1998).

\bibitem{Bertl:2006up}
W.~H. Bertl et~al., {A search for $\mu-e$ conversion in muonic gold},
  \emph{Eur. Phys. J.} {\bf C47}, \penalty0 337--346,  (2006).

\bibitem{Honecker:1996zf}
W.~Honecker et~al., {Improved Limit on the Branching Ratio of $\mu\to e$
  Conversion on Lead}, \emph{Phys. Rev. Lett.} {\bf 76}, \penalty0 200--203,
  (1996).

\bibitem{Escudero:2016odp}
L.~Escudero.
\newblock {Initial Probe of $\delta_{CP}$ by the T2K Experiment with
  $\nu_{\mu}$ Disappearance and $\nu_{e}$ Appearance}.
\newblock In eds. M.~Aguilar-Ben\'itez, J.~Fuster, S.~Mart\'i-Garc\'ia, and
  A.~Santamar\'ia, \emph{{Proceedings of the 37th International Conference on
  High Energy Physics (ICHEP 2014), 2--9 July 2014, Valencia, Spain, Nucl.
  Part. Phys. Proc.}}, vol. 273-275, pp. 1814--1819,  (2016).

\bibitem{Deppisch:2015qwa}
F.~F. Deppisch, P.~S. Bhupal~Dev, and A.~Pilaftsis, {Neutrinos and collider
  physics}, \emph{New J. Phys.} {\bf 17}\penalty0 (7), \penalty0 075019,
  (2015).

\bibitem{Bartoszek:2014mya}
L.~Bartoszek et~al.
\newblock {Mu2e Technical Design Report}.
\newblock Technical Report FERMILAB-TM-2594, FERMILAB-DESIGN-2014-01, Fermilab,
   (2014).
\newblock URL \url{http://arxiv.org/abs/arXiv:1501.05241}.

\bibitem{Kuno:2005mm}
Y.~Kuno.
\newblock {PRISM/PRIME}.
\newblock In eds. M.~Aoki, Y.~Iwashita, and M.~Kuze, \emph{{NuFact04:
  Proceedings of the 6th International Workshop on Neutrino Factories and
  Superbeams Osaka, 26 July--1 August 2014, Japan, Nucl. Phys. Proc. Suppl.}},
  vol. 149, pp. 376--378,  (2005).

\bibitem{Heurtier:2016iac}
L.~Heurtier and D.~Teresi, {Dark matter and observable lepton flavor
  violation}, \emph{Phys. Rev.} {\bf D94}\penalty0 (12), \penalty0 125022,
  (2016).

\bibitem{Hambye:2003ka}
T.~Hambye and G.~Senjanovi\'c, {Consequences of triplet seesaw for
  leptogenesis}, \emph{Phys. Lett.} {\bf B582}, \penalty0 73--81,  (2004).

\bibitem{Antusch:2004xy}
S.~Antusch and S.~F. King, {Type II leptogenesis and the neutrino mass scale},
  \emph{Phys. Lett.} {\bf B597}, \penalty0 199--207,  (2004).

\bibitem{Hambye:2005tk}
T.~Hambye, M.~Raidal, and A.~Strumia, {Efficiency and maximal CP-asymmetry of
  scalar triplet leptogenesis}, \emph{Phys. Lett.} {\bf B632}, \penalty0
  667--674,  (2006).

\bibitem{Chun:2006sp}
E.~J. Chun and S.~Scopel, {Analysis of leptogenesis in supersymmetric triplet
  seesaw model}, \emph{Phys. Rev.} {\bf D75}, \penalty0 023508,  (2007).

\bibitem{Hallgren:2007nq}
T.~H{\"a}llgren, T.~Konstandin, and T.~Ohlsson, {Triplet leptogenesis in
  left-right symmetric seesaw models}, \emph{JCAP}. {\bf 0801}, \penalty0 014,
  (2008).

\bibitem{Frigerio:2008ai}
M.~Frigerio, P.~Hosteins, S.~Lavignac, and A.~Romanino, {A new, direct link
  between the baryon asymmetry and neutrino masses}, \emph{Nucl. Phys.} {\bf
  B806}, \penalty0 84--102,  (2009).

\bibitem{Felipe:2013kk}
R.~Gonzalez~Felipe, F.~R. Joaquim, and H.~Serodio, {Flavoured CP asymmetries
  for type II seesaw leptogenesis}, \emph{Int. J. Mod. Phys.} {\bf A28},
  \penalty0 1350165,  (2013).

\bibitem{Sierra:2014tqa}
D.~Aristizabal~Sierra, M.~Dhen, and T.~Hambye, {Scalar triplet flavored
  leptogenesis: a systematic approach}, \emph{JCAP}. {\bf 1408}, \penalty0 003,
   (2014).

\bibitem{Lavignac:2015gpa}
S.~Lavignac and B.~Schmauch, {Flavour always matters in scalar triplet
  leptogenesis}, \emph{JHEP}. {\bf 05}, \penalty0 124,  (2015).

\bibitem{Magg:1980ut}
M.~Magg and C.~Wetterich, {Neutrino mass problem and gauge hierarchy},
  \emph{Phys. Lett.} {\bf B94}, \penalty0 61--64,  (1980).

\bibitem{Schechter:1980gr}
J.~Schechter and J.~W.~F. Valle, {Neutrino masses in $SU(2) \otimes U(1)$
  theories}, \emph{Phys. Rev.} {\bf D22}, \penalty0 2227,  (1980).

\bibitem{Lazarides:1980nt}
G.~Lazarides, Q.~Shafi, and C.~Wetterich, {Proton lifetime and fermion masses
  in an SO(10) model}, \emph{Nucl. Phys.} {\bf B181}, \penalty0 287--300,
  (1981).

\bibitem{Mohapatra:1980yp}
R.~N. Mohapatra and G.~Senjanovi\'c, {Neutrino masses and mixings in gauge
  models with spontaneous parity violation}, \emph{Phys. Rev.} {\bf D23},
  \penalty0 165,  (1981).

\bibitem{Bakshi:1962dv}
P.~M. Bakshi and K.~T. Mahanthappa, {Expectation Value Formalism in Quantum
  Field Theory. 1.}, \emph{J. Math. Phys.} {\bf 4}, \penalty0 1--11,  (1963).

\bibitem{Bakshi:1963bn}
P.~M. Bakshi and K.~T. Mahanthappa, {Expectation Value Formalism in Quantum
  Field Theory. 2.}, \emph{J. Math. Phys.} {\bf 4}, \penalty0 12--16,  (1963).

\bibitem{Cirigliano:2009yt}
V.~Cirigliano, C.~Lee, M.~J. Ramsey-Musolf, and S.~Tulin, {Flavored quantum
  Boltzmann equations}, \emph{Phys. Rev.} {\bf D81}, \penalty0 103503,  (2010).

\bibitem{Gonzalez-Garcia:2014bfa}
M.~C. Gonzalez-Garcia, M.~Maltoni, and T.~Schwetz, {Updated fit to three
  neutrino mixing: status of leptonic CP violation}, \emph{JHEP}. {\bf 11},
  \penalty0 052,  (2014).

\bibitem{Asaka:2005pn}
T.~Asaka and M.~Shaposhnikov, {The $\nu$MSM, dark matter and baryon asymmetry
  of the universe}, \emph{Phys. Lett.} {\bf B620}, \penalty0 17--26,  (2005).

\bibitem{Hambye:2016sby}
T.~Hambye and D.~Teresi, {Higgs Doublet Decay as the Origin of the Baryon
  Asymmetry}, \emph{Phys. Rev. Lett.} {\bf 117}\penalty0 (9), \penalty0 091801,
   (2016).

\bibitem{Hambye:2017elz}
T.~Hambye and D.~Teresi, {Baryogenesis from $L$-violating Higgs-doublet decay
  in the density-matrix formalism}, \emph{Phys. Rev.} {\bf D96}\penalty0 (1),
  \penalty0 015031,  (2017).

\bibitem{leptogenesis:A05}
E.~J. Chun, G.~Cveti\v{c}, P.~S.~B. Dev, M.~Drewes, C.~S. Fong, B.~Garbrecht,
  T.~Hambye, J.~Harz, P.~Hern\'andez, C.~S. Kim, E.~Molinaro, E.~Nardi,
  J.~Racker, N.~Rius, and J.~Zamora-Saa, {Probing leptogenesis},
  \emph{Int.~J.~Mod.~Phys.} {\bf A33}, \penalty0 1842005,  (2018).

\bibitem{Mohapatra:1974hk}
R.~N. Mohapatra and J.~C. Pati, {Left-right gauge symmetry and an
  {``}isoconjugate{''} model of CP violation}, \emph{Phys. Rev.} {\bf D11},
  \penalty0 566--571,  (1975).

\bibitem{Mohapatra:1974gc}
R.~N. Mohapatra and J.~C. Pati, {{``}Natural{''} left-right symmetry},
  \emph{Phys. Rev.} {\bf D11}, \penalty0 2558,  (1975).

\bibitem{Senjanovic:1975rk}
G.~Senjanovi\'c and R.~N. Mohapatra, {Exact left-right symmetry and spontaneous
  violation of parity}, \emph{Phys. Rev.} {\bf D12}, \penalty0 1502,  (1975).

\end{thebibliography}
\bibliographystyle{ws-rv-van-mod2}

\end{document}